\documentclass{article}
\usepackage{xcolor}
\definecolor{4DAF4A}{HTML}{4DAF4A}
\definecolor{984EA3}{HTML}{984EA3}
\definecolor{FF7F00}{HTML}{FF7F00}
\PassOptionsToPackage{numbers, compress}{natbib}
\usepackage[preprint]{neurips_2024}

\usepackage{neurips_2024}

\usepackage[utf8]{inputenc} 
\usepackage[T1]{fontenc}    
\usepackage{hyperref}       
\usepackage{url}            
\usepackage{booktabs}       
\usepackage{amsfonts}       
\usepackage{nicefrac}       
\usepackage{bm}             
\usepackage{xcolor}         
\usepackage{kotex}          
\usepackage{enumitem}       
\usepackage{amsmath}        
\usepackage{graphicx}
\usepackage[normalem]{ulem}  
\usepackage{multirow}    
\usepackage{booktabs}    
\usepackage{array}       
\usepackage{longtable}   
\usepackage{geometry}    
\usepackage{pdflscape}  
\usepackage{multirow}
\usepackage{float}
\usepackage{subcaption}
\usepackage{lmodern}
\usepackage{placeins}
\usepackage{microtype}

\definecolor{red}{RGB}{255, 0, 0}
\definecolor{greed}{RGB}{0, 255, 0}
\definecolor{blue}{RGB}{0, 0, 255}

\title{An Explainable Approach to Document-level Translation Evaluation with Topic Modeling}

\author{%
  Hyeokmin Lee\textsuperscript{1,2}\\
  \textsuperscript{1}{\scalebox{0.93}[1]{Intelligence and Interaction Research Center}}
  \\Korea Institute of Science and Technology
  \\
  \textsuperscript{2}Computer Science and Engineering
  \\Korea University
  \\Seongbuk-gu, Seoul 02792
  \\
  \texttt{hyeokminlee@kist.re.kr}
  \And
  Youngkyu Kim\textsuperscript{3}\\
  \textsuperscript{3}Center for Applied Scientific Computing 
  \\Lawrence Livermore National Laboratory
  \\Livermore, CA 94550
  \\
  \texttt{kim101@llnl.gov}
  \AND
  Byounghyun Yoo\textsuperscript{1,4,\thanks{Corresponding author.}}\\
  \textsuperscript{1}{\scalebox{0.93}[1]{Intelligence and Interaction Research Center}}
  \\Korea Institute of Science and Technology
  \\
  \textsuperscript{4}AI-Robotics, KIST School,
  \\Korea National University of Science and Technology,
  \\Seongbuk-gu, Seoul 02792
  \\
  \texttt{yoo@byoo.org}
}

\begin{document}

\maketitle

\begin{abstract}
  The advent of Neural Machine Translation (NMT) has expanded the scope of translation beyond isolated sentences, enabling context to be preserved across paragraphs and documents. However, current evaluation metrics largely remain restricted to the sentence level and typically depend on reference translations. Without references, existing metrics cannot provide a clear basis for their quality assessments. To address these limitations, we propose an evaluation framework that independently extracts and compares latent topic structures within source and translated texts. This framework utilises various topic modelling techniques, including LSA, LDA and BERTopic, to achieve this. Our methodology captures statistical frequency information and semantic context, providing a comprehensive evaluation of the entire document. It aligns key topic tokens from each language via a bilingual dictionary and quantifies thematic consistency through cosine similarity. This allows us to evaluate how faithfully the translation maintains the thematic integrity of the source text, even in the absence of reference translations. To this end, we used a large-scale dataset of 9.38 million Korean–English sentence pairs from AI-Hub, which includes pre-evaluated BLEU scores. We also calculated CometKiwi, a state-of-the-art, reference-free metric for this dataset, in order to conduct a comparative analysis with our proposed, topic-based framework. Through this analysis, we confirmed that, unlike existing metrics, our framework evaluates the differentiated attribute of document-level thematic units. Furthermore, visualising the key tokens that underpin the quantitative evaluation score provides clear insight into translation quality. Consequently, this study contributes to effectively complementing the existing translation evaluation system by proposing a new metric that intuitively identifies whether the document's theme has been preserved.

\end{abstract}

\section{Introduction}
Machine translation (MT) has made significant progress in recent years, primarily due to the introduction of neural machine translation (NMT) models \cite{bahdanau2014neural,vaswani2017attention}. These models have demonstrated exceptional ability in capturing complex language patterns and generating high-quality translations. In certain domains, NMT has approached or even surpassed human-level performance. For instance, Hassan et al. (2018) reported that NMT systems performed as well as or better than professional human translators in Chinese-English news translation tasks\cite{hassan2018achieving}.

NMT has been advancing beyond sentence-level translation to document-level translation, with recent studies focusing on enhancing translation consistency and cohesion by incorporating document-level context. Miculicich et al. (2018) proposed a model that captures document-level context in a structured manner by introducing hierarchical attention networks\cite{miculicich2018document}. This model improved BLEU scores by 0.92-1.80 points compared to the existing NMT baseline in English-German translation. Sun et al. (2020) experimented with directly inputting entire documents into the model (Document-to-document translation) and demonstrated that document-level NMT can work successfully through multi-resolution training techniques\cite{sun2020rethinking}. Their approach improved performance without modifying the model architecture or adding extra parameters, demonstrating its superiority across various evaluation metrics. In addition, various studies have been conducted to improve document-level machine translation, including memory-augmented translation \cite{zhang2022smdt}, lexical consistency modeling \cite{kang2021enhancing}, graph-based context integration \cite{chen2022effective}, and pre-trained language model fusion \cite{wang2023document}.

However, despite notable advances in document-level machine translation, most widely used evaluation metrics remain anchored at the sentence level, underscoring a gap between current practices and emerging translation capabilities. Traditional sentence-level metrics—such as BLEU \cite{papineni2002bleu} or METEOR \cite{banerjee2005meteor}—primarily rely on $N$-gram overlaps and synonym mappings at the sentence scale, thus limiting their ability to measure topic continuity, pronoun and tense consistency, or contextual flow across an entire text. In an effort to address these shortcomings, researchers have proposed various strategies—such as treating the entire document as one continuous sequence for BLEU calculation \cite{soricut2010trustrank}, aggregating sentence-level BLEU to approximate document-level performance \cite{laubli2018has}, or computing sentence-level METEOR and TER scores and using their statistical summaries as input features to a regression model that predicts document-level translation quality \cite{scarton2014document}. Yet, these approaches consistently struggle to capture cross-sentence contextual errors and long-range discourse structures, and they are often heavily dependent on reference translations, making them vulnerable to inaccuracies or stylistic mismatches in the reference text.

In light of these challenges, learned, advanced metrics have garnered sustained attention as viable alternatives for capturing contextually intricate aspects of translation quality and reducing reliance on reference translations. Consequently, recent research on MT evaluation has introduced various learned metrics built upon large pretrained language models, such as BLEURT \cite{sellam2020bleurt}, BERTScore \cite{zhang2019bertscore}, and GPT-based evaluators \cite{kocmi2023large}. Among these, COMET \cite{rei2020comet} has gained particular prominence by combining multilingual Transformer encoders (e.g., XLM-R) with human evaluation data to produce robust quality scores. A defining characteristic of COMET is that it utilizes both the source text and a reference translation, thereby integrating semantic and contextual cues from the source to more precisely evaluate translation fidelity. In fact, multiple studies have reported higher correlation with human judgments when source-side information is incorporated, as compared to metrics relying solely on references, and the WMT21 Metrics Shared Task also recognized COMET for its high evaluation accuracy \cite{freitag2021results}. However, although COMET partly alleviates reliance on reference translations, it still requires a reference to produce evaluation scores, which remains a persistent challenge shared by most evaluation metrics.

To enable translation evaluation in the absence of reference translations, CometKiwi, a reference-free extension of the original COMET metric, was recently proposed \cite{rei2021references,rei2022comet}. While the original COMET jointly encodes the source sentence, hypothesis (translated sentence), and reference translation through multilingual Transformer encoders, subsequently estimating translation quality via a regression head trained on human evaluation data, CometKiwi directly models semantic similarity between embeddings of the source sentence and the translation hypothesis without any references. This makes CometKiwi particularly practical in low-resource or real-world settings where obtaining reference translations is difficult. However, because CometKiwi does not leverage reference data, prior research has indicated that it struggles to capture detailed contextual structures, such as discourse-level coherence or long-range dependencies. Furthermore, similar to most neural-based metrics, CometKiwi functions as a black box—providing numerical quality scores without explicit explanations about how specific tokens or phrases contribute to those scores. To address this interpretability issue, X-COMET was recently introduced, employing visualization methods such as saliency maps and token-level attribution to explicitly highlight and explain the contributions of each input token to the final evaluation score \cite{guerreiro2024xcomet}. These visualization techniques offer transparency by directly showing users the tokens influencing model decisions, thus enabling interpretable evaluation results. Nonetheless, X-COMET still fundamentally relies on reference translations, limiting its direct application in reference-free evaluation scenarios. Considering this evolution from reference-based evaluation to reference-free evaluation, and recently toward providing explainability through X-COMET, there is now a clear motivation for developing a new evaluation paradigm capable of simultaneously explaining translation quality assessments without requiring reference translations, as current methods either lack interpretability or remain dependent on references.

In the field of MT, topic modeling has been leveraged as a key strategy to enhance domain adaptability such as the ability to handle specialized subject matter and inter-sentential coherence by analyzing word usage patterns and frequencies to automatically identify latent topics in text. For example, shen et al. (2023) LSA on a large-scale parallel corpus to extract domain-specific lexical knowledge, which they then organized into a specialized dictionary or domain-specific translation table to integrate into the translation system, thereby improving translation accuracy for specialized documents \cite{shen2023research}. Meanwhile, Su et al. (2015) proposed a Context-Aware Topic Model based on LDA, incorporating topic distributions derived at both the document and sentence levels as features in the decoding process of a hierarchical statistical MT engine, thus accounting for broader contextual information and increasing the accuracy of lexical selection \cite{su2015context}. Furthermore, Kuang et al. (2017) implemented LDA-derived topic information within a neural MT model via dynamic and topic-aware caches, allowing the decoder to reference this stored information in real time so as to maintain and refine previously inferred latent topics from preceding contexts—effectively reinforcing thematic consistency across consecutive sentences \cite{kuang2017modeling}. Although topic modeling shows strong potential to improve domain suitability and contextual awareness throughout the translation process, its direct application to translation evaluation from an explainability perspective remains limited.

In summary, most widely used metrics for machine translation evaluation share two fundamental limitations when applied at the document level. \textit{First}, these metrics heavily rely on reference translations, making the evaluation susceptible to stylistic variations or subjective biases present in the references. \textit{Second}, even recent reference-free approaches often operate as black-box models, providing quality scores without offering any clear rationale or explainability, which limits their transparency and usability in practical settings.

To simultaneously overcome these limitations, this paper is the first attempt to propose a topic-based approach for document-level translation evaluation that provides transparent and interpretable evaluation results without relying on reference translations and with minimal human intervention. Specifically, we collected approximately 9 million Korean-English parallel sentence pairs from the AI-Hub public data portal as of December 2023, each pair accompanied by precomputed BLEU scores. Using this dataset, we computed reference-free CometKiwi scores and systematically compared them with scores generated by three topic modeling methods as Latent Semantic Analysis (LSA), Latent Dirichlet Allocation (LDA), and BERTopic—applied within our proposed approach. Our evaluation framework independently applies topic modeling to the Korean source texts and their corresponding English translations to extract the most representative topics and key tokens for each document. Afterward, it aligns the tokens extracted from both languages based on a bilingual dictionary, and translation quality scores are calculated by comparing the cosine similarities of the tokens that most significantly contribute to the identified topics. Furthermore, our framework explicitly provides the key tokens from both Korean and English texts as evidence supporting the calculated scores, thus enabling clear interpretability of the evaluation outcomes. Consequently, this allows not only the interpretation of topic-based scores but also facilitates explanations for existing metrics such as BLEU and CometKiwi, thereby enhancing transparency and reliability in document-level translation evaluation.

The paper is structured as follows: Section~\ref{sec:materials} describes the data collection process and dataset characteristics. Section~\ref{sec:Methodology} introduces evaluation metrics, including the reference-based BLEU metric, the pretrained model-based CometKiwi metric, and the topic modeling techniques—LSA, LDA, and BERTopic—utilized in our study. Section~\ref{sec:Framework} proposes our evaluation framework, detailing methods for extracting representative topics and tokens, aligning tokens across languages, and computing translation quality scores via token similarity comparisons. Section~\ref{sec:Results} presents experimental results and explicitly analyzes how specific topic distributions and tokens contribute to translation quality scores. Finally, Section~\ref{sec:Conclusions} concludes by summarizing the findings and discussing limitations and future research directions.

\section{Materials}\label{sec:materials}
\subsection{Parallel Translation Data Sets}
Data collection was conducted from AI-Hub, Korea’s national platform that builds and continuously distributes public AI datasets while integrating key AI infrastructure, including datasets, software and APIs, algorithms, and computing resources needed to develop AI technologies, products, and services\cite{aihubportal}. The platform releases resources across multiple areas, including image recognition, machine reading, machine translation, and speech recognition, thereby supplying high-quality, large-scale data to Korea’s AI ecosystem spanning SMEs, research institutes, and individual practitioners. For this study, we searched the portal with the keyword "parallel corpus", selected only datasets that explicitly include Korean to English (KO to EN) sentence alignments, and downloaded all eligible releases available up to and including 31~Dec~2024.

To ensure consistency of the collected data, we selected datasets that provide a BLEU score computed with uniform $N$-gram weights and a 4-gram maximum, and we required that the BLEU score exceed 0.11 as a minimum quality threshold, following the Google Cloud Translate AutoML evaluation guide \cite{googlebleueval}. We included datasets with machine-generated references only when human inspection or post-editing was documented. For example, we excluded datasets whose reference translations were produced by machine translation when no human inspection or post-editing could be verified.

Independent studies have verified or characterized the quality of AI-Hub parallel corpora. For example, Park~et~al.\ conducted an empirical analysis of Korean public AI-Hub parallel corpora using Linguistic Inquiry and Word Count (LIWC) methodology and reported quality verification results \cite{park2021empirical}. Their study represents the first application of LIWC analysis to parallel corpora in the neural machine translation field and provided correlation analysis between LIWC features and MT performance. Additionally, they utilized AI-Hub Korean-English parallel data for various applications including translation quality assessment and machine translation system development. Related to quality estimation, synthetic datasets such as QUAK have been developed using automated methods that partially incorporate AI-Hub data, demonstrating the utility of these corpora for both supervised MT training and quality estimation research \cite{eo2022quak}.

\subsection{Dataset Characteristics}
The collected dataset is a Korean–English parallel corpus from the AI-Hub platform spanning 14 domains, comprising a total of 9,384,604 sentence pairs and covering the period from 2019 to 2024. The domains span a wide range of topics, including basic and technical sciences, humanities, Korean culture, daily life, social sciences, news, regulations, and conversational or spoken data. Here, a domain refers to an AI-Hub dataset category, such as Basic and Technical Sciences, Humanities, and Korean Culture. Each domain consists of multiple parallel translation rows, with each row containing a Korean text column and an aligned English translation column, which together form a one to one translation pair. In addition, each row contains a text unit of at least one sentence. Depending on how the dataset has been constructed, some rows are provided as document-like units, such as multi-sentence paragraphs or conversational segments. In this paper, we treat each row within a domain as a document and use it as the basic analysis unit for topic modeling.

Each domain-specific corpus is accompanied by metadata provided by AI-Hub, including the build year, download version date, the translation system used, and pre-computed BLEU scores. Notably, the BLEU values reported in this study were not recalculated by the authors. Because the original reference translations are not publicly available, recomputation is not possible; therefore, we directly adopt the BLEU scores reported in the official AI-Hub dataset documentation. Each dataset is openly available through the AI-Hub portal, and the download links are summarized in Appendix ~\ref{appendix:a}.  

\section{Dataset Characteristics}
Table~\ref{tab:Dataset} provides a summary of the AI-Hub corpora, including metadata and corpus size. The corpora were produced using four neural machine translation architectures, which differ by release period and domain. Specifically, Google Neural Machine Translation (GNMT) was used for several early datasets released in 2019, including Conversation, Korean culture, Municipal Websites, News, Regulations, and Spoken corpora, totaling 1,953,147 pairs with an average BLEU score of 0.3392. Transformer-based systems contributed to corpora such as Basic science, Humanities, Social science, and \textit{Broadcast content}, yielding 3,381,295 pairs with an average BLEU of 0.4982. OpenNMT was applied to Daily life and Technical science2, generating 2,700,162 pairs with the highest reported BLEU average of 0.7317. Finally, Google/AWS cloud-based AI models were used for the Professional Field dataset, comprising 1,350,000 pairs with a BLEU score of 0.5166.

All datasets were collected following standardized quality control procedures established by AI-Hub. BLEU scores were consistently computed using a 4-gram maximum configuration and a minimum threshold of 0.11, in line with evaluation guidelines. Furthermore, reference translations underwent human inspection or documented post-editing to ensure consistency and reliability across the corpora.

\setlength{\tabcolsep}{3.5pt}
\renewcommand{\arraystretch}{1.15}
\newcommand{\thickhline}{\noalign{\hrule height 1.0pt}}
\newcommand{\tightrow}{\\[-0.5ex]\hline}

\small
\begin{longtable}{
  >{\raggedright\arraybackslash}m{0.13\textwidth}
  >{\centering\arraybackslash}m{0.07\textwidth}
  >{\centering\arraybackslash}m{0.12\textwidth}
  >{\centering\arraybackslash}m{0.15\textwidth}
  >{\centering\arraybackslash}m{0.08\textwidth}
  >{\centering\arraybackslash}m{0.23\textwidth}
  >{\raggedleft\arraybackslash}m{0.10\textwidth}
}
\caption{Parallel corpora by domain}\label{tab:Dataset}\\
\thickhline
Domain & Build Year & Download Version Date & Translation Model & BLEU Score & Topic Distribution & Parallel Corpus Pairs \\
\thickhline
\endfirsthead


\hline
\multicolumn{7}{r}{Continued on next page} \\
\endfoot

\thickhline
\endlastfoot

Basic science      & 2022 & 2023-11-10 & Transformer & 0.3662 & Mathematics, Physics, Chemistry, Earth Science, Biology & 104,416 \tightrow
Broadcast content  & 2022 & 2023-10-30 & Transformer & 0.5136 & Documentary, Culture, Entertainment, Performance, Movie, Drama, Variety, Interview & 450,816 \tightrow
Conversation       & 2019 & 2019-12-31 & GNMT        & 0.2705 & Situation/Scenario-based dialogue set & 100,000 \tightrow
Daily life         & 2021 & 2022-07-29 & OpenNMT     & 0.7309 & Daily life, Overseas sales, Customer chat & 1,350,000 \tightrow
Humanities         & 2022 & 2024-04-08 & Transformer & 0.3254 & History/Archaeology, Politics/Administration, Economics/Management, Philosophy/Religion, Education & 119,587 \tightrow
Korean culture     & 2019 & 2019-12-31 & GNMT        & 0.2279 & Korean history, Cultural content & 100,646 \tightrow
Municipal Websites & 2019 & 2019-12-31 & GNMT        & 0.3970 & Government/Municipality websites, Publications & 450,816 \tightrow
News               & 2019 & 2019-12-31 & GNMT        & 0.3750 & News text & 801,387 \tightrow
Professional Field & 2020 & 2023-09-25 & Google/AWS Cloud Service AI Model & 0.5166 & Medical/Health, Finance/Stock, School notice, International sports, IT, Festival/Event, Court precedents, Local food/culture & 1,350,000 \tightrow
Regulations        & 2019 & 2019-12-31 & GNMT        & 0.4884 & Administrative rules, Local regulations & 100,298 \tightrow
Social science     & 2020 & 2022-01-24 & Transformer & 0.6600 & Law, Education, Economics, Culture, Tourism, Arts & 1,361,845 \tightrow
Spoken             & 2019 & 2019-12-31 & GNMT        & 0.2764 & Natural spoken language & 400,000 \tightrow
Technical science1 & 2020 & 2022-01-24 & Transformer & 0.6260 & ICT, Electrical, Electronics, Mechanical, Medical & 1,344,631 \tightrow
Technical science2 & 2021 & 2022-07-29 & OpenNMT     & 0.7325 & Technology, Economics, Politics, World, Climate & 1,350,162 \\
\end{longtable}
\normalsize

\begin{figure}[h!]
    \centering
    \begin{subfigure}{0.49\linewidth} 
        \centering
        \includegraphics[width=\linewidth]{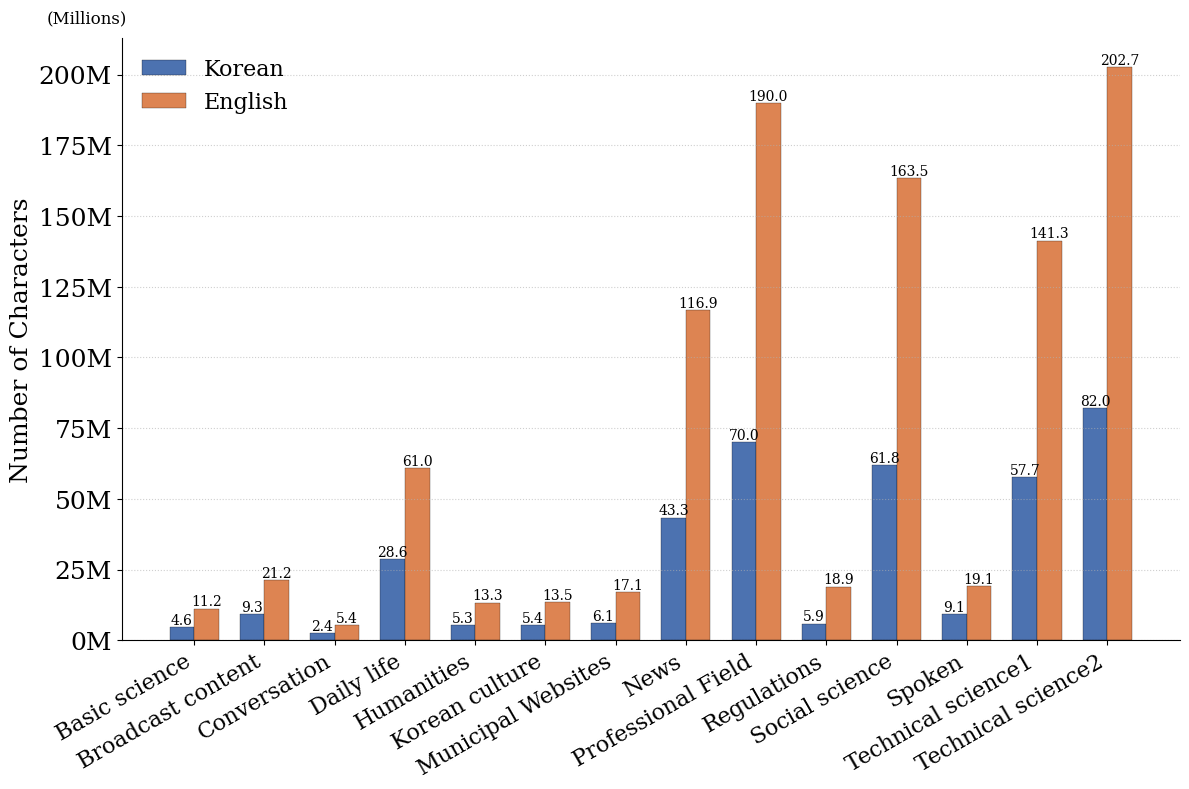}
        \caption{}
        \label{fig:total_characters}
    \end{subfigure}
    \hfill
    \begin{subfigure}{0.49\linewidth}
        \centering
        \includegraphics[width=\linewidth]{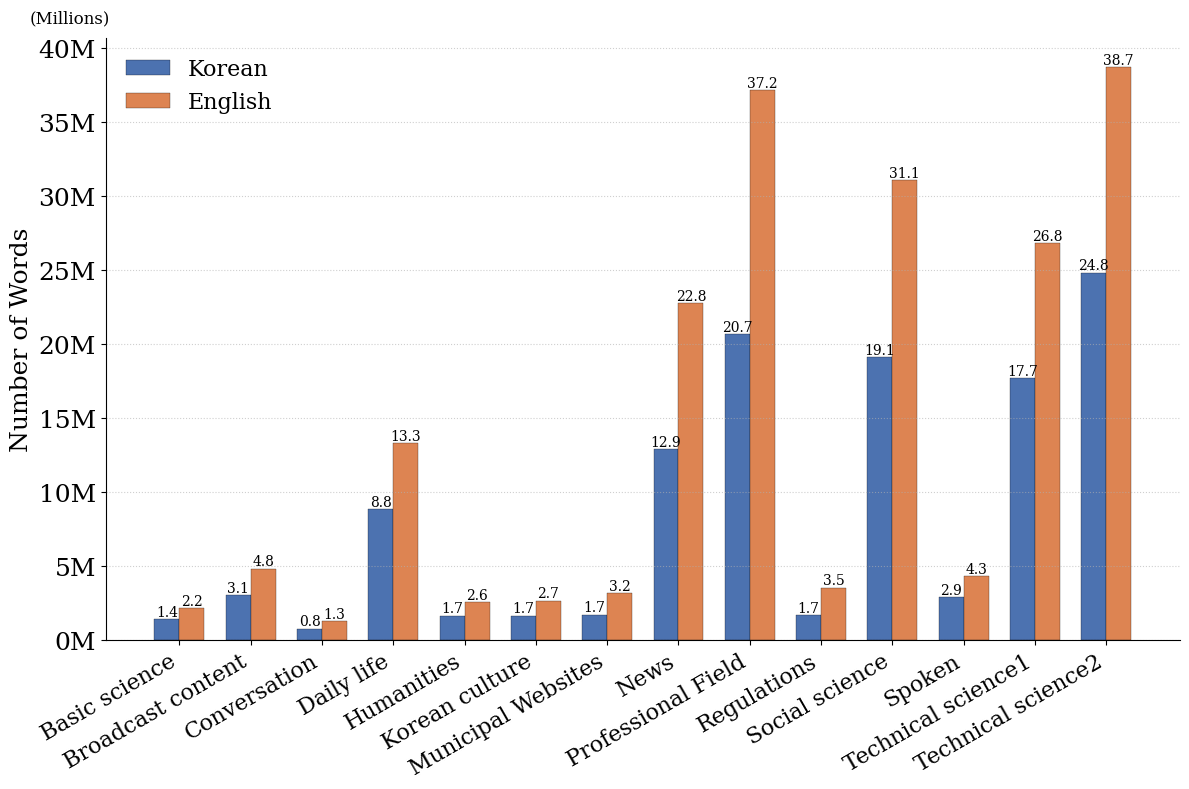}
        \caption{}
        \label{fig:total_words}
    \end{subfigure}
    \caption{Comparison of Korean and English corpus distribution across domains. (a) Total number of characters; (b) Total number of words;}
    \label{fig:char_word_comparison}
\end{figure}

Figure \ref{fig:char_word_comparison} reports the total number of characters and words across all documents in each of the 14 domains. Figure \ref{fig:char_word_comparison}(a) displays the total character counts per domain and shows that Technical science2 contains the largest volume of text with 202.7 million English characters and 82.0 million Korean characters, followed by Professional Field with 190.0 million English characters and 70.0 million Korean characters and then Social science with 163.5 million English characters and 61.8 million Korean characters. At the other extreme, the smallest domains are Conversation, which comprises 5.4 million English characters and 2.4 million Korean characters, and Korean culture, which comprises 5.4 million English characters and 2.6 million Korean characters. Figure \ref{fig:char_word_comparison}(b) shows the total word counts per domain and reveals the same ordering: Technical science2 leads with 38.7 million English words and 24.8 million Korean words, followed by Professional Field with 37.2 million English words and 20.7 million Korean words, then Social science with 31.1 million English words and 19.1 million Korean words.

Notably, when comparing domains by sentence pair count, English texts contain approximately one and a half to three times more characters and words than Korean texts. Daily life and Professional Field both include around 1.35 million sentence pairs, yet the Professional Field corpus contains more than three times as many English characters as Daily life—190 million versus 61 million. This disparity suggests that average sentence length differs substantially between domains. Technical and professional texts employ longer, information-dense sentences while conversational and daily-life texts consist of much shorter sentences. Such differences reflect the typological contrast between English, which encodes grammatical functions as separate words, and Korean, which conveys them through bound morphemes attached to stems \cite{comrie1989language, song2006korean}.

\begin{figure}[h!]
    \centering
    \begin{subfigure}{0.49\linewidth} 
        \centering
        \includegraphics[width=\linewidth]{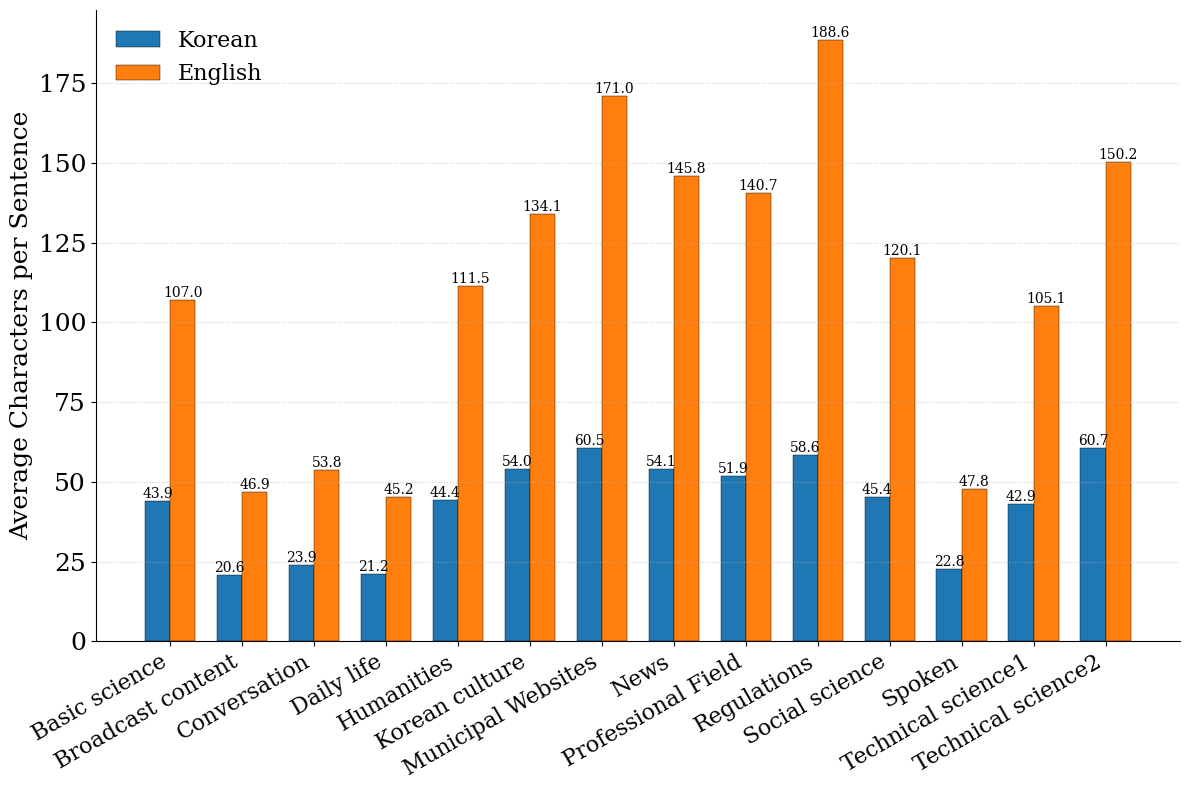}
        \caption{}
        \label{fig:avg_characters_sentence}
    \end{subfigure}
    \hfill
    \begin{subfigure}{0.49\linewidth}
        \centering
        \includegraphics[width=\linewidth]{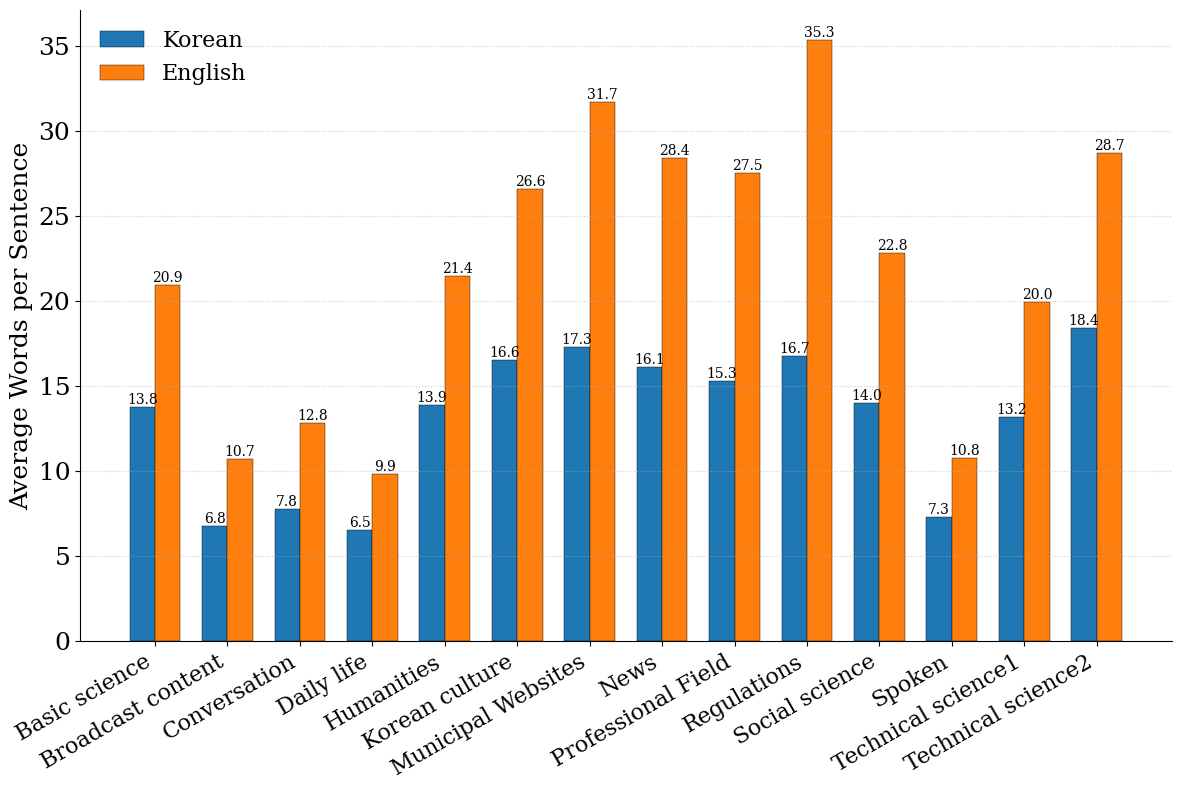}
        \caption{}
        \label{fig:avg_words_sentence}
    \end{subfigure}
    \caption{Comparison of sentence-level statistics between Korean and English across domains. (a) Average number of characters per sentence; (b) Average number of words per sentence.}
    \label{fig:avg_sentence_length}
\end{figure}

Figure \ref{fig:avg_sentence_length} provides insights into sentence-level characteristics across domains by examining the average number of characters and words per sentence. Figure \ref{fig:avg_sentence_length}(a) shows substantial variation in character density, with Social science exhibiting the highest average at 188.6 characters per English sentence and 58.6 characters per Korean sentence, followed by Professional Field with 171.0 English and 60.5 Korean characters. At the lower end, the Spoken domain records the shortest sentences with 47.6 English and 22.8 Korean characters, while Conversation averages 53.8 English and 23.9 Korean characters. Figure \ref{fig:avg_sentence_length}(b) presents a similar trend for word counts. Social science leads with 35.3 English and 16.7 Korean words per sentence, and Professional Field follows with 31.7 English and 17.3 Korean words. The shortest sentences appear in Spoken with 10.8 English and 7.3 Korean words and in Daily life with 9.9 English and 6.5 Korean words.

Sentence-level statistics, when compared with total counts of characters and words, reveal both consistency and divergence. The collected data shows that English sentences are systematically longer than Korean ones across all domains. It also shows that conversational and spoken texts consist of much shorter sentences, while technical and professional texts contain longer and more information-dense sentences.

\begin{figure}[h!]
    \centering
    \includegraphics[width=0.85\linewidth]{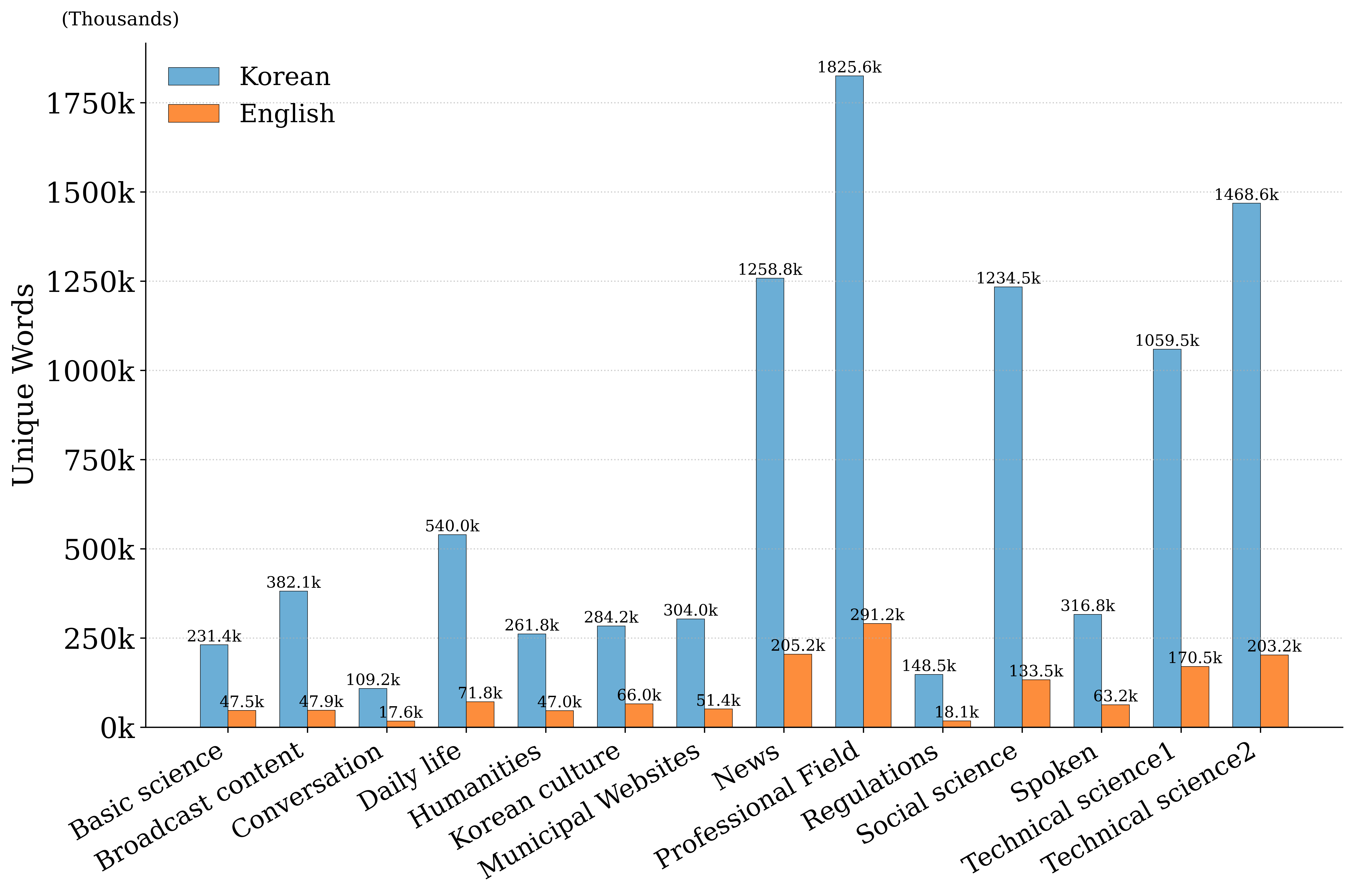}
    \caption{Comparison of unique words counts between Korean and English across domains.}
    \label{fig:unique_words}
\end{figure}

Figure \ref{fig:unique_words} shows that the number of unique words by domain diverges from both total character/word counts (Figure \ref{fig:char_word_comparison}) and sentence-level averages (Figure \ref{fig:avg_sentence_length}). Korean demonstrates substantially more unique tokens than English across nearly all domains—even where English has more total characters and words. The highest Korean unique word count appears in Professional Field with 1,825.6 thousand words, followed by Technical science2 with 1,468.6 thousand words, and News with 1,258.8 thousand words. In contrast, English shows much lower unique word counts, with Professional Field leading at only 291.2 thousand words, followed by Technical science1 with 170.5 thousand words, and Technical science2 with 203.2 thousand words. This pattern reflects how the counts were computed without removing inflectional endings or particles: Korean attaches case markers and verbal endings directly to stems (e.g.,haksaeng-i, haksaeng-eul, haksaeng-ege; handa, haetda, hagetta, hamyeonseo, haneunde), producing many distinct written forms, while English expresses much of this grammar with separate function words (e.g., the, to, will, do), so the vocabulary reuses a smaller set of word types even when overall volume is large. Consequently, unique-word statistics capture morphological productivity and stylistic variation rather than corpus size: conversational and spoken data can have short sentences yet still show high Korean type variety, while technical and professional domains may add English types mainly through specialized terminology.

\section{Methodology}\label{sec:Methodology}
This section introduces the evaluation metrics used in our study. We consider two families of metrics. First, a traditional reference based metric. BLEU measures clipped $N$-gram precision with a brevity penalty and is known to be sensitive to reference style and surface overlap. Second, learned metrics that can operate without references. COMETkiwi uses multilingual encoders to predict quality from source, hypothesis, and reference, while CometKiwi removes the reference and estimates quality from source and hypothesis only. These learned signals provide strong performance yet their internal reasoning is difficult to interpret. To complement them with document level semantics, we incorporate topic model signals. LSA projects a term document matrix into a low dimensional latent space and treats directions with large loadings as topics. LDA models each document as a mixture of topics and each topic as a distribution over words, yielding readable topic descriptors. BERTopic builds topics from contextual embeddings through clustering and derives labels using class based term weighting. Together, these signals capture surface overlap, learned semantic adequacy, and document level topical structure that are later analyzed for correlation and complementarity.

\subsection{Translation Quality Metrics}
\subsubsection{BLEU}
BLEU measures the overlap between a system hypothesis and one or more reference translations using modified $N$-gram precision up to a maximum order $N$ (typically $N=4$), combined with a brevity penalty to discourage overly short outputs \cite{papineni2002bleu}. Let $p_n$ denote the modified precision at order $n$, computed over hypothesis $n$-grams with clipping by the maximum count observed in any single reference translation:

\[
p_n \;=\; 
\frac{\sum_{g \in \mathrm{Hyp}_n} \mathrm{Count}_{\mathrm{clip}}(g)}
     {\sum_{g \in \mathrm{Hyp}_n} \mathrm{Count}(g)},
\qquad
\mathrm{Count}_{\mathrm{clip}}(g) \;=\;
\min\!\Big(\mathrm{Count}(g),\; \max_{r \in R}\mathrm{Count}_{r}(g)\Big).
\]

 Here, $g$ denotes an individual $n$-gram (token sequence) extracted from the hypothesis, $\mathrm{Hyp}_n$ is the multiset of such $n$-grams, $R$ is the set of reference translations, and $r \in R$ indexes a single reference translation. Let $\lvert h\rvert$ and $\lvert r\rvert$ denote the hypothesis length and the effective reference length, respectively. The brevity penalty is defined as follows:

\[
\mathrm{BP} \;=\;
\begin{cases}
1, & \lvert h\rvert > \lvert r\rvert,\\[2pt]
e^{\,1 - \lvert r\rvert / \lvert h\rvert}, & \lvert h\rvert \le \lvert r\rvert.
\end{cases}
\]

The effective reference length $\lvert r\rvert$ is taken to be the length of the reference translation closest to $\lvert h\rvert$, choosing the shorter one in case of ties. And, the BLEU score is the brevity-penalized geometric mean of the precisions:

\[
\mathrm{BLEU} = \mathrm{BP}\,\exp\!\left(\sum_{n=1}^{N} w_n \log p_n\right),\quad
\text{where } w_n \ge 0 \text{ and } \sum_{n=1}^{N} w_n = 1.
\]

The clipped count $\mathrm{Count}_{\mathrm{clip}}$ limits the credit for a hypothesis $n$-gram to its maximum frequency in any single reference translation, preventing repeated $n$-grams from receiving excessive credit. Intuitively, $p_n$ measures the fraction of hypothesis $N$-grams attested in the references, $\mathrm{BP}$ penalizes overly short hypotheses, and BLEU aggregates $p_1,\ldots,p_N$ via a geometric mean. While simple and reproducible, BLEU relies only on precision (not recall), is sensitive to reference choice and style, and does not directly capture synonymy, paraphrase, or discourse-level coherence.

\subsubsection{CometKiwi}
CometKiwi is a reference-free quality estimation model that predicts a scalar score from a source–hypothesis pair without using human references \cite{unbabel2020wmt20cometqeda}. Given a source sentence \(s\) and a system output \(h\), both are encoded with a multilingual Transformer (XLM-R) to obtain sentence-level embeddings by pooling token representations. A comparison feature is then constructed by concatenating the pooled vectors and simple elementwise operators:

\[
\mathbf{x} = \big[\, \hat{\mathbf{h}} \;;\; \hat{\mathbf{s}} \;;\; \lvert \hat{\mathbf{h}} - \hat{\mathbf{s}} \rvert \;;\; \hat{\mathbf{h}} \odot \hat{\mathbf{s}} \,\big],
\]

Here, $s$ denotes the source sentence, and $h$ denotes the hypothesis sentence, i.e., the system-generated translation output. In this expression, \(\hat{\mathbf{h}}\) and \(\hat{\mathbf{s}}\) are the pooled embeddings of \(h\) and \(s\), and \(\odot\) denotes elementwise multiplication. A small feed-forward regressor \(f(\cdot)\) maps \(\mathbf{x}\) to a quality score \(\hat{y} = f(\mathbf{x})\), trained against human direct-assessment targets \(y\) using mean squared error and, optionally, a pairwise ranking objective,

\[
\mathcal{L}_{\mathrm{MSE}} = \big(\hat{y} - y\big)^2,
\qquad
\mathcal{L}_{\mathrm{rank}} = \max \big(0,\, m - f(s,h^{+}) + f(s,h^{-}) \big).
\]

In the ranking objective, $h^{+}$ and $h^{-}$ denote two hypotheses selected for comparison among multiple candidate hypotheses for the same source sentence $s$, where $h^{+}$ is designated as the hypothesis that should receive a higher score than $h^{-}$. The scoring function $f(s,h)$ is a scalar score that represents the quality of the $(s,h)$ pair, and $m$ is a margin that specifies the minimum separation between the two scores. Accordingly, the ranking loss trains the model such that $f(s,h^{+})$ is at least $m$ larger than $f(s,h^{-})$. At inference time, CometKiwi outputs sentence-level scores that can be averaged to document or system level. Despite limited interpretability typical of neural metrics, it provides an effective reference-free complement to reference-based metrics such as BLEU while maintaining strong correlations with human evaluation.

\subsection{Topic Models}
Topic modeling is an unsupervised machine learning technique designed to discover latent semantic structures, referred to as topics, within a large collection of documents. By analyzing the co-occurrence patterns or semantic embeddings of words, these models map high-dimensional text data into a lower-dimensional topic space, where each document is represented as a mixture of underlying themes. In the context of this study, topic modeling serves as the core mechanism for extracting thematic representations independently from source and translated texts, thereby enabling a comparative analysis of document-level consistency without reliance on reference translations. To ensure a comprehensive evaluation that captures both statistical frequency and contextual nuances, we employ three distinct paradigms: algebraic (LSA), probabilistic (LDA), and embedding-based (BERTopic).

\subsubsection{Latent Semantic Analysis (LSA)}
LSA is a mathematical technique for uncovering hidden semantic relationships between terms and documents\cite{landauer1998introduction}. We represent each document as a Term Frequency-Inverse Document Frequency (TF--IDF) vector, where TF--IDF assigns a higher weight to terms that are frequent in a document but relatively rare across the corpus. Using these weights, we construct the TF--IDF matrix $A$. Specifically, for a token $t$ and a document $d$, we define the term frequency $\mathrm{tf}(t,d)$ as the number of occurrences of $t$ in $d$. We use the following inverse document frequency ($\mathrm{idf}$):

\begin{equation}
\mathrm{idf}(t)=\log\!\left(\frac{N_{\text{doc}}}{1+\mathrm{df}(t)}\right),
\end{equation}

where $N_{\text{doc}}$ is the number of documents and $\mathrm{df}(t)$ is the number of documents that contain $t$,  compute the TF--IDF weight as follows:

\begin{equation}
\mathrm{tfidf}(t,d)=\mathrm{tf}(t,d)\cdot \mathrm{idf}(t),
\end{equation}

LSA applies Singular Value Decomposition (SVD), a matrix factorization method that expresses $A$ as the product of three component matrices:

\begin{equation}
    A = U \Sigma V^T
\end{equation}

In this decomposition, \(U\) represents an orthogonal matrix whose columns correspond to term vectors in the reduced semantic space, \(\Sigma\) is a diagonal matrix containing singular values that indicate the importance of each latent dimension, and \(V^T\) is an orthogonal matrix whose rows correspond to document vectors in the same space.  
By keeping only the top \(k\) singular values in \(\Sigma\), LSA effectively reduces dimensionality, filters noise, and preserves the most relevant semantic structure. Through this process, it captures underlying thematic patterns and associations between words and documents that are not directly observable in raw frequency data.

\subsubsection{Latent Dirichlet Allocation (LDA)}

LDA is a generative probabilistic model in which each document is a mixture over \(K\) latent topics and each topic is a multinomial distribution over the vocabulary\cite{blei2003latent}. For document \(d\), a topic mixture \(\theta_d \sim \mathrm{Dirichlet}(\alpha)\) is drawn; for each position \(n\), a topic assignment \(z_{dn} \sim \mathrm{Multinomial}(\theta_d)\) is sampled; the word \(w_{dn}\) is then generated from the topic-specific distribution \(\phi_{z_{dn}}\), with \(\phi_k \sim \mathrm{Dirichlet}(\eta)\).
The joint distribution under hyperparameters \(\alpha,\eta\) factorizes as:

\[
p(W,Z,\Theta,\Phi \mid \alpha,\eta)
= \Big(\prod_{k=1}^{K} p(\phi_k \mid \eta)\Big)
  \Big(\prod_{d=1}^{D} p(\theta_d \mid \alpha)
     \prod_{n=1}^{N_d} p(z_{dn}\mid \theta_d)\,p(w_{dn}\mid \phi_{z_{dn}})\Big)
\]

Here, \(D\) denotes the number of documents in the domain, with document index \(d \in \{1,\ldots,D\}\). Document \(d\) contains \(N_d\) token positions, indexed by \(n \in \{1,\ldots,N_d\}\), and \(K\) denotes the number of latent topics. The observed token \(w_{dn}\) is the word token at position \(n\) in document \(d\), and the latent variable \(z_{dn} \in \{1,\ldots,K\}\) is the topic index assigned to that token. \(W\) denotes the collection of all observed tokens \(w_{dn}\) in the domain, \(Z\) the collection of all token-level topic assignments \(z_{dn}\), \(\Theta\) the collection of all document-level topic mixtures \(\theta_d\), and \(\Phi\) the collection of all topic-word distributions \(\phi_k\). The hyperparameters \(\alpha\) and \(\eta\) are Dirichlet prior parameters for \(\theta_d\) and \(\phi_k\), respectively, controlling the concentration (e.g., sparsity) of the corresponding distributions. Specifically, \(\alpha\) controls the prior concentration of document-level topic proportions \(\theta_d\), while \(\eta\) controls the prior concentration of topic-level word distributions \(\phi_k\).

This factorization mirrors the generative process. First, draw each topic-word distribution \(\phi_k\). Second, draw each document-topic mixture \(\theta_d\). Third, for each token position \(n\) in document \(d\), draw a topic label \(z_{dn}\) and then generate the observed token \(w_{dn}\) from \(\phi_{z_{dn}}\). Here, to infer the latent topic structures (\(Z, \Theta, \Phi\)) from the observed documents (\(W\)), we employ a variational inference approach that maximizes the Evidence Lower Bound (ELBO) as follows:

\[
\log p(W\mid \alpha,\eta)
\;\ge\;
\mathbb{E}_q\!\big[\log p(W,Z,\Theta,\Phi\mid \alpha,\eta)\big]
- \mathbb{E}_q\!\big[\log q(Z,\Theta,\Phi)\big],
\]

In the above inequality, the first term on the right-hand side represents the expected log joint probability under the variational distribution \(q\), serving as a goodness-of-fit metric that indicates how precisely the model explains the structural relationship between the observed data and latent variables. Here, \(q(Z,\Theta,\Phi)\) denotes a variational distribution (approximate posterior) used to approximate the true posterior \(p(Z,\Theta,\Phi \mid W,\alpha,\eta)\). Conversely, the second term corresponds to the entropy of the variational distribution, which acts as a regularization term to prevent the inferred latent distribution from converging too deterministically and to encourage the preservation of uncertainty. Consequently, maximiullback--Leibler (KL) divergence between the approximate distributizing the ELBO is mathematically equivalent to minimizing the Kon \(q\) and the true posterior distribution \(p\). This process seeks an optimal balance point that maximizes explanatory power for the data while simultaneously securing the model's generalization performance.

This optimization process is performed via variational inference or collapsed Gibbs sampling, estimating interpretable per-document topic mixtures \(\theta_d\) and per-topic word probabilities \(\phi_k\). In this study, we configured the prior probability parameters \(\alpha\) and \(\eta\) to be automatically adjusted based on the statistical characteristics of the corpus. This contributes to optimizing model sparsity and enhancing interpretability by naturally excluding unnecessary topics that fail to reflect the data distribution or lack information content.

\subsubsection{BERTopic}
BERTopic operates as a modular framework that integrates document embedding, dimensionality reduction, clustering, and topic representation extraction into a unified pipeline\cite{grootendorst2022bertopic}. In this study, we optimized each stage to derive robust topic structures. First, in the embedding stage, we employed language-specific pre-trained encoders \( f_{\theta} \) to precisely capture the semantic nuances of each language, rather than relying on a generic multilingual model. Specifically, Korean texts were mapped into high-dimensional embedding space using \texttt{klue/roberta-base} \cite{park2021klue}, while English texts utilized \texttt{sentence-transformers/all-MiniLM-L6-v2} \cite{reimers2019sentence}. To mitigate the curse of dimensionality inherent in these vectors, we applied the UMAP algorithm, which compresses the data while preserving its topological structure. Subsequently, we utilized HDBSCAN, a density-based clustering method, to form high-quality topic clusters by effectively filtering out noise. Finally, topic representation extraction, constitutes a method that completes the final topic representation by calculating class-based TF-IDF scores to select the most representative keywords for each cluster $k$.

\[
\mathrm{c\mbox{-}tfidf}(t,k)
\;=\;
\frac{\mathrm{tf}(t,k)}{\sum_{t'} \mathrm{tf}(t',k)} \cdot
\log\!\Big(\frac{N}{\mathrm{df}(t)}\Big),
\]

Here, \( \mathrm{tf}(t, k) \) denotes the frequency of term \( t \) within cluster \( k \), \( N \) represents the total number of documents, and \( \mathrm{df}(t) \) is the document frequency of term \( t \) across the entire corpus. This metric treats the entire cluster as a single class to calculate weights. Consequently, by assigning higher weights to terms that are frequent within a specific cluster but rare in the global corpus, it generates descriptors that clearly define the thematic nature of each topic.

This pipeline leverages context-aware embeddings to effectively capture semantic nuances and synonym relationships often missed by traditional frequency-based models. Furthermore, by quantitatively calculating the importance of terms within each topic via class-based TF-IDF, it maximizes the interpretability of the results. Consequently, for tasks involving the comparative analysis of thematic structures across heterogeneous languages, this framework offers a distinct advantage in elucidating deep semantic connectivity beyond simple lexical matching.

\section{Framework}\label{sec:Framework}
Our proposed framework constitutes the core contribution of this study as it establishes a systematic pipeline for explainable document-level translation evaluation. Designed to move beyond existing sentence-level metrics, this framework leverages topic modeling to capture thematic consistency across both source and translated texts. The overall process consists of three major stages including Tokenization, Topic Extraction, and Matching Topics.

In the first stage, tokenization, we utilize language-specific analyzers as a preprocessing step to selectively extract only nouns—the lexical units that best encapsulate the core semantics of a document—thereby minimizing analytical noise. The second stage, topic extraction, employs LSA, LDA, and BERTopic to independently identify the single dominant topic that best explains each document within the source and translated corpora. In the final stage, Matching Topics, we utilize a bilingual dictionary to map Korean topic keywords into English. We then quantify the thematic correlation between the source and translated texts by calculating the cosine similarity between topic vectors constructed from the word importance weights derived by each model.

Through this pipeline, our framework goes beyond simple score generation to provide interpretable evidence in the form of topic–token pairs. This offers a transparent explanation of how the translation preserves or distorts thematic content, thereby complementing reference-free black-box neural metrics and traditional quantitative metrics that rely solely on numerical scores. The integration of such token-level evidence, thematic abstraction, and cross-lingual matching forms the foundation for explainable, reference-free translation evaluation at the document scale.

\subsection{Tokenization}\label{sec:Tokenization}
In this study, tokenization was performed using MeCab for Korean texts and spaCy for English texts, extracting only nouns from the tokenized output. MeCab is a dictionary-based morphological analyzer built on Conditional Random Fields (CRFs), originally developed for Japanese and adapted for Korean as MeCab-ko \cite{park2014konlpy}. It is particularly well-suited for Korean, an agglutinative language, because it precisely segments morphemes including stems, endings, and particles, enabling accurate extraction of nouns and compound nominal expressions from complex morphological structures. For English, spaCy provides a rule-based tokenizer with robust sentence segmentation and part-of-speech tagging capabilities, enabling consistent noun extraction from large-scale corpora \cite{vasiliev2020natural}.

Nouns are the lexical units that most strongly encode thematic and semantic content in documents, serving as the primary carriers of topic information in probabilistic topic models \cite{martin2015more}. In contrast, function words, particles, and inflectional morphemes contribute minimal semantic information while introducing data sparsity through morphological variation. By filtering out non-noun tokens, the signal-to-noise ratio is improved, concentrating the vocabulary on content-bearing units that define document themes. Empirical studies have demonstrated that noun-only corpora produce equivalent or improved topic semantic coherence compared to modeling full texts, with a 6\% improvement in observed coherence and an 8\% improvement in word intrusion detection \cite{savoy2013authorship}. Furthermore, noun-only approaches reduce model training time by decreasing vocabulary size without sacrificing interpretability. For cross-lingual analysis, this approach also mitigates morphological asymmetry between Korean and English: while English expresses grammatical markers as independent tokens, Korean incorporates them into word-internal morphology, making noun-centered comparison more balanced across languages.

\begin{figure}[h!]
    \centering
    \begin{subfigure}{0.48\linewidth} 
        \centering
        \includegraphics[width=\linewidth]{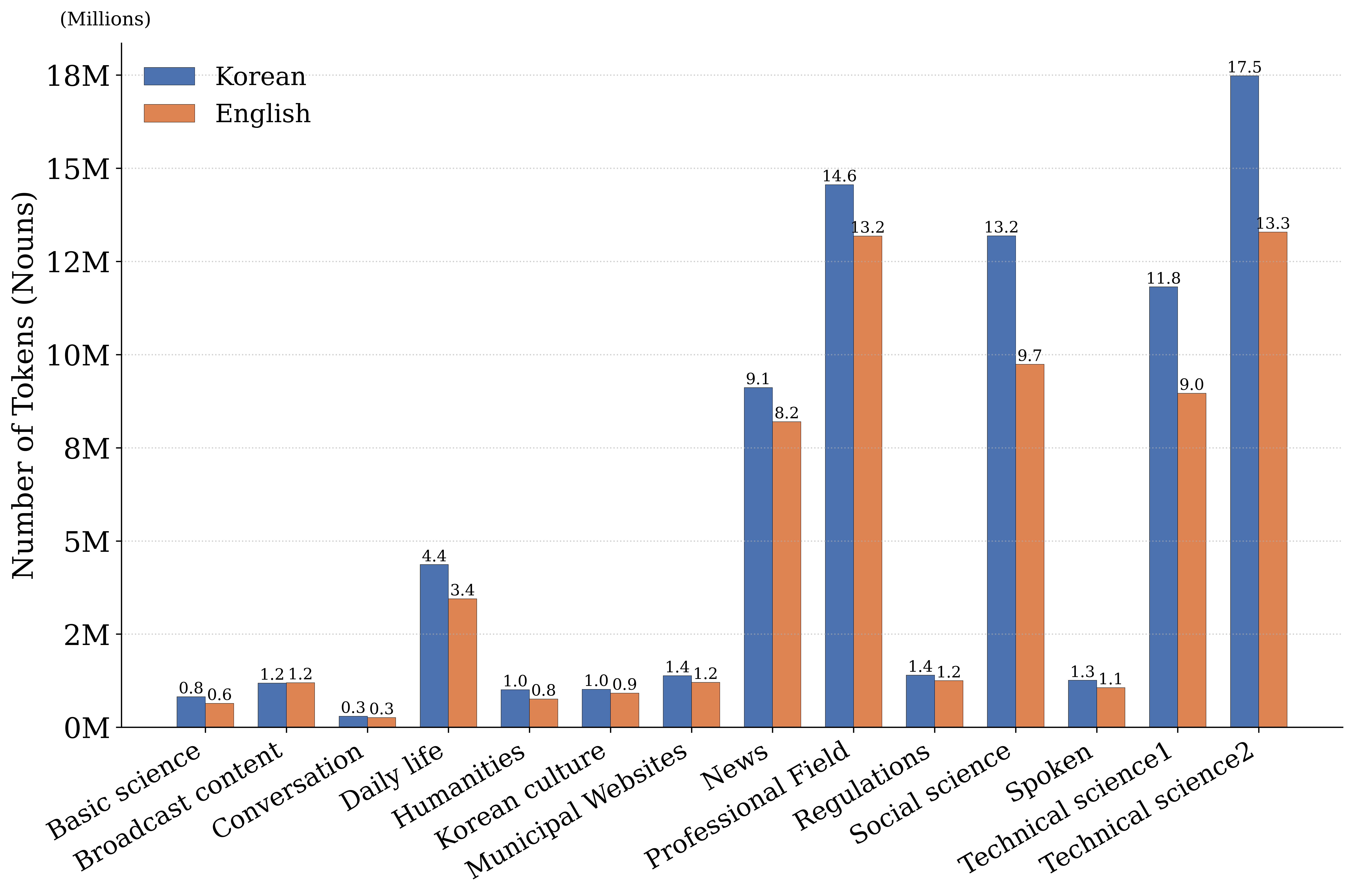}
        \caption{}
        \label{fig:total_tokens}
    \end{subfigure}
    \hfill
    \begin{subfigure}{0.48\linewidth}
        \centering
        \includegraphics[width=\linewidth]{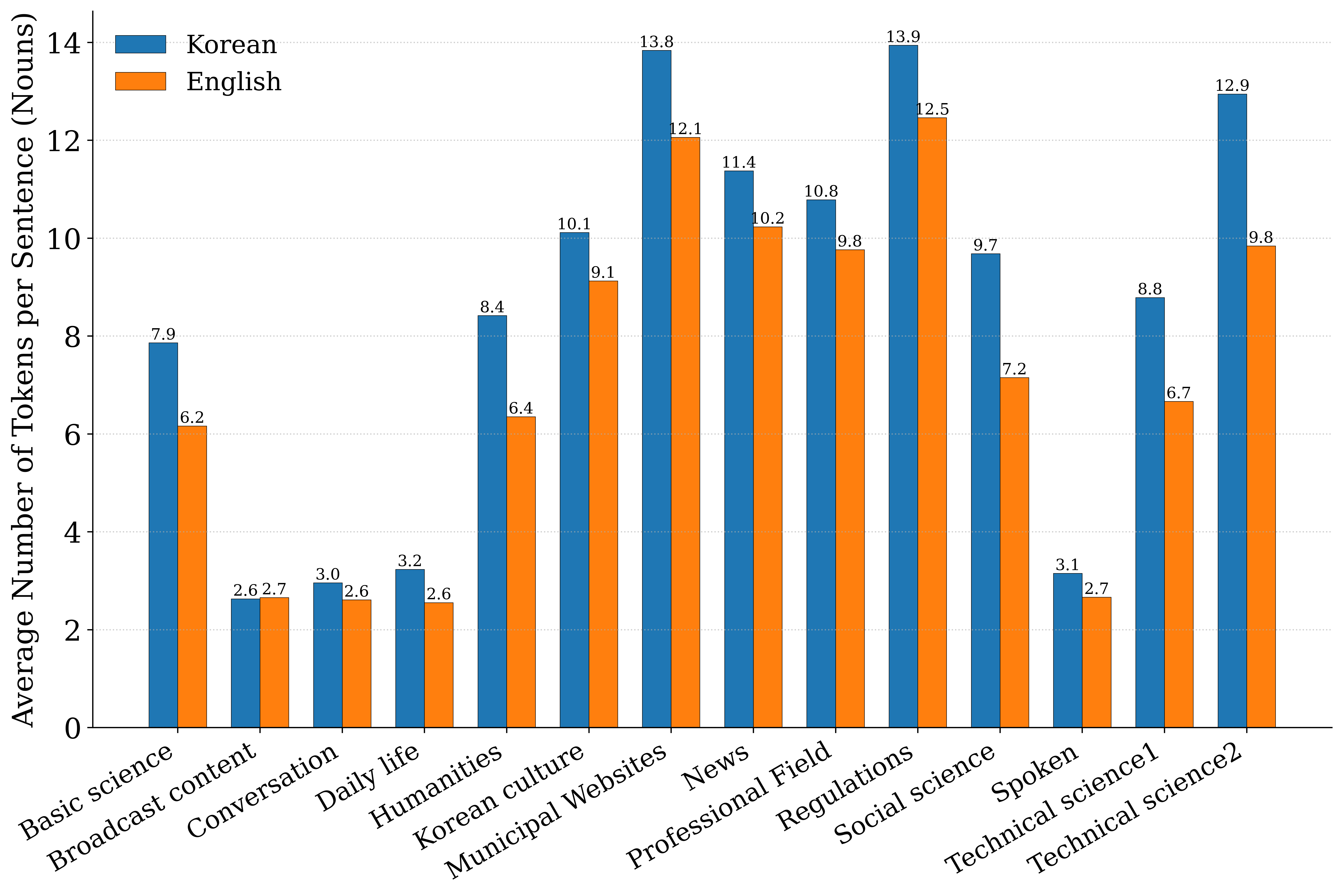}
        \caption{}
        \label{fig:avg_tokens}
    \end{subfigure}

    \vspace{0.8em} 

    \begin{subfigure}{0.48\linewidth}
        \centering
        \includegraphics[width=\linewidth]{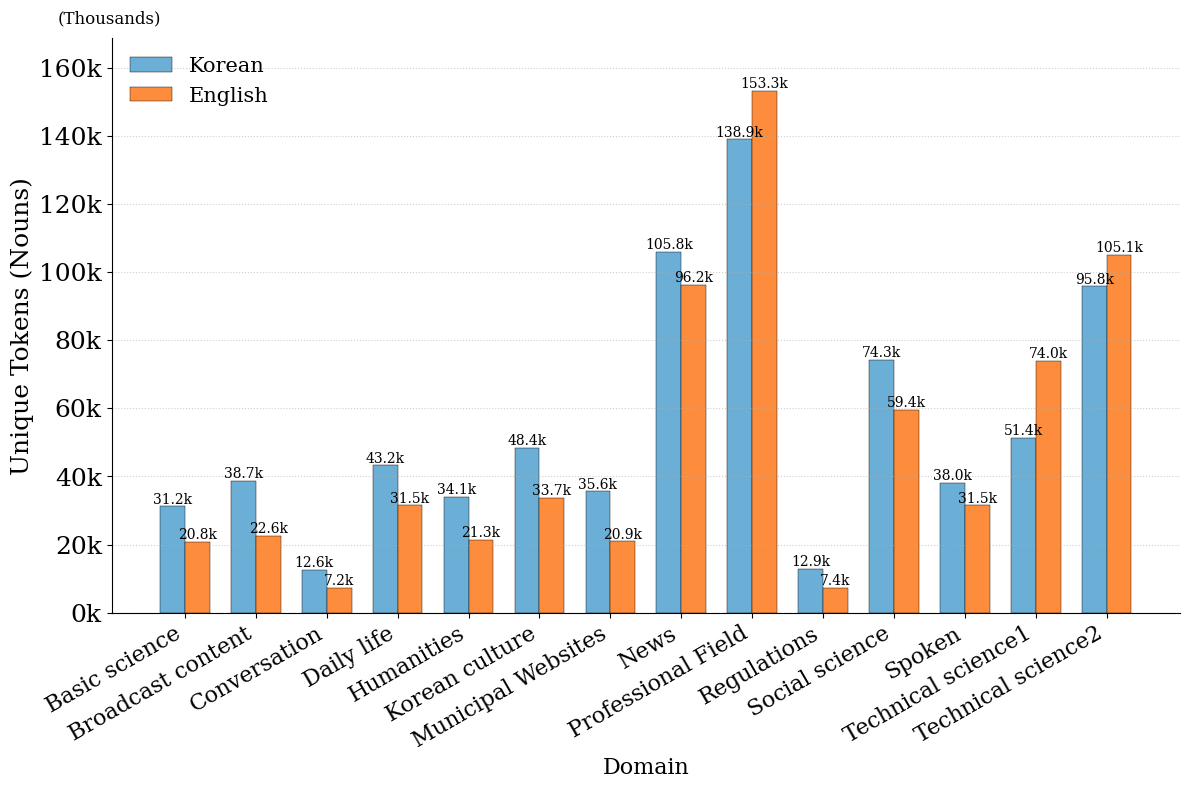}
        \caption{}
        \label{fig:unique_tokens}
    \end{subfigure}

    \caption{Comparison of Korean and English noun-token statistics across domains. (a) Number of Tokens; (b) Average Tokens per Sentence; (c) Unique Tokens by Domain;}
    \label{fig:noun_token_overview}
\end{figure}

Figure~\ref{fig:noun_token_overview} summarizes noun based token statistics across the fourteen domains. Figure~\ref{fig:noun_token_overview}\subref{fig:total_tokens} shows the total number of noun tokens and confirms that Korean contains more noun tokens than English in most domains. For example, the Technical science2 domain includes approximately 17.5 million Korean noun tokens and 13.3 million English noun tokens, and the Regulations domain includes about 14.6 million and 13.2 million noun tokens in Korean and English, respectively. Figure~\ref{fig:noun_token_overview}\subref{fig:avg_tokens} presents the average number of nouns per sentence and shows that Korean consistently exhibits higher values than English across all domains. In Municipal Websites and Regulations, Korean contains 13.8 and 13.9 nouns per sentence, whereas English contains 12.1 and 12.5, and \textit{conversation} and \textit{speech} oriented domains such as \textit{Broadcast content} and \textit{Spoken} contain only about 2.6 to 3.1 nouns per sentence and are therefore much shorter. Figure~\ref{fig:noun_token_overview}\subref{fig:unique_tokens} reports the number of noun types, that is, the number of distinct noun tokens, and shows that Korean has a richer noun vocabulary in most domains. In Daily life and Municipal Websites, for instance, Korean contains 43,200 and 48,400 distinct nouns, while English contains 31,500 and 35,600. In contrast, in Professional Field the Korean corpus contains 138,900 noun types and the English corpus contains 153,300, and in Technical science2 the Korean and English corpora contain 95,800 and 105,100 noun types, which indicates that in domains with a high proportion of specialized terminology the English side has a slightly larger number of distinct nouns.

In the preceding section~\ref{sec:materials}, the character and word level statistics showed that in most domains the English translations were substantially longer than the Korean originals. For example, the total number of words in Daily life was about 8.8 million in Korean and 13.3 million in English, and in Regulations about 20.7 million in Korean and 37.2 million in English, so that the English side was roughly one and a half to two times larger. The average number of words per sentence also showed the same tendency, with 13.8 and 20.9 words in \textit{Basic science} and 14.0 and 35.3 words in \textit{Social scienc}e for Korean and English, respectively. This can be interpreted as a consequence of grammatical function words such as articles, prepositions, pronouns, auxiliaries, adverbs, and adjectives being realized as independent tokens in English, which greatly increases total length. By contrast, in the noun based statistics in Figure~\ref{fig:noun_token_overview} Korean now has about ten to thirty percent more noun tokens than English in many domains, and the average number of nouns per sentence, as seen in Municipal Websites and Regulations, is more densely distributed on the Korean side. The number of distinct noun tokens is also higher for Korean in many domains, while English slightly surpasses Korean in specialized domains, yet the gap between the two languages is much smaller than at the level of all words. In other words, the length increase observed on the English side arises mainly from function words and grammatical markers rather than from content words, and when the representation is restricted to nouns, the density and distribution of thematic content conveyed by the two languages converge to far more comparable levels. These distributions quantitatively show how densely and how diversely nouns that primarily carry semantic information appear in each domain, and, on this basis, the subsequent topic modeling stage removes function words and morphological markers that are unnecessary for interpretation and focuses on content words so that the core content of documents can be identified more clearly.

\subsection{Topic Extraction}\label{sec:Topic_Extraction}

After noun-based tokenization, we extract topics from the Korean source documents and the English translated documents using LSA, LDA, and BERTopic for each domain. Within a domain, topic modeling is performed over its documents, and for each domain we then select a single representative topic that most strongly reflects the domain’s core theme. In LSA, we build a TF--IDF representation for each domain (i.e., using the collection of documents in that domain) and apply SVD. The representative topic is the latent axis with the largest contribution at the domain level. In LDA, we fit the model on the documents in each domain and obtain a topic distribution for each document, and the representative topic is the largest proportion at the domain level. In BERTopic, we cluster documents within each domain based on embeddings. We then choose the representative topic as the largest cluster, namely the topic assigned to the largest number of documents in that domain.

The representative topic is summarized by the top 20 tokens and their weights for interpretability and comparability, and we represent each topic with its top-word list and compare topic-to-topic correspondences across domains accordingly \cite{niekler2012matching}. In addition, from the perspective of quantifying cross-lingual topic similarity via vector-based measures, we follow prior work by treating topic representations as weighted token vectors and computing cosine similarity \cite{reber2019overcoming}. The meaning of token weights differs by model. In LSA, token weights are the term loadings within a topic component. In LDA, they are the conditional probabilities of words given a topic. In BERTopic, they are the c-TF-IDF scores used to rank words that characterize a topic.

\begin{figure}[h!]
    \centering
    \begin{subfigure}{0.48\linewidth}
        \centering
        \includegraphics[width=\linewidth]{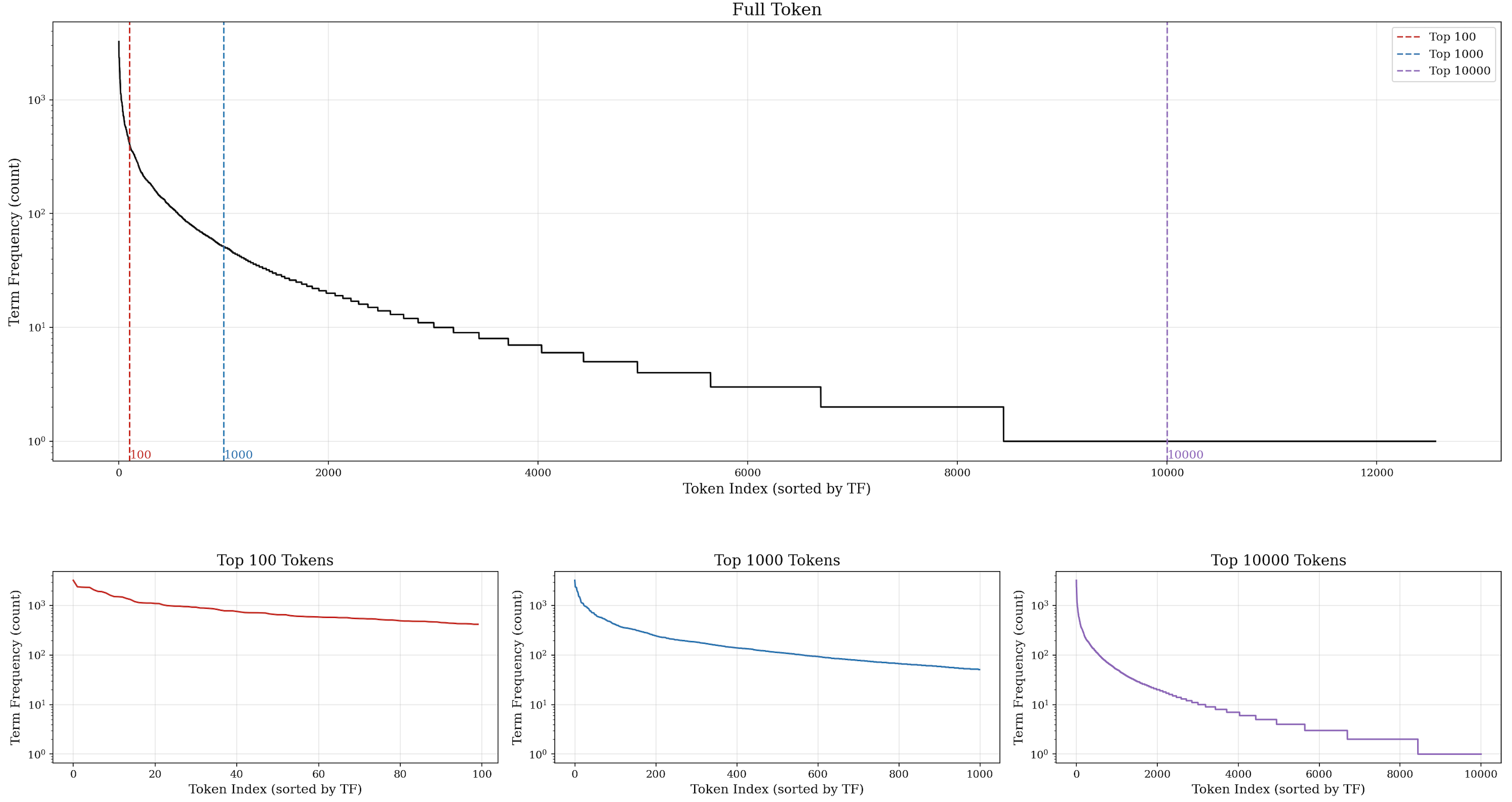}
        \caption{}
        \label{fig:maxfeat_conv_ko}
    \end{subfigure}
    \hfill
    \begin{subfigure}{0.48\linewidth}
        \centering
        \includegraphics[width=\linewidth]{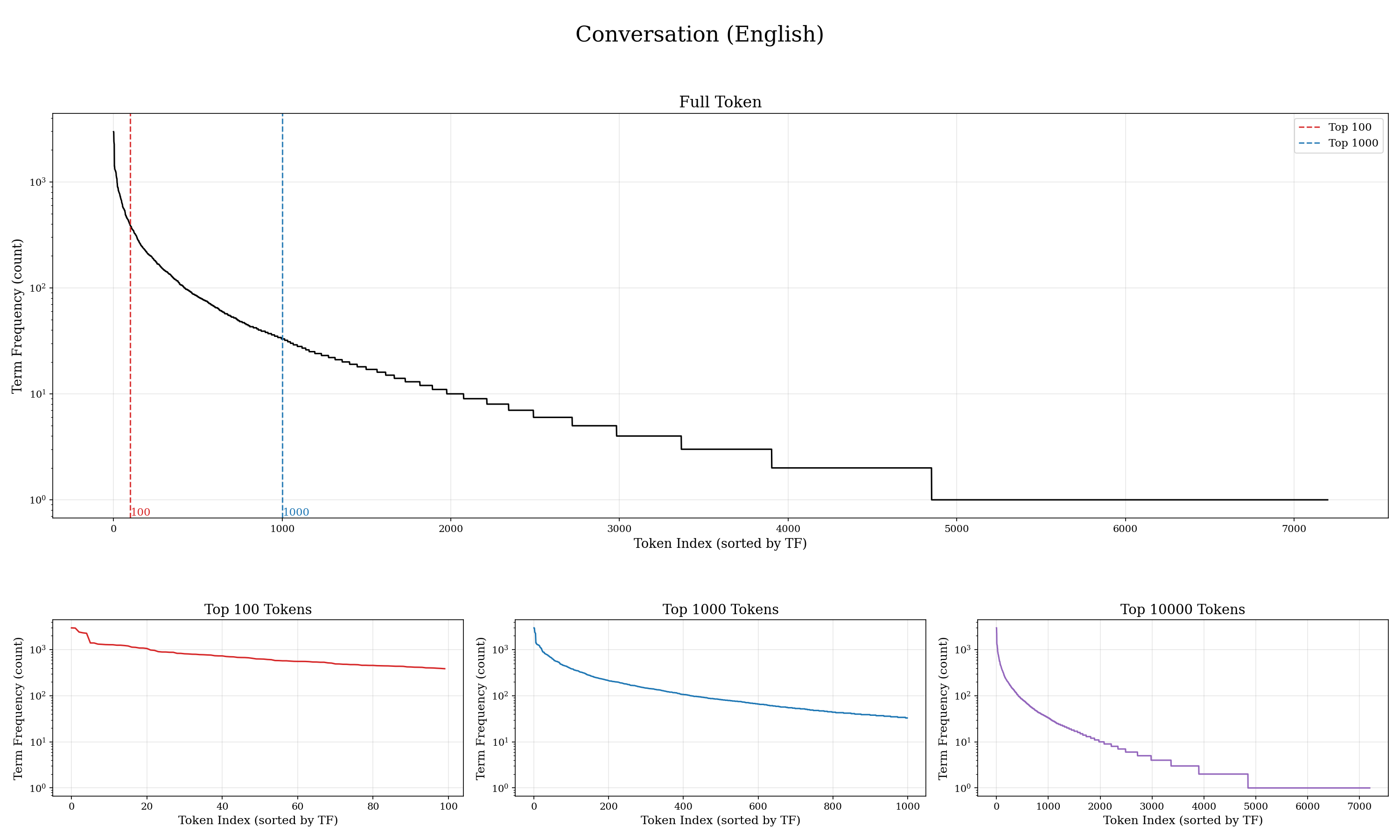}
        \caption{}
        \label{fig:maxfeat_conv_en}
    \end{subfigure}

    \caption{Example of term frequency distributions of noun tokens in the Conversation domain. (a) Korean corpus; (b) English corpus.}
    \label{fig:maxfeat_conversation}
\end{figure}

In LSA, we first select the top \(N\) noun tokens that define the TF--IDF vocabulary used to build the topic space. Figure~\ref{fig:maxfeat_conversation} illustrates this selection procedure for the Conversation domain. The upper panel shows the full TF curve of noun tokens, and the lower panels zoom into three ranges corresponding to the top 100, 1{,}000, and 10{,}000 nouns. If we were to use the entire curve, the vocabulary would contain extremely frequent tokens whose term frequencies are orders of magnitude larger than typical topical nouns, as well as extremely rare tokens that occur only once or twice. To avoid this imbalance, we define a minimum TF cutoff relative to the maximum TF in each domain and language. Let \(\mathrm{tf}_{\max}\) be the maximum raw term frequency among noun tokens in a given domain and language, and let \(a\) be the integer obtained by rounding down \(\log_{10}(\mathrm{tf}_{\max})\). We retain only noun tokens with \(\mathrm{tf} \ge 10^{a-2}\) and set (Add. A3) \texttt{max\_features} to the largest rank that satisfies this cutoff. (add.A3) For example, in Figure~\ref{fig:maxfeat_conv_ko}, when \(a=3\), we include noun tokens down to \(\mathrm{tf} \ge 10^{1}\). In practice, for each domain and language we sort all nouns by raw term frequency and inspect plots such as Figure~\ref{fig:maxfeat_conversation}. This cutoff keeps noun tokens that are used often enough to represent topical content, while discarding rarely used noun tokens that mostly contribute noise. The domain- and language-specific \texttt{max\_features} configurations used for LSA are documented in a separate Mendeley dataset\cite{lee2025mendeley}.

\begin{figure}[h!]
    \centering
    \includegraphics[width=0.9\linewidth]{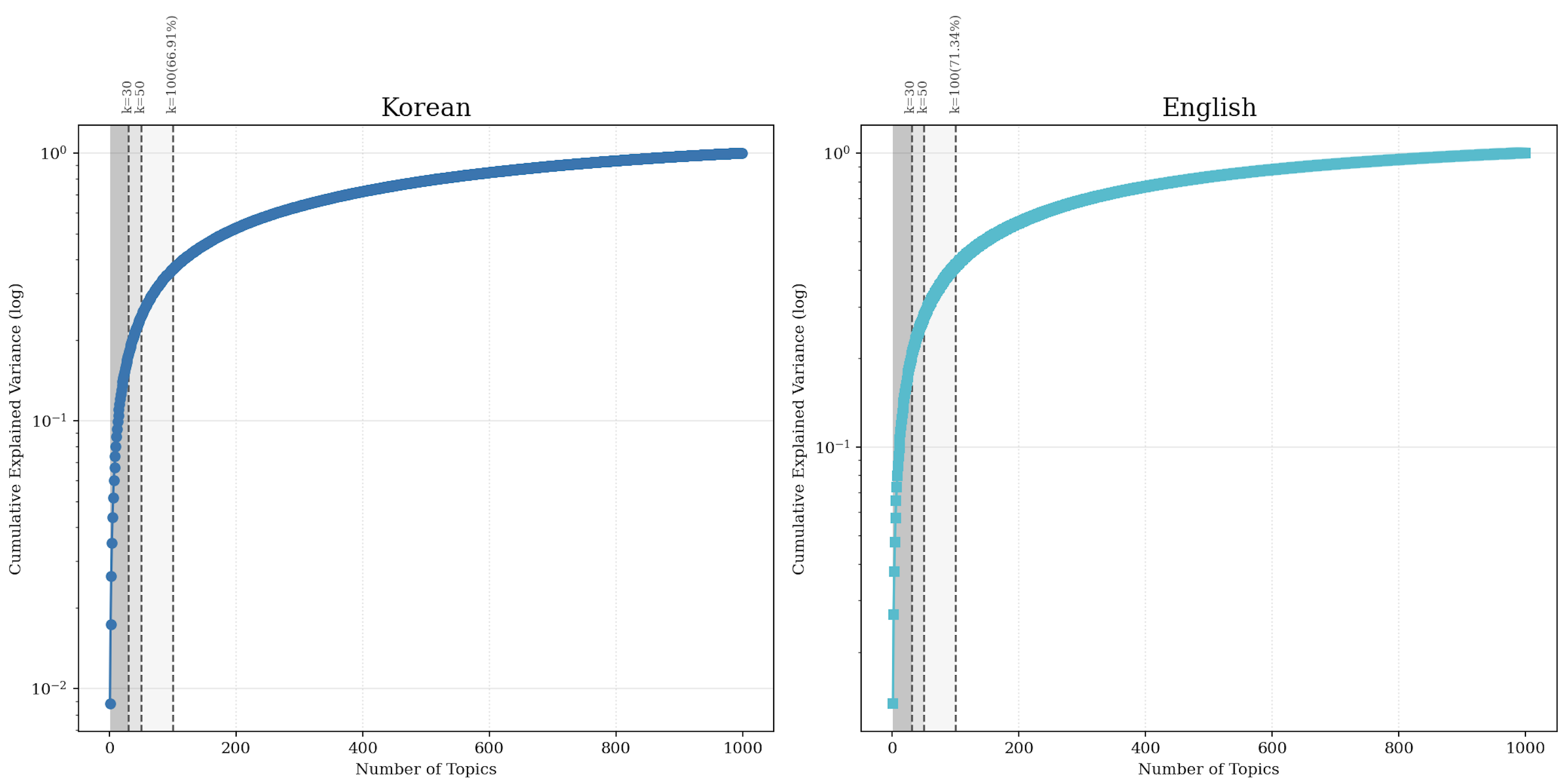}
    \caption{Cumulative explained variance of LSA singular values for the Conversation domain computed from 1{,}000 components, the shaded bands indicate representative truncation ranges at $k = 30$, $k = 50$, and $k = 100$.}
    \label{fig:lsa_full_conversation}
\end{figure}

In the next step of LSA, we apply truncated SVD to the TF--IDF matrix constructed in the previous stage. For each domain and language, we compute 1{,}000 singular values $\sigma_{1}, \ldots, \sigma_{1000}$, convert each $\sigma_i$ into an explained variance $\lambda_i = \sigma_i^{2} / \sum_{j}\sigma_j^{2}$, and then take the cumulative sum to obtain a cumulative explained--variance curve. Figure~\ref{fig:lsa_full_conversation} illustrates this for the Conversation domain, showing the cumulative explained variance over all 1{,}000 components for the Korean and English corpora. In the initial part of the curve, a small number of components rapidly accounts for a large proportion of the variance, after which the slope becomes gradually flatter and additional components contribute only marginal gains. We mark three representative candidate truncation ranks, $k = 30$, $k = 60$, and $k = 100$, with shaded vertical bands to indicate that most of the variance has already been captured within this range. This analysis confirms that it is unnecessary to increase the LSA topic dimensionality into the hundreds; a few dozen latent semantic axes are sufficient to explain the domain-specific semantic structure. The full singular-value spectra and cumulative explained--variance curves for all domain--language pairs are likewise documented in the accompanying Mendeley dataset\cite{lee2025mendeley}.

To determine the final number of topics used in LSA, we restricted the candidate numbers of components to \(k \in \{1,\dots,30\}\) to facilitate interpretation of the resulting latent semantic axes. When the number of topics becomes too large, individual axes become difficult to interpret, and we therefore restricted model selection to this range. Following the Kneedle algorithm proposed by Satopaa et al.~\cite{satopaa2011finding}, we first normalized the cumulative explained variance of the top \(k\) components to the interval \([0,1]\) and denoted it by \(f(k)\) for \(k = 1,\ldots,30\). We then defined a straight line \(g(k)\) that connects the start and end points of this interval, computed the difference \(d(k) = f(k) - g(k)\), and took the knee point \(k^{\ast} = \arg\max_{k} d(k)\) as the optimal number of topics for each domain--language pair. Intuitively, \(k^{\ast}\) corresponds to the point that deviates most from the straight line connecting the endpoints, which Kneedle uses the point that deviates most from the straight line connecting the endpoints, which Kneedle uses to approximate the point of maximum curvature in discrete data; equivalently, it marks where the curve begins to flatten out, where the rate of increase of explained variance starts to decrease\cite{lee2025mendeley}.

\begin{figure}[h!]
    \centering
    \includegraphics[width=0.9\linewidth]{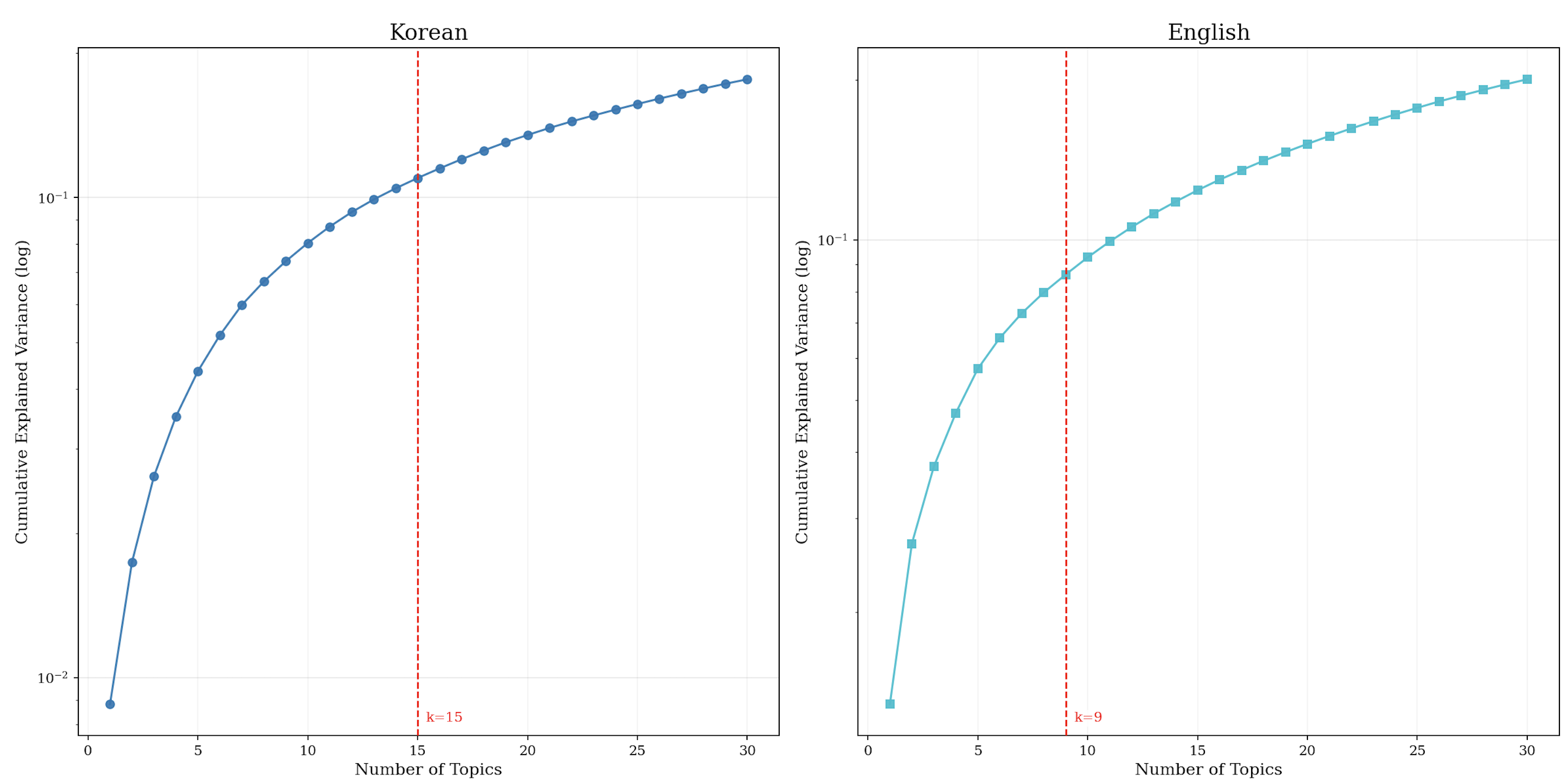}
    \caption{Cumulative explained variance for the first 30 LSA components in the Conversation domain. The red dashed lines indicate the knee points \(k^{\ast}\) selected by the Kneedle algorithm for the Korean corpus (left) and the English corpus (right).}
    \label{fig:lsa_30_conversation}
\end{figure}

In LDA, we estimated domain and language specific topic models over the same noun-based vocabularies and determined the optimal number of topics $k$ based on how the topic coherence (coherence score, $C_v$) changes with $k$. LDA requires hyperparameters for the document–topic distributions ($\alpha$) and the topic–word distributions ($\eta$, or $\beta$). In this study, we followed the implementation in the Gensim library and set both priors to symmetric Dirichlet distributions whose components are $1/k$ for a model with $k$ topics\cite{kuo2023handbook}. This configuration avoids artificial prior bias toward particular topics or words and keeps the model structure simple for fair comparison across languages.

To select the optimal $k$, we sequentially trained models for each domain and language with $k \leq 30$ and computed the coherence score $C_v$ proposed by Newman et al.\ for each setting\cite{newman2010automatic}. The $C_v$ metric quantifies the interpretability of a topic by evaluating how strongly the top words of that topic are semantically connected in the underlying text. Concretely, it constructs Normalized Pointwise Mutual Information(NPMI) based context vectors from sliding-window co-occurrence statistics and then averages the cosine similarities between all pairs of these vectors, that is, $C_v(k) \approx \frac{1}{\lvert P_t \rvert} \sum_{(w_i, w_j) \in P_t} \cos\bigl(v_i, v_j\bigr)$, where $P_t$ is the set of top word pairs for topic $t$, and $v_i$ and $v_j$ are the NPMI-based context vectors for words $w_i$ and $w_j$, respectively.

\begin{figure}[h!]
\centering
\includegraphics[width=0.9\linewidth]{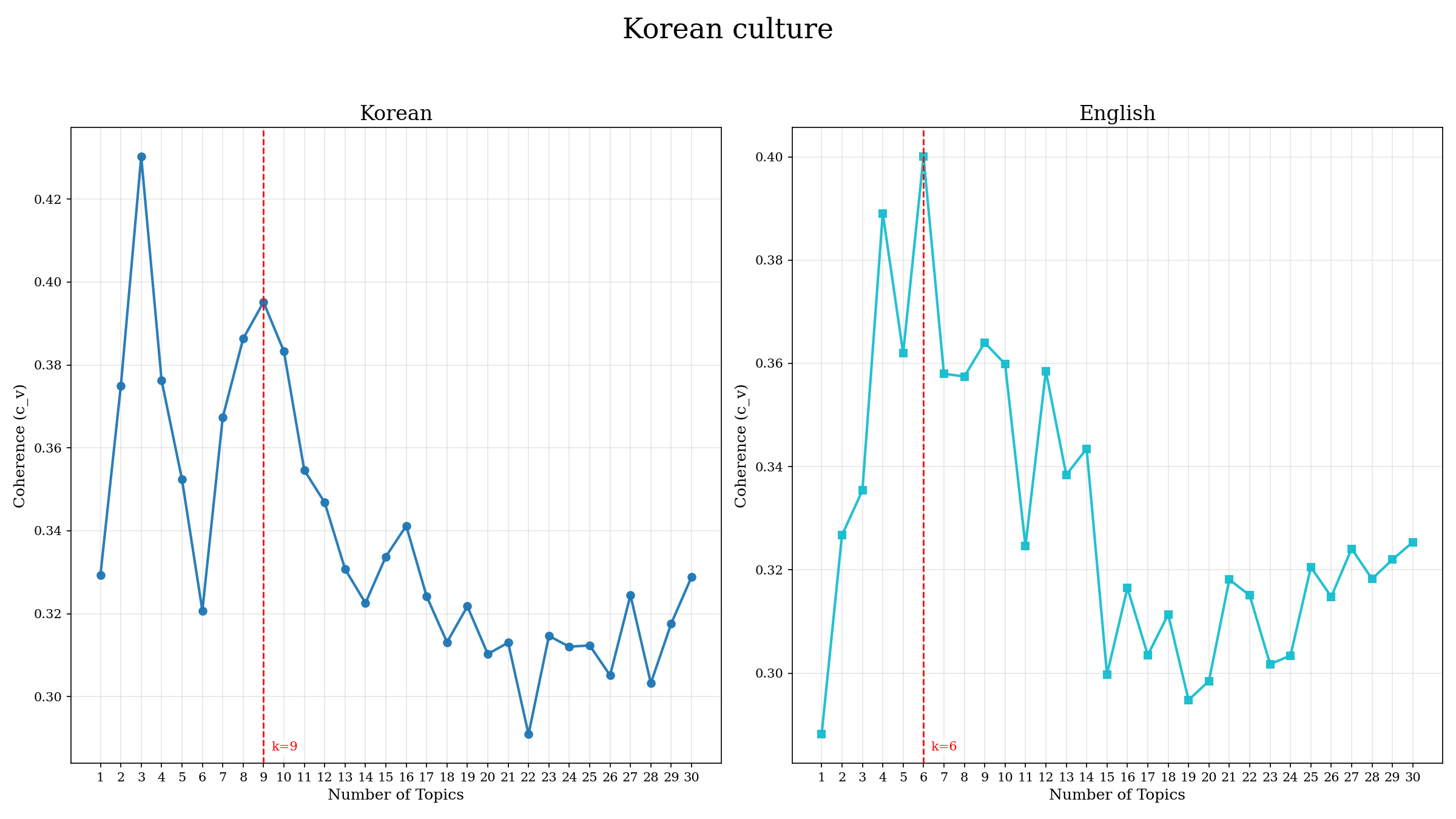}
\caption{LDA topic coherence (\(C_v\)) as a function of the number of topics in the Korean-culture domain. The red dashed line indicates the selected number of topics \(k^{\ast}\) for the Korean corpus (left) and the English corpus (right).}
\label{fig:lda_coherence_koreanculture}
\end{figure}

When the number of topics $k$ becomes excessively large, the $C_v$ score tends to plateau or even decrease, indicating that topics become fragmented and less interpretable. Therefore, as in the LSA analysis, we applied the Kneedle algorithm\cite{satopaa2011finding} to the normalized $C_v(k)$ curve, detected the knee point of the curve, and adopted the corresponding value as the optimal number of topics $k$. Figure~\ref{fig:lda_coherence_koreanculture} illustrates this for the Korean Culture domain, showing the $C_v(k)$ curves for the Korean and English corpora and the final numbers of topics $k$ selected by the algorithm\cite{lee2025mendeley}.

In BERTopic, we do not pre-specify the number of topics $k$; instead, topics are extracted by density-based clustering of document embeddings, and the clusters that emerge from the data are regarded as topics. We first convert each document into a dense vector representation using a sentence-embedding model, and then apply UMAP (Uniform Manifold Approximation and Projection) to reduce the embedding dimensionality while preserving the structure of the high-dimensional space and facilitating clustering. Next, we cluster the reduced embeddings using HDBSCAN (Hierarchical Density-Based Spatial Clustering of Applications with Noise), where the number of clusters is automatically determined by the clustering outcome rather than set a priori. HDBSCAN assigns a cluster label to each document, and the label $-1$ indicates noise or outliers that do not stably belong to any cluster; we exclude such documents from subsequent analysis. Accordingly, for each remaining document, we use the cluster label automatically assigned by HDBSCAN as the document's representative topic under a single-topic assignment. For interpretability, each topic is summarized by ranking words using BERTopic's c-TF-IDF (class-based TF-IDF) term-importance scores and retaining the top tokens with their corresponding weights. In addition, to ensure representativeness for cross-lingual comparison and to simplify the unit of analysis, we select a single representative topic for each domain-language pair by choosing, among all non-noise topics, the topic that contains the largest number of documents, i.e., the largest cluster. Here, topic influence is defined by cluster size, and this selection is independent of the numerical topic label or its creation order. The selected representative topic is likewise summarized by the top tokens and their c-TF-IDF weights and is used in subsequent cross-lingual topic-similarity computation.

\subsection{Matching Topics}\label{sec:Matching_Topics}
This section establishes cross-lingual correspondences between Korean and English topics using the representative topic extracted for each domain–language pair in the previous section, summarized by the top 20 tokens and their weights. To this end, we perform dictionary-based lexical alignment by mapping Korean tokens into the English token space using a Korean–English bilingual dictionary, and then quantify topical similarity by computing cosine similarity between the aligned token–weight vectors. Because the resulting similarity score is directly determined by the aligned token pairs and their weights, it can be interpreted as an explainable metric in which the specific tokens contributing to the score are traceable. This alignment and similarity comparison enable us to measure how consistently the thematic structure and the emphasis of key lexical items are preserved across languages: higher similarity indicates that the translation faithfully maintains the source document’s topical content and distribution, whereas lower similarity suggests topical drift or shifts in emphasis.

We apply a rule-based lexical alignment procedure to align Korean tokens with English tokens. Specifically, for each of the top 20 tokens in the Korean representative topic, we construct a token-level bilingual mapping by querying the (Add.Q4) dictionary-style word lookup in Google Translate for an English candidate. We use the primary candidate as the English mapping and confirm a match only when this selected candidate appears in the top 20 token list of the English representative topic. The alignment is constrained to a one-to-one correspondence between Korean and English tokens; therefore, we apply the following additional rules to handle unmatched cases and duplicate mappings. First, when multiple Korean tokens translate to the same English token, we resolve this conflict by prioritizing the Korean token with the higher rank within the representative topic. Specifically, we confirm the match for the highest-ranking token and leave any lower-ranking tokens unmatched if they correspond to an already assigned English token, thereby ensuring a one-to-one alignment. Second, some tokens remain unmatched because the English translation generated by the dictionary lookup does not appear in the English topic list; this issue is particularly prevalent with proper nouns. Therefore, we implemented a post-hoc verification step that identifies proper nouns and attempts an additional matching process to account for common variations. Specifically, this method matches tokens despite romanization discrepancies, such as the list containing Kim Yuna while the dictionary returns Kim Yeona, and also matches differing acronym usage, where the list uses SNU while the dictionary outputs Seoul National University. Furthermore, it matches minor formatting differences like COVID-19 versus COVID19, as well as variations in conventional naming, such as the list using Blue House while the dictionary provides Cheong Wa Dae. In contrast, tokens that are not proper nouns and lack overlap with the English token list are left unmatched without further correction.

After alignment, we quantify the topical similarity between the Korean and English representative topics by computing cosine similarity using only the matched token pairs and their weights, while retaining the top-20 token list as a fixed-length (20-dimensional) representation. Because our lexical alignment enforces a one-to-one correspondence, the translated Korean topic vector $\bm{\beta}^{\mathrm{kr}}$ is represented in the English token space as a sparse vector with at most 20 dimensions. Tokens that fail to be matched are not removed; instead, we assign a weight of \(0\) to the corresponding dimensions, thereby fixing the vector length to 20 while ensuring that only matched token pairs contribute non-zero entries to the similarity computation. We then compute the cosine similarity between $\bm{\beta}^{\mathrm{kr}}$ and the English representative topic vector \(\bm{\alpha}^{\mathrm{en}}\). Token weights are taken directly from each model’s output. In LSA, we use the absolute values of token coefficients in the selected truncated-SVD components as weights. In LDA we use posterior topic–word probabilities as weights. In BERTopic, we use c-TF--IDF scores as weights. Cosine similarity measures the directional agreement of the weight distributions over the one-to-one aligned dimensions. Because the score is computed directly from the aligned token pairs and their weights, it can be interpreted as an explainable metric in which the token-level evidence contributing to the score is traceable.

The cross-lingual topic matching process consists of three stages, summarized as follows:




\begin{enumerate}
    \item \textbf{Input Definition:} 
    First, we define the representative topics for the source (Korean) and target (English) languages as top-20 token-weight pairs:
    \begin{equation}
        \begin{aligned}
            T^{\mathrm{kr}} &= (\mathbf{t}^{\mathrm{kr}}, \boldsymbol{\alpha}^{\mathrm{kr}})
            = \bigl((t^{\mathrm{kr}}_1, \alpha^{\mathrm{kr}}_1), \dots, (t^{\mathrm{kr}}_{20}, \alpha^{\mathrm{kr}}_{20})\bigr), \\
            T^{\mathrm{en}} &= (\mathbf{t}^{\mathrm{en}}, \boldsymbol{\alpha}^{\mathrm{en}})
            = \bigl((t^{\mathrm{en}}_1, \alpha^{\mathrm{en}}_1), \dots, (t^{\mathrm{en}}_{20}, \alpha^{\mathrm{en}}_{20})\bigr)
        \end{aligned}
    \end{equation}
    where, $\mathbf{t}^{\mathrm{kr}}$ is the ordered vector of the top-20 Korean tokens and $\boldsymbol{\alpha}^{\mathrm{kr}}$ is the corresponding weight vector, with $t^{\mathrm{kr}}_j$ and $\alpha^{\mathrm{kr}}_j$ denoting the $j$-th token and its weight, respectively. Likewise, $\mathbf{t}^{\mathrm{en}}$ is the ordered vector of the top-20 English tokens and $\boldsymbol{\alpha}^{\mathrm{en}}$ is the corresponding weight vector, with $t^{\mathrm{en}}_j$ and $\alpha^{\mathrm{en}}_j$ denoting the $j$-th token and its weight, respectively.

    \item \textbf{Alignment and Vector Construction:} 
    We map the Korean topic into the English vector space to create an aligned vector $\bm{\beta}^{\mathrm{kr}} \in \mathbb{R}^{20}$. The value of the $j$-th dimension is determined by lexical alignment:
    \begin{equation}
        \beta^{\mathrm{kr}}_j =
        \begin{cases}
           \alpha^{\mathrm{kr}}_i & \text{if } \mathrm{Trans}(t^{\mathrm{kr}}_i) = t^{\mathrm{en}}_j \text{ (matched)} \\
           0 & \text{otherwise}
        \end{cases}
    \end{equation}
    where, $\alpha^{\mathrm{kr}}_i$ is the weight of the Korean token $t^{\mathrm{kr}}_i$. When $t^{\mathrm{kr}}_i$ is matched to $t^{\mathrm{en}}_j$, the corresponding entry $\beta^{\mathrm{kr}}_j$ takes the Korean weight $\alpha^{\mathrm{kr}}_i$, otherwise $\beta^{\mathrm{kr}}_j$ is set to $0$. Furthermore, the matching is one-to-one, meaning that for each $j$ there exists at most one such index $i$.

    \item \textbf{Similarity Computation:} 
    Finally, we quantify the topical similarity by computing the cosine similarity between the constructed vectors. The score is derived solely from the aligned token pairs:
    \begin{equation}
        \mathrm{Sim}\!\left(\bm{\beta}^{\mathrm{kr}}, \bm{\alpha}^{\mathrm{en}}\right)
        =
        \frac{\sum_{j=1}^{20} \beta^{\mathrm{kr}}_j \alpha^{\mathrm{en}}_j}
        {\sqrt{\sum_{j=1}^{20} (\beta^{\mathrm{kr}}_j)^2}\;
         \sqrt{\sum_{j=1}^{20} (\alpha^{\mathrm{en}}_j)^2}}
    \end{equation}
\end{enumerate}

\section{Results \& Discussions}\label{sec:Results}
In this section, we present a comparative analysis of translation quality metrics, including BLEU as a traditional reference-based metric, CometKiwi as a supervised metric built on pretrained language models, and LSA, LDA, and BERTopic as topic modeling-based approaches. The evaluation scores were obtained using the proposed framework and compared across domain-specific datasets. In particular, based on the representative tokens extracted through topic modeling, we aim to interpret not only the topic-based scores but also the score patterns observed in traditional metrics such as BLEU and CometKiwi.

\subsection{Metric Distribution Overview}\label{subsec:metric_overview}
\begin{figure}[H] 
    \centering
    \includegraphics[width=\linewidth]{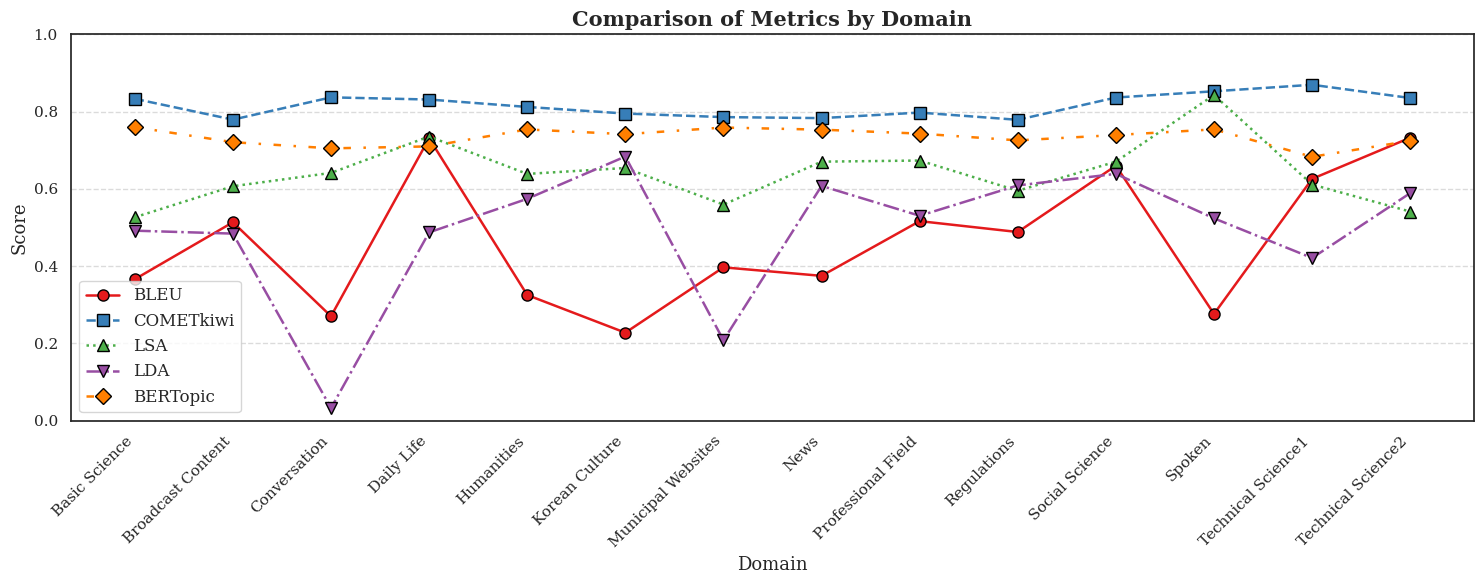}
    \caption{Comparison of Translation Quality Evaluation Metrics by Domain}
    \label{fig:Comparisonof}
\end{figure}

Figure \ref{fig:Comparisonof} depicts the distributions of the five evaluation metrics. BLEU records the lowest scores overall and, exhibiting considerable variability across datasets, extends across an interval from 0.228 to 0.733 (mean = 0.465, SD = 0.17). In contrast, CometKiwi occupies the narrowest range—0.779 to 0.870—showing the least inter-dataset dispersion; its mean is 0.816 with an SD of 0.03, and values are consistently high. Among the topic-model metrics, LSA shows intermediate variability, averaging 0.641 (SD = 0.08). LDA yields the lowest minimum (0.033) and the widest span, reaching 0.684; its mean of 0.492 and SD = 0.17 reflects variability comparable to BLEU. BERTopic, meanwhile, exhibits the narrowest spread among the topic-based metrics, returning values between 0.683 and 0.760, with a mean of 0.734 (SD = 0.02), a pattern analogous to the tight range of CometKiwi. Collectively, these observations indicate that BLEU and LDA are most sensitive to dataset differences, CometKiwi and BERTopic are relatively insensitive, and LSA lies between these two groups in terms of variability.

\subsection{Explainability from Topic Models}\label{subsec:topic_explainability}

Table \ref{tab:domains} presents the top 20 Korean and English tokens extracted by the topic models (LSA, LDA, and BERTopic) for each dataset. These tokens, assigned the highest topic weights, serve as the key lexical signals that most strongly influence the topic distributions within each document set. In what follows, we draw on this token-level information to interpret the score patterns observed in Figure \ref{fig:Comparisonof} and to explore how the proposed document-level translation evaluation framework offers an evidence-driven approach to explainability.

\subsubsection{BLEU Score Variability Inference}\label{subsubsec:bleu_variability}
BLEU achieved markedly low scores in \textit{Korean culture} (0.2279), \textit{Conversation} (0.2705), \textit{Spoken} (0.2764), and \textit{Humanities} (0.3254), while recording substantially higher scores in \textit{Technical science2} (0.7325), \textit{Daily life} (0.7309), \textit{Social science} (0.6600), and \textit{Technical science1} (0.6260). This score distribution pattern stems from the linguistic characteristics and translation complexity inherent in each domain type. Low scores appear to result from domains with strong cultural specificity such as \textit{Korean culture}, as well as \textit{Conversation} and \textit{Spoken} domains that are characterized by emotional expressions, context-dependent vocabulary, and relatively short sentence lengths due to their colloquial nature. In contrast, \textit{Technical science} domains with standardized technical terminology and clear one-to-one correspondence relationships, and \textit{Daily life} domains dominated by formalized expressions in commercial sectors, achieved high scores due to well-established translation conventions.

\textbf{Challenges in Conversational and Spoken Domains}: The \textit{Conversation} domain is anchored in commerce and service scenarios, with recurring nouns such as "제품 (product)", "주문 (order)", "예약(reservation)", "가격(price)", "배송(delivery)", "환불(refund)", and "결제 charge)". At the same time, interpersonal and organizational actors such as "사람(person)" and "회사(company)" appear prominently, and situational nouns like "문제 (problem)" frequently co-occur with practical items and settings (e.g., "음식(food)", "포장(packaging)", "화장실(restroom)", "커피(coffee)", "도착(arrival)"). This mixture suggests that the domain is not purely transactional. It blends customer–provider interaction with everyday contextual details, yielding high lexical variety even within a single dialogue. The \textit{Spoken} domain, in contrast, is organized around personal and relational vocabulary such as "사람(person)", "친구(friend)", "사랑(love)", "행복(happiness)", and "마음(mind)", alongside common topical references including "영화(movie)", "음식(food)", and "나라(country)". Across both domains, the repeated but context-sensitive use of umbrella nouns like "사람(person)" amplifies translation variability (e.g., "person/people/someone/individual"), and affect-related terms such as "마음(mind)" can be rendered as "mind/heart/feeling" depending on discourse intent. Combined with the typically short, colloquial utterances in these datasets, such variability reduces stable $N$-gram overlap and can systematically depress BLEU scores despite semantically appropriate translations.

\textbf{Cultural and Humanities Domain Complexities}: \textit{Korean culture} faces significant difficulties with culture-specific terminology translation, featuring culturally-bound concepts such as "한류(hallyu)", "조선(joseon)", "드라마(drama)", "공연(performance)", "영화(film)", "영화제(film festival)", "스타(star)", and "부산(Busan)" that resist direct translation — "한류(hallyu)" can be rendered as "hallyu," "Korean Wave," or "K-culture" depending on target audience familiarity, while "조선(joseon)" may appear as "joseon," "chosun," or "korean dynasty" based on historical context preferences. \textit{Humanities} encounters distinct challenges with abstract societal concepts including "교육(education)", "사회(society)", "경제(economy)", "정부(government)", and "정책(policy)", which demonstrate high translation variability—"정부(government)" can be translated as "government," "administration," or "state" depending on political discourse context, and "교육(education)" may appear as "education," "schooling," or "learning" based on academic versus practical contexts. Both domains share a common vulnerability through geographic and political entity terminology, as reveals frequent tokens including "한국(korea)", "중국(china)", "일본(japan)" from \textit{Korean culture} and "미국(usa)", "북한(northkorea)", "독일(germany)" from \textit{Humanities}. Geographic entities present inherent translation challenges due to regional naming conventions and political contexts—different translation standards may prefer formal names versus common usage, transliteration versus translation, or contemporary versus historical designations. The fundamental limitation lies in BLEU's single-reference dependency, where only one translation choice is considered correct regardless of equally valid alternatives. Cultural adaptation requirements further compound this issue, as culture-specific concepts from \textit{Korean culture} such as "한류(hallyu)" and "조선(joseon)" require cultural knowledge for appropriate English rendering, while abstract societal concepts from \textit{Humanities} like "사회(society)" and "정부(government)" demand contextual interpretation. This systematic translation complexity across cultural terminology, abstract concepts, and geographic entities creates a fundamental incompatibility with BLEU's rigid matching methodology, causing the metric to systematically undervalue translations that prioritize cultural authenticity and contextual appropriateness over lexical consistency.

\textbf{BLEU's Strength in Standardized Domains}: \textit{Technical science1} and \textit{Technical science2} domains benefit significantly from internationally standardized technical vocabulary that exhibits minimal cross-linguistic variation. Table 2 reveals frequent tokens such as "장치(device)", "데이터(data)", "정보(information)", "시스템(system)", "프로세서(processor)", and "디스플레이(display)" that maintain direct one-to-one correspondence across languages. These terms, largely derived from English technical vocabulary or established as global standards, resist cultural interpretation and maintain consistent translation patterns regardless of context. \textit{Social science} demonstrates similar advantages through academic terminology standardization. Extracted tokens including "연구(research)", "교육(education)", "분석(analysis)", "규정(regulation)", and "방법(method)" represent established scholarly vocabulary with conventional translation patterns developed through decades of academic exchange and international collaboration. \textit{Daily life} achieves high BLEU scores through commercial terminology that has been standardized by global business practices. Table 2 shows consistent extraction of commerce-related tokens such as "제품(product)", "주문(order)", "배송(delivery)", "고객(customer)", "가격(price)", and "회사(company)" that follow established business translation conventions across international markets. This terminological consistency across technical, academic, and commercial domains enables higher $N$-gram overlap rates, aligning perfectly with BLEU's evaluation methodology and resulting in elevated scores that reflect the metric's preference for direct lexical matching over cultural adaptation or contextual interpretation.

\subsection{Robustness of Embedding-based Metrics}\label{subsubsec:embedding_robustness}
 The remarkable consistency observed in CometKiwi and BERTopic score across diverse datasets suggests that embedding-based evaluation approaches demonstrate inherent robustness to domain-specific variations. Unlike traditional lexical matching methods, these metrics appear to capture semantic relationships that transcend surface-level linguistic differences across domains.

\textbf{Semantic Abstraction in Pre-trained Metrics}: BERTopic consistently selects higher-order, domain-agnostic tokens, which explains its low cross-domain score variance observed in Figure \ref{fig:Comparisonof}. In \textit{Korean culture}, for example, LSA and LDA focus on fine-grained exports such as "한류(hallyu)", "드라마(drama)", and "영화제(film festival)", whereas BERTopic generalises them to broader tokens like "culture", "performance", and "concert" that recur in other media-oriented corpora. Within the commercial \textit{Conversation} dataset, LSA/LDA split the customer journey into "주문(order)", "결제(payment)", and "배송(delivery)", but BERTopic replaces these with universal actors and states—"problem", "company", and "person"—yielding a dialogue template transferable to any service scenario. A similar shift appears in the policy-heavy \textit{Humanities} corpus: concrete entities such as "미국(USA)", "정부(government)", and "정책(policy)" are abstracted by BERTopic into systemic categories like "society", "economy", "nation", and "situation". The same pattern is visible in technical data: LSA/LDA in \textit{Technical science1} enumerate component-level nouns ("장치(device)", "신호(signal)", "프로세서(processor)"), whereas BERTopic lifts them to infrastructure-level terms such as "system", "information", and "data"; in \textit{Basic science} it moves from laboratory units ("원자(atom)", "분자(molecule)") to discipline-wide notions ("theory", "space", "result"). Even in the \textit{News} corpus, where LSA/LDA fragment events with "대통령(president)", "삼성(Samsung)", and "투자(investment)", BERTopic again gravitates to umbrella economics tokens like "market", "business", and "economy". Across the 14 domains, "problem" appears in seven datasets, while "person" and "company" each occur in five. Because such umbrella-level tokens recur so often, a greater proportion of terms overlaps when one domain’s topic vector is compared with another’s; this broader lexical overlap blurs domain boundaries and largely explains BERTopic’s tightly clustered scores. For the same reason—working in an abstract embedding space rather than matching surface words—CometKiwi can also be expected to yield a similarly narrow score distribution across datasets.

\subsubsection{Lexical Topic Model Variability}\label{subsubsec:classical_topic_models}

LSA and LDA demonstrate high inter-domain variability as lexical-based topic modeling approaches, in contrast to the stable performance of embedding-based metrics. While LSA achieves a relatively high score (0.843) in the \textit{Spoken} domain, LDA records an extremely low score near zero (0.033) in the \textit{Conversation} domain, clearly revealing the domain-specific sensitivity of frequency-based topic modeling techniques. This extreme difference in score distribution stems from the interaction between each method's unique token processing approach and the linguistic characteristics of colloquial text.

\textbf{Lexical Distribution Characteristics in Colloquial Domains}: \textit{Spoken} domain is related to the consistency of token distribution confirmed in Table 2. In the \textit{Spoken} domain, emotional and relational core vocabulary such as "사람(person)", "친구(friend)", "한국(korea)", "사랑(love)", and "행복(happiness)" appear repeatedly, and these tokens maintain considerable frequency throughout the document, enabling the formation of a stable semantic space during LSA's SVD process. Although the number of tokens per sentence is relatively small due to the colloquial nature, it is judged that sufficient signals could be secured in the co-occurrence frequency matrix due to the repeated use of core emotional expressions and person-centered vocabulary. LSA's linear dimensionality reduction characteristics effectively capture these repetitive lexical patterns, and it is analyzed that this enabled the generation of relatively stable topic vectors in colloquial texts with emotion-centered consistent discourse structure.

\textbf{Limitations of Probabilistic Topic Assignment and Commercial Dialogue Complexity}: The primary reason for the extremely low LDA-based topic similarity score in the \textit{Conversation} domain lies in the thematic diversity and frequent contextual shifts inherent in commercial dialogues. As shown in Table \ref{tab:domains}, the \textit{Conversation} domain contains a mixture of diverse business terminology spanning the entire customer service process, including "제품(product)", "주문(order)", "예약(reservation)", "신청(application)", "거래(transaction)", "결제(payment)", "문의(inquiry)", and "접수(reception)". Since each dialogue addresses different products, services, and problem situations, it becomes difficult to form consistent topic distributions. LDA's Dirichlet distribution assumption presupposes that each document consists of a small number of dominant topics, but commercial dialogues involve multi-stage business processes such as order-payment-delivery-inquiry-resolution that unfold sequentially within a single conversation, making it challenging to maintain thematic consistency. The context-dependent nature of customer-agent interactions particularly complicates this, as identical vocabulary can carry different semantic weights depending on the situation. In this domain, a single dialogue often shifts sequentially across subtopics such as ordering, payment, delivery, inquiry, and resolution, so the representative topic is less likely to converge to a small set of dominant themes and is instead prone to being dispersed across multiple themes. When many subtopics alternate within a single dialogue, the representative topic tends to spread across multiple themes rather than concentrating on a few dominant ones. As a result, even if the source and the translation are both coherent within their own flow, small differences in which subtopics are emphasized can lead to a low topic-based similarity score.

\clearpage
\newgeometry{left=1.2cm,right=1.2cm,top=1.0cm,bottom=1.0cm}

\begin{landscape}
\setlength{\LTpre}{0pt}
\setlength{\LTpost}{0pt}
\setlength{\tabcolsep}{2pt}
\setlength{\aboverulesep}{0pt}
\setlength{\belowrulesep}{0pt}
\setlength{\cmidrulesep}{0pt}
\fontsize{6.5}{7.2}\selectfont
\renewcommand{\arraystretch}{1.1}

\scriptsize
\renewcommand{\arraystretch}{1.15}

\begin{longtable}{p{2.5cm} >{\arraybackslash}p{3cm} >{\arraybackslash}p{3cm} >{\arraybackslash}p{3cm} >{\arraybackslash}p{3cm} >{\arraybackslash}p{3cm} >{\arraybackslash}p{3cm}}

\caption{Explaining Translation Evaluation Metrics via Topic Modeling (LSA, LDA, BERTopic)- Top 20 Korean and English Tokens per Domain\label{tab:domains}}\\

\toprule
\multirow{2}{*}{\textbf{Domain Name}} 
 & \multicolumn{2}{c}{\textbf{LSA}} 
 & \multicolumn{2}{c}{\textbf{LDA}} 
 & \multicolumn{2}{c}{\textbf{BERTopic}} \\
\cmidrule(lr){2-3}\cmidrule(lr){4-5}\cmidrule(lr){6-7}
 & \textbf{KR} & \textbf{ENG} & \textbf{KR} & \textbf{ENG} & \textbf{KR} & \textbf{ENG} \\
\midrule
\endfirsthead

\multicolumn{7}{c}{\tablename\ \thetable\ -- \textit{Continued from previous page}}\\
\toprule
\multirow{2}{*}{\textbf{Domain Name}} 
 & \multicolumn{2}{c}{\textbf{LSA}} 
 & \multicolumn{2}{c}{\textbf{LDA}} 
 & \multicolumn{2}{c}{\textbf{BERTopic}} \\
\cmidrule(lr){2-3}\cmidrule(lr){4-5}\cmidrule(lr){6-7}
 & \textbf{KR} & \textbf{ENG} & \textbf{KR} & \textbf{ENG} & \textbf{KR} & \textbf{ENG} \\
\midrule
\endhead

\midrule
\multicolumn{7}{r}{\textit{Continued on next page}}\\
\endfoot

\bottomrule
\endlastfoot

\textbf{Basic Science}
& \textbf{\textcolor{4DAF4A}{에너지}}, 물질, 연구, \textbf{\textcolor{4DAF4A}{이론}}, 문제, 생각, \textbf{\textcolor{4DAF4A}{결과}}, 화학, 방법, \textbf{\textcolor{4DAF4A}{원자}}, \textbf{\textcolor{4DAF4A}{분자}}, \textbf{\textcolor{4DAF4A}{입자}}, 사실, 예측, \textbf{\textcolor{4DAF4A}{우주}}, 기술, 실험, \textbf{\textcolor{4DAF4A}{원소}}, 반응, \textbf{\textcolor{4DAF4A}{사람}}
& \textbf{\textcolor{4DAF4A}{energy}}, \textbf{\textcolor{4DAF4A}{person}}, \textbf{\textcolor{4DAF4A}{theory}}, \textbf{\textcolor{4DAF4A}{result}}, cell, \textbf{\textcolor{4DAF4A}{atom}}, light, \textbf{\textcolor{4DAF4A}{molecule}}, \textbf{\textcolor{4DAF4A}{particle}}, electron, system, carbon, \textbf{\textcolor{4DAF4A}{element}}, temperature, wave, quantum, datum, earth, \textbf{\textcolor{4DAF4A}{space}}, hydrogen
& \textbf{\textcolor{984EA3}{이론}}, 설명, 상대성, 질량, 예측, 발견, 줄기세포, 사실, \textbf{\textcolor{984EA3}{물리학}}, \textbf{\textcolor{984EA3}{우주}}, 중력, 개념, 블랙홀, \textbf{\textcolor{984EA3}{입자}}, 역학, \textbf{\textcolor{984EA3}{방정식}}, 원리, \textbf{\textcolor{984EA3}{모형}}, 존재, 차원
& \textbf{\textcolor{984EA3}{theory}}, \textbf{\textcolor{984EA3}{particle}}, force, \textbf{\textcolor{984EA3}{space}}, acid, quantum, object, \textbf{\textcolor{984EA3}{physics}}, energy, gravity, motion, \textbf{\textcolor{984EA3}{equation}}, \textbf{\textcolor{984EA3}{model}}, einstein, newton, alcohol, spin, mechanics, radiation, matter
& \textbf{\textcolor{FF7F00}{물질}}, \textbf{\textcolor{FF7F00}{원자}}, \textbf{\textcolor{FF7F00}{연구}}, 가능, \textbf{\textcolor{FF7F00}{입자}}, 상태, \textbf{\textcolor{FF7F00}{화학}}, \textbf{\textcolor{FF7F00}{원소}}, \textbf{\textcolor{FF7F00}{방법}}, \textbf{\textcolor{FF7F00}{문제}}, \textbf{\textcolor{FF7F00}{사실}}, \textbf{\textcolor{FF7F00}{기술}}, 존재, \textbf{\textcolor{FF7F00}{실험}}, 발견, \textbf{\textcolor{FF7F00}{설명}}, \textbf{\textcolor{FF7F00}{예측}}, \textbf{\textcolor{FF7F00}{정도}}, \textbf{\textcolor{FF7F00}{우주}}, 변화
& \textbf{\textcolor{FF7F00}{substance}}, \textbf{\textcolor{FF7F00}{atom}}, result, \textbf{\textcolor{FF7F00}{research}}, \textbf{\textcolor{FF7F00}{particle}}, theory, \textbf{\textcolor{FF7F00}{chemistry}}, \textbf{\textcolor{FF7F00}{element}}, \textbf{\textcolor{FF7F00}{method}}, \textbf{\textcolor{FF7F00}{problem}}, \textbf{\textcolor{FF7F00}{technology}}, \textbf{\textcolor{FF7F00}{fact}}, earth, \textbf{\textcolor{FF7F00}{experiment}}, electron, \textbf{\textcolor{FF7F00}{explanation}}, \textbf{\textcolor{FF7F00}{prediction}}, \textbf{\textcolor{FF7F00}{degree}}, \textbf{\textcolor{FF7F00}{space}}, study \\

\textbf{Broadcast Content}
& \textbf{\textcolor{4DAF4A}{사람}}, 사랑, \textbf{\textcolor{4DAF4A}{그림}}, \textbf{\textcolor{4DAF4A}{한국}}, \textbf{\textcolor{4DAF4A}{엄마}}, 시작, \textbf{\textcolor{4DAF4A}{이야기}}, 모습, 남편, \textbf{\textcolor{4DAF4A}{음식}}, \textbf{\textcolor{4DAF4A}{나라}}, \textbf{\textcolor{4DAF4A}{작업}}, \textbf{\textcolor{4DAF4A}{얼굴}}, 친구, 결혼, 행복, 아들, 건강, 사진, \textbf{\textcolor{4DAF4A}{이유}}
& \textbf{\textcolor{4DAF4A}{person}}, life, \textbf{\textcolor{4DAF4A}{food}}, money, \textbf{\textcolor{4DAF4A}{work}}, house, \textbf{\textcolor{4DAF4A}{mother}}, \textbf{\textcolor{4DAF4A}{picture}}, \textbf{\textcolor{4DAF4A}{korea}}, heart, \textbf{\textcolor{4DAF4A}{country}}, hand, eye, \textbf{\textcolor{4DAF4A}{reason}}, \textbf{\textcolor{4DAF4A}{face}}, water, \textbf{\textcolor{4DAF4A}{story}}, family, body, mountain
& \textbf{\textcolor{984EA3}{시작}}, \textbf{\textcolor{984EA3}{엄마}}, 사랑, 지원, 부부, 산업, 아침, 할머니, \textbf{\textcolor{984EA3}{회사}}, 사업, \textbf{\textcolor{984EA3}{활동}}, 효과, 작품, 선원, \textbf{\textcolor{984EA3}{환자}}, 촬영, 생산, 연락, \textbf{\textcolor{984EA3}{치료}}, \textbf{\textcolor{984EA3}{집중}}
& \textbf{\textcolor{984EA3}{start}}, water, wife, rice, fish, friend, \textbf{\textcolor{984EA3}{mother}}, \textbf{\textcolor{984EA3}{activity}}, \textbf{\textcolor{984EA3}{company}}, teacher, parent, winter, result, \textbf{\textcolor{984EA3}{patient}}, road, phone, honey, \textbf{\textcolor{984EA3}{treatment}}, environment, \textbf{\textcolor{984EA3}{focus}}
& \textbf{\textcolor{FF7F00}{사람}}, 생각, \textbf{\textcolor{FF7F00}{엄마}}, 마음, \textbf{\textcolor{FF7F00}{얘기}}, \textbf{\textcolor{FF7F00}{아내}}, 시작, \textbf{\textcolor{FF7F00}{나무}}, \textbf{\textcolor{FF7F00}{남편}}, \textbf{\textcolor{FF7F00}{느낌}}, 작업, \textbf{\textcolor{FF7F00}{문제}}, \textbf{\textcolor{FF7F00}{그림}}, \textbf{\textcolor{FF7F00}{아들}}, \textbf{\textcolor{FF7F00}{여행}}, 얼굴, \textbf{\textcolor{FF7F00}{친구}}, \textbf{\textcolor{FF7F00}{노래}}, \textbf{\textcolor{FF7F00}{언니}}, \textbf{\textcolor{FF7F00}{준비}}
& \textbf{\textcolor{FF7F00}{person}}, today, \textbf{\textcolor{FF7F00}{mother}}, \textbf{\textcolor{FF7F00}{song}}, \textbf{\textcolor{FF7F00}{story}}, \textbf{\textcolor{FF7F00}{wife}}, voice, \textbf{\textcolor{FF7F00}{tree}}, \textbf{\textcolor{FF7F00}{husband}}, brother, \textbf{\textcolor{FF7F00}{feeling}}, \textbf{\textcolor{FF7F00}{problem}}, \textbf{\textcolor{FF7F00}{picture}}, \textbf{\textcolor{FF7F00}{son}}, \textbf{\textcolor{FF7F00}{travel}}, woman, \textbf{\textcolor{FF7F00}{friend}}, \textbf{\textcolor{FF7F00}{sister}}, \textbf{\textcolor{FF7F00}{preparation}}, life \\

\textbf{Conversation}
& \textbf{\textcolor{4DAF4A}{제품}}, \textbf{\textcolor{4DAF4A}{사람}}, 확인, \textbf{\textcolor{4DAF4A}{주문}}, \textbf{\textcolor{4DAF4A}{예약}}, \textbf{\textcolor{4DAF4A}{회사}}, 필요, 구매, 버스, \textbf{\textcolor{4DAF4A}{음식}}, 전화, \textbf{\textcolor{4DAF4A}{문제}}, 상품, \textbf{\textcolor{4DAF4A}{가격}}, 준비, 감사, \textbf{\textcolor{4DAF4A}{배송}}, \textbf{\textcolor{4DAF4A}{환불}}, 결제, 고객
& \textbf{\textcolor{4DAF4A}{product}}, \textbf{\textcolor{4DAF4A}{person}}, \textbf{\textcolor{4DAF4A}{company}}, room, \textbf{\textcolor{4DAF4A}{price}}, sale, store, \textbf{\textcolor{4DAF4A}{order}}, size, color, \textbf{\textcolor{4DAF4A}{problem}}, \textbf{\textcolor{4DAF4A}{refund}}, discount, meeting, \textbf{\textcolor{4DAF4A}{food}}, \textbf{\textcolor{4DAF4A}{delivery}}, \textbf{\textcolor{4DAF4A}{reservation}}, exchange, charge, business
& 사진, 포장, 물건, 음료, 이름, 종류, 주소, 업무, 발송, 소리, 촬영, 계획, 효과, 야근, \textbf{\textcolor{984EA3}{시장}}, 도움, 홈페이지, 손님, \textbf{\textcolor{984EA3}{신발}}, 무료
& price, store, problem, discount, office, home, \textbf{\textcolor{984EA3}{shoes}}, \textbf{\textcolor{984EA3}{market}}, museum, bathroom, entrance, corner, foot, copy, noodle, salary, bread, library, cheese, clothing
& \textbf{\textcolor{FF7F00}{사람}}, \textbf{\textcolor{FF7F00}{감사}}, \textbf{\textcolor{FF7F00}{도서}}, \textbf{\textcolor{FF7F00}{제품}}, 주문, \textbf{\textcolor{FF7F00}{친구}}, \textbf{\textcolor{FF7F00}{회의}}, \textbf{\textcolor{FF7F00}{회사}}, \textbf{\textcolor{FF7F00}{걱정}}, 준비, \textbf{\textcolor{FF7F00}{문제}}, \textbf{\textcolor{FF7F00}{포장}}, 운동, 사이즈, \textbf{\textcolor{FF7F00}{화장실}}, 가방, 추가, 배송, \textbf{\textcolor{FF7F00}{커피}}, \textbf{\textcolor{FF7F00}{도착}}
& \textbf{\textcolor{FF7F00}{person}}, \textbf{\textcolor{FF7F00}{thanks}}, \textbf{\textcolor{FF7F00}{book}}, \textbf{\textcolor{FF7F00}{product}}, photo, \textbf{\textcolor{FF7F00}{friend}}, \textbf{\textcolor{FF7F00}{meeting}}, \textbf{\textcolor{FF7F00}{company}}, \textbf{\textcolor{FF7F00}{worry}}, item, part, \textbf{\textcolor{FF7F00}{problem}}, professor, \textbf{\textcolor{FF7F00}{packaging}}, moment, picture, \textbf{\textcolor{FF7F00}{restroom}}, \textbf{\textcolor{FF7F00}{coffee}}, \textbf{\textcolor{FF7F00}{arrival}}, presentation \\

\textbf{Daily Life}
& \textbf{\textcolor{4DAF4A}{제품}}, \textbf{\textcolor{4DAF4A}{회사}}, \textbf{\textcolor{4DAF4A}{주문}}, 소개, 귀사, \textbf{\textcolor{4DAF4A}{고객}}, 구매, \textbf{\textcolor{4DAF4A}{배송}}, 제공, \textbf{\textcolor{4DAF4A}{가격}}, 생각, 안녕, 관심, 판매, \textbf{\textcolor{4DAF4A}{문제}}, 귀하, 출시, 생산, 당사
& \textbf{\textcolor{4DAF4A}{product}}, \textbf{\textcolor{4DAF4A}{company}}, \textbf{\textcolor{4DAF4A}{price}}, \textbf{\textcolor{4DAF4A}{customer}}, quality, \textbf{\textcolor{4DAF4A}{order}}, \textbf{\textcolor{4DAF4A}{problem}}, person, service, design, sample, \textbf{\textcolor{4DAF4A}{delivery}}, question, information, catalog, stock, market, material, color, email
& \textbf{\textcolor{984EA3}{주문}}, 귀사, 연락, 확인, 배송, \textbf{\textcolor{984EA3}{사람}}, \textbf{\textcolor{984EA3}{상품}}, 귀하, 말씀, 메일, \textbf{\textcolor{984EA3}{사항}}, 요청, 부탁, \textbf{\textcolor{984EA3}{정보}}, \textbf{\textcolor{984EA3}{고객}}, 관심, 상황, 전화, 회신, \textbf{\textcolor{984EA3}{가격}}
& \textbf{\textcolor{984EA3}{order}}, \textbf{\textcolor{984EA3}{person}}, quality, item, store, \textbf{\textcolor{984EA3}{matters}}, power, \textbf{\textcolor{984EA3}{information}}, taste, ingredient, \textbf{\textcolor{984EA3}{customer}}, \textbf{\textcolor{984EA3}{price}}, idea, catalog, inconvenience, list, opportunity, \textbf{\textcolor{984EA3}{product}}, supplier, sound
& \textbf{\textcolor{FF7F00}{제품}}, \textbf{\textcolor{FF7F00}{연락}}, \textbf{\textcolor{FF7F00}{주문}}, 생각, 가능, \textbf{\textcolor{FF7F00}{문제}}, \textbf{\textcolor{FF7F00}{회사}}, \textbf{\textcolor{FF7F00}{확인}}, 사용, \textbf{\textcolor{FF7F00}{고객}}, \textbf{\textcolor{FF7F00}{배송}}, 사람, 필요, \textbf{\textcolor{FF7F00}{메일}}, \textbf{\textcolor{FF7F00}{회신}}, \textbf{\textcolor{FF7F00}{귀하}}, \textbf{\textcolor{FF7F00}{준비}}, \textbf{\textcolor{FF7F00}{상품}}, 부탁, \textbf{\textcolor{FF7F00}{얘기}}
& \textbf{\textcolor{FF7F00}{product}}, \textbf{\textcolor{FF7F00}{contact}}, thing, \textbf{\textcolor{FF7F00}{order}}, \textbf{\textcolor{FF7F00}{problem}}, \textbf{\textcolor{FF7F00}{company}}, column, \textbf{\textcolor{FF7F00}{confirmation}}, \textbf{\textcolor{FF7F00}{dear}}, \textbf{\textcolor{FF7F00}{customer}}, \textbf{\textcolor{FF7F00}{delivery}}, character, \textbf{\textcolor{FF7F00}{reply}}, kind, \textbf{\textcolor{FF7F00}{preparation}}, list, detail, \textbf{\textcolor{FF7F00}{email}}, language, \textbf{\textcolor{FF7F00}{story}} \\

\textbf{Humanities}
& \textbf{\textcolor{4DAF4A}{사람}}, \textbf{\textcolor{4DAF4A}{교육}}, 생각, \textbf{\textcolor{4DAF4A}{사회}}, \textbf{\textcolor{4DAF4A}{문제}}, 필요, \textbf{\textcolor{4DAF4A}{학교}}, \textbf{\textcolor{4DAF4A}{미국}}, 인간, \textbf{\textcolor{4DAF4A}{세계}}, 경제, 과정, 존재, \textbf{\textcolor{4DAF4A}{일본}}, \textbf{\textcolor{4DAF4A}{한국}}, \textbf{\textcolor{4DAF4A}{학생}}, \textbf{\textcolor{4DAF4A}{나라}}, 마음, 연구, \textbf{\textcolor{4DAF4A}{사실}}
& \textbf{\textcolor{4DAF4A}{person}}, \textbf{\textcolor{4DAF4A}{education}}, \textbf{\textcolor{4DAF4A}{school}}, child, life, \textbf{\textcolor{4DAF4A}{student}}, teacher, \textbf{\textcolor{4DAF4A}{world}}, \textbf{\textcolor{4DAF4A}{korea}}, \textbf{\textcolor{4DAF4A}{usa}}, system, \textbf{\textcolor{4DAF4A}{country}}, \textbf{\textcolor{4DAF4A}{society}}, government, order, \textbf{\textcolor{4DAF4A}{japan}}, \textbf{\textcolor{4DAF4A}{fact}}, business, issue, \textbf{\textcolor{4DAF4A}{problem}}
& \textbf{\textcolor{984EA3}{미국}}, 경제, 지역, \textbf{\textcolor{984EA3}{국가}}, \textbf{\textcolor{984EA3}{정부}}, \textbf{\textcolor{984EA3}{중국}}, \textbf{\textcolor{984EA3}{정책}}, \textbf{\textcolor{984EA3}{북한}}, \textbf{\textcolor{984EA3}{독일}}, 국제, \textbf{\textcolor{984EA3}{한국}}, 세계, 유럽, 무역, 전쟁, \textbf{\textcolor{984EA3}{대통령}}, 협상, 상황, 영국, 위기
& \textbf{\textcolor{984EA3}{korea}}, \textbf{\textcolor{984EA3}{usa}}, japan, \textbf{\textcolor{984EA3}{government}}, \textbf{\textcolor{984EA3}{nation}}, \textbf{\textcolor{984EA3}{northkorea}}, family, \textbf{\textcolor{984EA3}{china}}, \textbf{\textcolor{984EA3}{policy}}, member, conflict, peace, \textbf{\textcolor{984EA3}{germany}}, national, \textbf{\textcolor{984EA3}{president}}, organization, minister, century, issue, security
& \textbf{\textcolor{FF7F00}{교육}}, \textbf{\textcolor{FF7F00}{사회}}, \textbf{\textcolor{FF7F00}{생각}}, \textbf{\textcolor{FF7F00}{문제}}, \textbf{\textcolor{FF7F00}{경제}}, \textbf{\textcolor{FF7F00}{국가}}, 인간, 연구, \textbf{\textcolor{FF7F00}{세계}}, 기업, 한국, \textbf{\textcolor{FF7F00}{교사}}, \textbf{\textcolor{FF7F00}{정부}}, \textbf{\textcolor{FF7F00}{시장}}, \textbf{\textcolor{FF7F00}{미국}}, 사용, \textbf{\textcolor{FF7F00}{활동}}, \textbf{\textcolor{FF7F00}{지역}}, \textbf{\textcolor{FF7F00}{상황}}, 관계
& \textbf{\textcolor{FF7F00}{education}}, \textbf{\textcolor{FF7F00}{society}}, \textbf{\textcolor{FF7F00}{thinking}}, \textbf{\textcolor{FF7F00}{economy}}, \textbf{\textcolor{FF7F00}{nation}}, \textbf{\textcolor{FF7F00}{problem}}, \textbf{\textcolor{FF7F00}{teacher}}, student, \textbf{\textcolor{FF7F00}{world}}, system, \textbf{\textcolor{FF7F00}{unitedstates}}, \textbf{\textcolor{FF7F00}{market}}, \textbf{\textcolor{FF7F00}{government}}, product, child, order, \textbf{\textcolor{FF7F00}{activity}}, \textbf{\textcolor{FF7F00}{region}}, study, \textbf{\textcolor{FF7F00}{situation}} \\

\textbf{Korean Culture} 
& \textbf{\textcolor{4DAF4A}{한국}}, 관습, \textbf{\textcolor{4DAF4A}{영화}}, \textbf{\textcolor{4DAF4A}{사람}}, \textbf{\textcolor{4DAF4A}{행사}}, \textbf{\textcolor{4DAF4A}{드라마}}, 한류, 관심, \textbf{\textcolor{4DAF4A}{공연}}, \textbf{\textcolor{4DAF4A}{음식}}, 인기, 중국, \textbf{\textcolor{4DAF4A}{일본}}, \textbf{\textcolor{4DAF4A}{세계}}, 지역, \textbf{\textcolor{4DAF4A}{한국어}}, \textbf{\textcolor{4DAF4A}{음악}}, 전통, \textbf{\textcolor{4DAF4A}{학생}}, 작품
& \textbf{\textcolor{4DAF4A}{person}}, \textbf{\textcolor{4DAF4A}{korea}}, \textbf{\textcolor{4DAF4A}{korean}}, \textbf{\textcolor{4DAF4A}{event}}, culture, festival, \textbf{\textcolor{4DAF4A}{film}}, wave, country, performance, fan, \textbf{\textcolor{4DAF4A}{food}}, \textbf{\textcolor{4DAF4A}{drama}}, \textbf{\textcolor{4DAF4A}{music}}, \textbf{\textcolor{4DAF4A}{world}}, \textbf{\textcolor{4DAF4A}{student}}, movie, song, \textbf{\textcolor{4DAF4A}{japan}}, \textbf{\textcolor{4DAF4A}{concert}}
& \textbf{\textcolor{984EA3}{한국}}, \textbf{\textcolor{984EA3}{영화}}, \textbf{\textcolor{984EA3}{한류}}, \textbf{\textcolor{984EA3}{드라마}}, \textbf{\textcolor{984EA3}{중국}}, 인기, 현지, \textbf{\textcolor{984EA3}{관심}}, 시장, \textbf{\textcolor{984EA3}{스타}}, 미국, \textbf{\textcolor{984EA3}{세계}}, 최근, 대만, 홍콩, 방송, \textbf{\textcolor{984EA3}{영화제}}, 기사, 독일, 프로그램
& \textbf{\textcolor{984EA3}{korea}}, korean, \textbf{\textcolor{984EA3}{film}}, \textbf{\textcolor{984EA3}{drama}}, culture, person, \textbf{\textcolor{984EA3}{movie}}, country, fan, \textbf{\textcolor{984EA3}{china}}, \textbf{\textcolor{984EA3}{star}}, group, media, \textbf{\textcolor{984EA3}{world}}, japan, \textbf{\textcolor{984EA3}{hallyu}}, interest, content, industry, \textbf{\textcolor{984EA3}{attention}}
& \textbf{\textcolor{FF7F00}{한국}}, \textbf{\textcolor{FF7F00}{행사}}, \textbf{\textcolor{FF7F00}{문화}}, \textbf{\textcolor{FF7F00}{공연}}, \textbf{\textcolor{FF7F00}{사람}}, 지역, 드라마, 영화, \textbf{\textcolor{FF7F00}{한류}}, \textbf{\textcolor{FF7F00}{조선}}, \textbf{\textcolor{FF7F00}{세계}}, \textbf{\textcolor{FF7F00}{음식}}, 전통, \textbf{\textcolor{FF7F00}{스타}}, 무대, \textbf{\textcolor{FF7F00}{중국}}, \textbf{\textcolor{FF7F00}{대회}}, \textbf{\textcolor{FF7F00}{일본}}, 마을, \textbf{\textcolor{FF7F00}{부산}}
& \textbf{\textcolor{FF7F00}{korea}}, \textbf{\textcolor{FF7F00}{performance}}, \textbf{\textcolor{FF7F00}{concert}}, \textbf{\textcolor{FF7F00}{culture}}, \textbf{\textcolor{FF7F00}{person}}, \textbf{\textcolor{FF7F00}{food}}, music, \textbf{\textcolor{FF7F00}{joseon}}, \textbf{\textcolor{FF7F00}{hallyu}}, center, country, song, \textbf{\textcolor{FF7F00}{star}}, \textbf{\textcolor{FF7F00}{china}}, school, \textbf{\textcolor{FF7F00}{world}}, \textbf{\textcolor{FF7F00}{competition}}, \textbf{\textcolor{FF7F00}{japan}}, singer, \textbf{\textcolor{FF7F00}{busan}} \\

\textbf{Municipal Websites}
& \textbf{\textcolor{4DAF4A}{경기도}}, \textbf{\textcolor{4DAF4A}{사업}}, \textbf{\textcolor{4DAF4A}{지원}}, 지역, 기업, \textbf{\textcolor{4DAF4A}{교육}}, \textbf{\textcolor{4DAF4A}{센터}}, 사회, \textbf{\textcolor{4DAF4A}{개발}}, \textbf{\textcolor{4DAF4A}{주민}}, 계획, 문화, 기관, 운영, \textbf{\textcolor{4DAF4A}{시설}}, 환경, 경제, \textbf{\textcolor{4DAF4A}{시장}}, 기술, 여성
& \textbf{\textcolor{4DAF4A}{gyeonggi}}, province, government, project, city, person, company, \textbf{\textcolor{4DAF4A}{center}}, governor, \textbf{\textcolor{4DAF4A}{business}}, provincial, korea, \textbf{\textcolor{4DAF4A}{support}}, \textbf{\textcolor{4DAF4A}{resident}}, \textbf{\textcolor{4DAF4A}{development}}, service, \textbf{\textcolor{4DAF4A}{education}}, office, \textbf{\textcolor{4DAF4A}{market}}, \textbf{\textcolor{4DAF4A}{facility}}
& 지역, 시설, 도로, \textbf{\textcolor{984EA3}{경기도}}, 버스, 평택, 공원, 공간, 위치, 교통, \textbf{\textcolor{984EA3}{서울}}, 공사, 차량, \textbf{\textcolor{984EA3}{주민}}, 인구, 효과, 운행, \textbf{\textcolor{984EA3}{화성}}, 수도, 감소
& \textbf{\textcolor{984EA3}{gyeonggi}}, province, government, governor, policy, director, committee, organization, president, \textbf{\textcolor{984EA3}{hwaseong}}, council, chairman, ceremony, city, administration, \textbf{\textcolor{984EA3}{seoul}}, cooperation, association, \textbf{\textcolor{984EA3}{resident}}, office
& \textbf{\textcolor{FF7F00}{경기도}}, \textbf{\textcolor{FF7F00}{사업}}, \textbf{\textcolor{FF7F00}{지원}}, \textbf{\textcolor{FF7F00}{지역}}, 관리, \textbf{\textcolor{FF7F00}{기업}}, \textbf{\textcolor{FF7F00}{교육}}, 참여, \textbf{\textcolor{FF7F00}{추진}}, \textbf{\textcolor{FF7F00}{센터}}, 대상, \textbf{\textcolor{FF7F00}{주민}}, \textbf{\textcolor{FF7F00}{계획}}, 진행, \textbf{\textcolor{FF7F00}{시설}}, 실시, \textbf{\textcolor{FF7F00}{기관}}, 사회, 문화, \textbf{\textcolor{FF7F00}{개발}}
& \textbf{\textcolor{FF7F00}{gyeonggi}}, \textbf{\textcolor{FF7F00}{business}}, \textbf{\textcolor{FF7F00}{region}}, \textbf{\textcolor{FF7F00}{support}}, government, \textbf{\textcolor{FF7F00}{company}}, \textbf{\textcolor{FF7F00}{education}}, \textbf{\textcolor{FF7F00}{promotion}}, \textbf{\textcolor{FF7F00}{center}}, \textbf{\textcolor{FF7F00}{resident}}, investment, result, \textbf{\textcolor{FF7F00}{planning}}, addition, \textbf{\textcolor{FF7F00}{facility}}, office, \textbf{\textcolor{FF7F00}{institution}}, information, \textbf{\textcolor{FF7F00}{development}}, system \\

\textbf{News}
& \textbf{\textcolor{4DAF4A}{미국}}, \textbf{\textcolor{4DAF4A}{대통령}}, \textbf{\textcolor{4DAF4A}{한국}}, \textbf{\textcolor{4DAF4A}{정부}}, \textbf{\textcolor{4DAF4A}{시장}}, \textbf{\textcolor{4DAF4A}{기업}}, 대표, \textbf{\textcolor{4DAF4A}{중국}}, 지역, \textbf{\textcolor{4DAF4A}{서울}}, 문제, \textbf{\textcolor{4DAF4A}{사람}}, 경제, \textbf{\textcolor{4DAF4A}{북한}}, 조사, \textbf{\textcolor{4DAF4A}{경찰}}, \textbf{\textcolor{4DAF4A}{사업}}, 상황, 필요, \textbf{\textcolor{4DAF4A}{의원}}
& \textbf{\textcolor{4DAF4A}{korea}}, \textbf{\textcolor{4DAF4A}{person}}, \textbf{\textcolor{4DAF4A}{northkorea}}, party, \textbf{\textcolor{4DAF4A}{company}}, \textbf{\textcolor{4DAF4A}{government}}, \textbf{\textcolor{4DAF4A}{president}}, \textbf{\textcolor{4DAF4A}{seoul}}, \textbf{\textcolor{4DAF4A}{market}}, city, service, country, \textbf{\textcolor{4DAF4A}{china}}, \textbf{\textcolor{4DAF4A}{police}}, \textbf{\textcolor{4DAF4A}{chairman}}, \textbf{\textcolor{4DAF4A}{business}}, \textbf{\textcolor{4DAF4A}{usa}}, park, school, member
& \textbf{\textcolor{984EA3}{시장}}, 국내, \textbf{\textcolor{984EA3}{기업}}, 금융, 세계, \textbf{\textcolor{984EA3}{투자}}, \textbf{\textcolor{984EA3}{가격}}, \textbf{\textcolor{984EA3}{사업}}, 규모, 중국, \textbf{\textcolor{984EA3}{제품}}, \textbf{\textcolor{984EA3}{은행}}, \textbf{\textcolor{984EA3}{판매}}, 해외, 자동차, 미국, \textbf{\textcolor{984EA3}{삼성}}, 상승, 분석, \textbf{\textcolor{984EA3}{업계}}
& \textbf{\textcolor{984EA3}{company}}, \textbf{\textcolor{984EA3}{market}}, \textbf{\textcolor{984EA3}{industry}}, \textbf{\textcolor{984EA3}{business}}, \textbf{\textcolor{984EA3}{product}}, \textbf{\textcolor{984EA3}{sale}}, \textbf{\textcolor{984EA3}{investment}}, technology, service, stock, \textbf{\textcolor{984EA3}{price}}, \textbf{\textcolor{984EA3}{bank}}, \textbf{\textcolor{984EA3}{samsung}}, customer, fund, consumer, profit, card, contract, employee
& \textbf{\textcolor{FF7F00}{대표}}, \textbf{\textcolor{FF7F00}{기업}}, 지역, \textbf{\textcolor{FF7F00}{한국}}, \textbf{\textcolor{FF7F00}{서울}}, \textbf{\textcolor{FF7F00}{대통령}}, \textbf{\textcolor{FF7F00}{정부}}, \textbf{\textcolor{FF7F00}{시장}}, \textbf{\textcolor{FF7F00}{사업}}, 의원, 경찰, \textbf{\textcolor{FF7F00}{문제}}, \textbf{\textcolor{FF7F00}{경제}}, 경기, \textbf{\textcolor{FF7F00}{사람}}, \textbf{\textcolor{FF7F00}{중국}}, \textbf{\textcolor{FF7F00}{사회}}, 사건, \textbf{\textcolor{FF7F00}{수사}}, 금융
& \textbf{\textcolor{FF7F00}{representative}}, \textbf{\textcolor{FF7F00}{company}}, \textbf{\textcolor{FF7F00}{korea}}, \textbf{\textcolor{FF7F00}{seoul}}, \textbf{\textcolor{FF7F00}{government}}, \textbf{\textcolor{FF7F00}{president}}, province, \textbf{\textcolor{FF7F00}{business}}, \textbf{\textcolor{FF7F00}{market}}, director, country, \textbf{\textcolor{FF7F00}{economy}}, \textbf{\textcolor{FF7F00}{problem}}, expert, \textbf{\textcolor{FF7F00}{person}}, \textbf{\textcolor{FF7F00}{china}}, \textbf{\textcolor{FF7F00}{investigation}}, stock, \textbf{\textcolor{FF7F00}{society}}, bank \\

\textbf{Professional Field}
& \textbf{\textcolor{4DAF4A}{마을}}, \textbf{\textcolor{4DAF4A}{환자}}, \textbf{\textcolor{4DAF4A}{코로나}}, 지역, \textbf{\textcolor{4DAF4A}{병원}}, \textbf{\textcolor{4DAF4A}{사람}}, \textbf{\textcolor{4DAF4A}{치료}}, \textbf{\textcolor{4DAF4A}{국내}}, \textbf{\textcolor{4DAF4A}{서울}}, \textbf{\textcolor{4DAF4A}{시장}}, 검사, \textbf{\textcolor{4DAF4A}{감염}}, 바이러스, 금융, \textbf{\textcolor{4DAF4A}{학교}}, 신종, 의료, 대표, \textbf{\textcolor{4DAF4A}{기업}}, 운영
& \textbf{\textcolor{4DAF4A}{person}}, \textbf{\textcolor{4DAF4A}{patient}}, \textbf{\textcolor{4DAF4A}{korea}}, \textbf{\textcolor{4DAF4A}{hospital}}, \textbf{\textcolor{4DAF4A}{school}}, \textbf{\textcolor{4DAF4A}{company}}, center, disease, \textbf{\textcolor{4DAF4A}{covid}}, \textbf{\textcolor{4DAF4A}{market}}, health, service, \textbf{\textcolor{4DAF4A}{village}}, \textbf{\textcolor{4DAF4A}{seoul}}, \textbf{\textcolor{4DAF4A}{treatment}}, child, student, government, \textbf{\textcolor{4DAF4A}{infection}}, bank
& 금융, 시장, \textbf{\textcolor{984EA3}{투자}}, \textbf{\textcolor{984EA3}{기업}}, \textbf{\textcolor{984EA3}{은행}}, 원고, \textbf{\textcolor{984EA3}{서비스}}, 대출, 가능, \textbf{\textcolor{984EA3}{사업}}, 기술, 상품, \textbf{\textcolor{984EA3}{회사}}, 자금, \textbf{\textcolor{984EA3}{개발}}, \textbf{\textcolor{984EA3}{보험}}, \textbf{\textcolor{984EA3}{고객}}, 자산, \textbf{\textcolor{984EA3}{주식}}, 주택
& \textbf{\textcolor{984EA3}{company}}, \textbf{\textcolor{984EA3}{bank}}, \textbf{\textcolor{984EA3}{service}}, fund, \textbf{\textcolor{984EA3}{investment}}, \textbf{\textcolor{984EA3}{business}}, industry, management, financial, \textbf{\textcolor{984EA3}{insurance}}, information, \textbf{\textcolor{984EA3}{stock}}, \textbf{\textcolor{984EA3}{customer}}, government, credit, security, \textbf{\textcolor{984EA3}{development}}, policy, economy, sector
& \textbf{\textcolor{FF7F00}{경우}}, 지역, 마을, \textbf{\textcolor{FF7F00}{환자}}, \textbf{\textcolor{FF7F00}{금융}}, \textbf{\textcolor{FF7F00}{경기}}, \textbf{\textcolor{FF7F00}{선수}}, \textbf{\textcolor{FF7F00}{원고}}, 가능, \textbf{\textcolor{FF7F00}{병원}}, \textbf{\textcolor{FF7F00}{사건}}, \textbf{\textcolor{FF7F00}{코로나}}, \textbf{\textcolor{FF7F00}{시장}}, \textbf{\textcolor{FF7F00}{대표}}, \textbf{\textcolor{FF7F00}{기업}}, \textbf{\textcolor{FF7F00}{사업}}, \textbf{\textcolor{FF7F00}{한국}}, 지원, 서비스, 운영
& \textbf{\textcolor{FF7F00}{case}}, \textbf{\textcolor{FF7F00}{korea}}, \textbf{\textcolor{FF7F00}{patient}}, \textbf{\textcolor{FF7F00}{company}}, \textbf{\textcolor{FF7F00}{finance}}, team, \textbf{\textcolor{FF7F00}{manuscript}}, \textbf{\textcolor{FF7F00}{player}}, \textbf{\textcolor{FF7F00}{game}}, \textbf{\textcolor{FF7F00}{market}}, \textbf{\textcolor{FF7F00}{hospital}}, \textbf{\textcolor{FF7F00}{incident}}, \textbf{\textcolor{FF7F00}{representative}}, \textbf{\textcolor{FF7F00}{covid}}, disease, \textbf{\textcolor{FF7F00}{business}}, lotte, school, seoul, city \\

\textbf{Regulations}
& \textbf{\textcolor{4DAF4A}{위원회}}, 구청장, \textbf{\textcolor{4DAF4A}{운영}}, \textbf{\textcolor{4DAF4A}{지원}}, \textbf{\textcolor{4DAF4A}{사항}}, \textbf{\textcolor{4DAF4A}{의원}}, \textbf{\textcolor{4DAF4A}{조례}}, 규정, \textbf{\textcolor{4DAF4A}{관리}}, 사업, \textbf{\textcolor{4DAF4A}{위원장}}, 예산, 기관, 위탁, \textbf{\textcolor{4DAF4A}{시설}}, 심의, 조제, \textbf{\textcolor{4DAF4A}{서울특별시}}, \textbf{\textcolor{4DAF4A}{공무원}}, \textbf{\textcolor{4DAF4A}{도지사}}
& head, \textbf{\textcolor{4DAF4A}{member}}, \textbf{\textcolor{4DAF4A}{committee}}, article, \textbf{\textcolor{4DAF4A}{chairperson}}, \textbf{\textcolor{4DAF4A}{matters}}, \textbf{\textcolor{4DAF4A}{ordinance}}, metropolitan, government, \textbf{\textcolor{4DAF4A}{operation}}, \textbf{\textcolor{4DAF4A}{seoul}}, \textbf{\textcolor{4DAF4A}{management}}, \textbf{\textcolor{4DAF4A}{facility}}, \textbf{\textcolor{4DAF4A}{support}}, council, project, \textbf{\textcolor{4DAF4A}{official}}, center, education, \textbf{\textcolor{4DAF4A}{governor}}
& \textbf{\textcolor{984EA3}{서식}}, 별지, \textbf{\textcolor{984EA3}{구청장}}, \textbf{\textcolor{984EA3}{제출}}, \textbf{\textcolor{984EA3}{납부}}, 징수, 신청, 기간, 재산, \textbf{\textcolor{984EA3}{관리}}, 시설, 허가, \textbf{\textcolor{984EA3}{융자}}, 물품, 작성, 승인, 사업, \textbf{\textcolor{984EA3}{기금}}, 교부, 차량
& \textbf{\textcolor{984EA3}{mayor}}, \textbf{\textcolor{984EA3}{form}}, \textbf{\textcolor{984EA3}{submission}}, \textbf{\textcolor{984EA3}{payment}}, \textbf{\textcolor{984EA3}{fund}}, report, audit, \textbf{\textcolor{984EA3}{loan}}, result, request, investigation, budget, matter, receipt, official, \textbf{\textcolor{984EA3}{management}}, proposal, subsidy, agency, account
& \textbf{\textcolor{FF7F00}{구청장}}, 필요, \textbf{\textcolor{FF7F00}{사항}}, \textbf{\textcolor{FF7F00}{조례}}, \textbf{\textcolor{FF7F00}{지원}}, 규정, \textbf{\textcolor{FF7F00}{위원회}}, \textbf{\textcolor{FF7F00}{운영}}, 관리, 해당, \textbf{\textcolor{FF7F00}{기관}}, \textbf{\textcolor{FF7F00}{사업}}, \textbf{\textcolor{FF7F00}{서울특별시}}, \textbf{\textcolor{FF7F00}{도지사}}, \textbf{\textcolor{FF7F00}{공무원}}, 업무, \textbf{\textcolor{FF7F00}{시설}}, 목적, \textbf{\textcolor{FF7F00}{서식}}, \textbf{\textcolor{FF7F00}{예산}}
& \textbf{\textcolor{FF7F00}{matter}}, \textbf{\textcolor{FF7F00}{committee}}, \textbf{\textcolor{FF7F00}{mayor}}, \textbf{\textcolor{FF7F00}{ordinance}}, \textbf{\textcolor{FF7F00}{support}}, meeting, director, \textbf{\textcolor{FF7F00}{operation}}, fund, \textbf{\textcolor{FF7F00}{institution}}, \textbf{\textcolor{FF7F00}{business}}, \textbf{\textcolor{FF7F00}{governor}}, \textbf{\textcolor{FF7F00}{seoul}}, traffic, plan, \textbf{\textcolor{FF7F00}{official}}, duty, \textbf{\textcolor{FF7F00}{facility}}, \textbf{\textcolor{FF7F00}{form}}, \textbf{\textcolor{FF7F00}{budget}} \\

\textbf{Social Science}
& \textbf{\textcolor{4DAF4A}{연구}}, \textbf{\textcolor{4DAF4A}{교육}}, \textbf{\textcolor{4DAF4A}{경우}}, \textbf{\textcolor{4DAF4A}{결과}}, 필요, \textbf{\textcolor{4DAF4A}{분석}}, \textbf{\textcolor{4DAF4A}{문제}}, 학습, \textbf{\textcolor{4DAF4A}{학생}}, 사회, \textbf{\textcolor{4DAF4A}{기업}}, 영향, \textbf{\textcolor{4DAF4A}{관계}}, \textbf{\textcolor{4DAF4A}{교사}}, \textbf{\textcolor{4DAF4A}{활동}}, \textbf{\textcolor{4DAF4A}{학교}}, \textbf{\textcolor{4DAF4A}{효과}}, \textbf{\textcolor{4DAF4A}{방법}}, 규정, 의미
& study, \textbf{\textcolor{4DAF4A}{education}}, \textbf{\textcolor{4DAF4A}{case}}, \textbf{\textcolor{4DAF4A}{school}}, \textbf{\textcolor{4DAF4A}{student}}, \textbf{\textcolor{4DAF4A}{result}}, system, \textbf{\textcolor{4DAF4A}{company}}, \textbf{\textcolor{4DAF4A}{effect}}, person, \textbf{\textcolor{4DAF4A}{problem}}, \textbf{\textcolor{4DAF4A}{teacher}}, \textbf{\textcolor{4DAF4A}{research}}, \textbf{\textcolor{4DAF4A}{method}}, korea, \textbf{\textcolor{4DAF4A}{analysis}}, level, content, \textbf{\textcolor{4DAF4A}{relationship}}, \textbf{\textcolor{4DAF4A}{activity}}
& \textbf{\textcolor{984EA3}{사회}}, \textbf{\textcolor{984EA3}{문화}}, 사람, \textbf{\textcolor{984EA3}{표현}}, 생각, \textbf{\textcolor{984EA3}{예술}}, 인간, \textbf{\textcolor{984EA3}{공연}}, \textbf{\textcolor{984EA3}{음악}}, \textbf{\textcolor{984EA3}{작품}}, \textbf{\textcolor{984EA3}{의미}}, 공간, 존재, 이해, \textbf{\textcolor{984EA3}{전통}}, \textbf{\textcolor{984EA3}{역사}}, 시대, \textbf{\textcolor{984EA3}{자연}}, 무용, \textbf{\textcolor{984EA3}{연극}}
& \textbf{\textcolor{984EA3}{performance}}, \textbf{\textcolor{984EA3}{work}}, \textbf{\textcolor{984EA3}{art}}, \textbf{\textcolor{984EA3}{culture}}, \textbf{\textcolor{984EA3}{society}}, \textbf{\textcolor{984EA3}{music}}, term, \textbf{\textcolor{984EA3}{meaning}}, world, \textbf{\textcolor{984EA3}{tradition}}, element, reality, \textbf{\textcolor{984EA3}{nature}}, power, \textbf{\textcolor{984EA3}{history}}, body, character, \textbf{\textcolor{984EA3}{theater}}, \textbf{\textcolor{984EA3}{expression}}, sense
& \textbf{\textcolor{FF7F00}{경우}}, \textbf{\textcolor{FF7F00}{연구}}, \textbf{\textcolor{FF7F00}{교육}}, \textbf{\textcolor{FF7F00}{기업}}, 문제, \textbf{\textcolor{FF7F00}{규정}}, \textbf{\textcolor{FF7F00}{분석}}, \textbf{\textcolor{FF7F00}{행위}}, 사회, \textbf{\textcolor{FF7F00}{관계}}, 가능, \textbf{\textcolor{FF7F00}{학생}}, \textbf{\textcolor{FF7F00}{국가}}, \textbf{\textcolor{FF7F00}{정보}}, 의미, \textbf{\textcolor{FF7F00}{영향}}, \textbf{\textcolor{FF7F00}{지역}}, 활동, \textbf{\textcolor{FF7F00}{학습}}, 효과
& \textbf{\textcolor{FF7F00}{case}}, \textbf{\textcolor{FF7F00}{research}}, system, \textbf{\textcolor{FF7F00}{company}}, \textbf{\textcolor{FF7F00}{education}}, \textbf{\textcolor{FF7F00}{analysis}}, \textbf{\textcolor{FF7F00}{regulation}}, \textbf{\textcolor{FF7F00}{action}}, school, \textbf{\textcolor{FF7F00}{relationship}}, \textbf{\textcolor{FF7F00}{information}}, \textbf{\textcolor{FF7F00}{student}}, \textbf{\textcolor{FF7F00}{nation}}, content, method, \textbf{\textcolor{FF7F00}{influence}}, \textbf{\textcolor{FF7F00}{region}}, level, \textbf{\textcolor{FF7F00}{learning}}, value \\

\textbf{Spoken}
& \textbf{\textcolor{4DAF4A}{사람}}, \textbf{\textcolor{4DAF4A}{친구}}, \textbf{\textcolor{4DAF4A}{한국}}, \textbf{\textcolor{4DAF4A}{사랑}}, 필요, 마음, 행복, \textbf{\textcolor{4DAF4A}{사진}}, 여자, 남자, 이야기, \textbf{\textcolor{4DAF4A}{영화}}, 문제, \textbf{\textcolor{4DAF4A}{영어}}, 여행, \textbf{\textcolor{4DAF4A}{이유}}, 노력, \textbf{\textcolor{4DAF4A}{나라}}, \textbf{\textcolor{4DAF4A}{음식}}, \textbf{\textcolor{4DAF4A}{세상}}
& \textbf{\textcolor{4DAF4A}{person}}, \textbf{\textcolor{4DAF4A}{friend}}, \textbf{\textcolor{4DAF4A}{korea}}, life, product, place, \textbf{\textcolor{4DAF4A}{world}}, work, charge, \textbf{\textcolor{4DAF4A}{country}}, company, \textbf{\textcolor{4DAF4A}{food}}, family, \textbf{\textcolor{4DAF4A}{photo}}, \textbf{\textcolor{4DAF4A}{english}}, \textbf{\textcolor{4DAF4A}{reason}}, money, \textbf{\textcolor{4DAF4A}{film}}, house, \textbf{\textcolor{4DAF4A}{love}}
& \textbf{\textcolor{984EA3}{제품}}, 확인, 요청, \textbf{\textcolor{984EA3}{데이터}}, 아래, 이름, 첨부, 판매, 경험, 파일, 기술, 연구, 생산, \textbf{\textcolor{984EA3}{영어}}, 인터넷, 귀사, \textbf{\textcolor{984EA3}{이미지}}, 공간, 환불, 발전
& \textbf{\textcolor{984EA3}{product}}, friend, \textbf{\textcolor{984EA3}{english}}, sample, \textbf{\textcolor{984EA3}{data}}, china, member, quality, application, environment, decision, skill, opportunity, department, post, interview, town, \textbf{\textcolor{984EA3}{image}}, region, comment
& \textbf{\textcolor{FF7F00}{사람}}, \textbf{\textcolor{FF7F00}{친구}}, \textbf{\textcolor{FF7F00}{사진}}, \textbf{\textcolor{FF7F00}{사랑}}, 행복, \textbf{\textcolor{FF7F00}{마음}}, \textbf{\textcolor{FF7F00}{영어}}, 필요, 제품, \textbf{\textcolor{FF7F00}{문제}}, 연락, \textbf{\textcolor{FF7F00}{오늘}}, 사용, \textbf{\textcolor{FF7F00}{감사}}, \textbf{\textcolor{FF7F00}{영화}}, \textbf{\textcolor{FF7F00}{남자}}, \textbf{\textcolor{FF7F00}{여자}}, \textbf{\textcolor{FF7F00}{여행}}, \textbf{\textcolor{FF7F00}{공부}}, 학교
& \textbf{\textcolor{FF7F00}{person}}, \textbf{\textcolor{FF7F00}{friend}}, \textbf{\textcolor{FF7F00}{photo}}, \textbf{\textcolor{FF7F00}{love}}, life, \textbf{\textcolor{FF7F00}{english}}, book, \textbf{\textcolor{FF7F00}{problem}}, student, \textbf{\textcolor{FF7F00}{mind}}, \textbf{\textcolor{FF7F00}{today}}, university, teacher, \textbf{\textcolor{FF7F00}{thanks}}, \textbf{\textcolor{FF7F00}{movie}}, \textbf{\textcolor{FF7F00}{man}}, \textbf{\textcolor{FF7F00}{woman}}, \textbf{\textcolor{FF7F00}{study}}, lunch, \textbf{\textcolor{FF7F00}{travel}} \\

\textbf{Technical Science1}
& \textbf{\textcolor{4DAF4A}{장치}}, 포함, \textbf{\textcolor{4DAF4A}{발명}}, \textbf{\textcolor{4DAF4A}{정보}}, 형성, \textbf{\textcolor{4DAF4A}{데이터}}, 연구, \textbf{\textcolor{4DAF4A}{영역}}, \textbf{\textcolor{4DAF4A}{신호}}, \textbf{\textcolor{4DAF4A}{방법}}, 경우, \textbf{\textcolor{4DAF4A}{환자}}, 제공, \textbf{\textcolor{4DAF4A}{제어}}, 동작, 구성, \textbf{\textcolor{4DAF4A}{입력}}, \textbf{\textcolor{4DAF4A}{디스플레이}}, 결과, 위치
& \textbf{\textcolor{4DAF4A}{device}}, embodiment, unit, \textbf{\textcolor{4DAF4A}{information}}, \textbf{\textcolor{4DAF4A}{invention}}, user, \textbf{\textcolor{4DAF4A}{display}}, \textbf{\textcolor{4DAF4A}{data}}, image, \textbf{\textcolor{4DAF4A}{control}}, layer, \textbf{\textcolor{4DAF4A}{method}}, operation, \textbf{\textcolor{4DAF4A}{signal}}, diagram, \textbf{\textcolor{4DAF4A}{input}}, \textbf{\textcolor{4DAF4A}{patient}}, system, \textbf{\textcolor{4DAF4A}{area}}, example
& 정보, \textbf{\textcolor{984EA3}{데이터}}, \textbf{\textcolor{984EA3}{장치}}, 영상, 포함, 저장, \textbf{\textcolor{984EA3}{입력}}, 이미지, \textbf{\textcolor{984EA3}{메모리}}, 제공, 생성, 통신, \textbf{\textcolor{984EA3}{디바이스}}, \textbf{\textcolor{984EA3}{프로세서}}, 서버, 수행, 상기, \textbf{\textcolor{984EA3}{디스플레이}}, 처리, \textbf{\textcolor{984EA3}{신호}}
& \textbf{\textcolor{984EA3}{device}}, \textbf{\textcolor{984EA3}{signal}}, circuit, voltage, \textbf{\textcolor{984EA3}{memory}}, \textbf{\textcolor{984EA3}{display}}, operation, output, \textbf{\textcolor{984EA3}{input}}, \textbf{\textcolor{984EA3}{data}}, \textbf{\textcolor{984EA3}{processor}}, reference, touch, content, panel, frequency, controller, transmission, item, node
& 상기, \textbf{\textcolor{FF7F00}{제어}}, \textbf{\textcolor{FF7F00}{장치}}, 실시, 포함, \textbf{\textcolor{FF7F00}{환자}}, \textbf{\textcolor{FF7F00}{연구}}, \textbf{\textcolor{FF7F00}{정보}}, \textbf{\textcolor{FF7F00}{경우}}, \textbf{\textcolor{FF7F00}{결과}}, \textbf{\textcolor{FF7F00}{발명}}, \textbf{\textcolor{FF7F00}{형성}}, 영역, \textbf{\textcolor{FF7F00}{신호}}, 발생, \textbf{\textcolor{FF7F00}{데이터}}, \textbf{\textcolor{FF7F00}{방법}}, \textbf{\textcolor{FF7F00}{치료}}, \textbf{\textcolor{FF7F00}{분석}}, 상태
& \textbf{\textcolor{FF7F00}{device}}, \textbf{\textcolor{FF7F00}{control}}, unit, \textbf{\textcolor{FF7F00}{research}}, display, system, \textbf{\textcolor{FF7F00}{patient}}, \textbf{\textcolor{FF7F00}{information}}, \textbf{\textcolor{FF7F00}{case}}, \textbf{\textcolor{FF7F00}{result}}, \textbf{\textcolor{FF7F00}{invention}}, \textbf{\textcolor{FF7F00}{formation}}, power, \textbf{\textcolor{FF7F00}{signal}}, \textbf{\textcolor{FF7F00}{data}}, \textbf{\textcolor{FF7F00}{method}}, surface, \textbf{\textcolor{FF7F00}{treatment}}, \textbf{\textcolor{FF7F00}{analysis}}, material \\

\textbf{Technical Science2}
& \textbf{\textcolor{4DAF4A}{정보}}, 상기, \textbf{\textcolor{4DAF4A}{데이터}}, \textbf{\textcolor{4DAF4A}{장치}}, 포함, \textbf{\textcolor{4DAF4A}{입력}}, 경우, 연구, 제공, 분석, 중국, 지역, 결과, 경제, 저장, \textbf{\textcolor{4DAF4A}{미국}}, \textbf{\textcolor{4DAF4A}{디스플레이}}, 변화, 정부, \textbf{\textcolor{4DAF4A}{시스템}}
& \textbf{\textcolor{4DAF4A}{device}}, \textbf{\textcolor{4DAF4A}{data}}, user, \textbf{\textcolor{4DAF4A}{information}}, unit, image, \textbf{\textcolor{4DAF4A}{input}}, \textbf{\textcolor{4DAF4A}{system}}, \textbf{\textcolor{4DAF4A}{display}}, korea, \textbf{\textcolor{4DAF4A}{usa}}, country, control, touch, value, object, signal, content, server, operation
& \textbf{\textcolor{984EA3}{경제}}, 기업, \textbf{\textcolor{984EA3}{사회}}, 정책, \textbf{\textcolor{984EA3}{국가}}, \textbf{\textcolor{984EA3}{정치}}, \textbf{\textcolor{984EA3}{시장}}, \textbf{\textcolor{984EA3}{관계}}, 문제, 정부, 북한, 제도, 필요, 중요, \textbf{\textcolor{984EA3}{중국}}, \textbf{\textcolor{984EA3}{전략}}, 체제, \textbf{\textcolor{984EA3}{국제}}, 금융
& \textbf{\textcolor{984EA3}{economy}}, problem, right, \textbf{\textcolor{984EA3}{politics}}, \textbf{\textcolor{984EA3}{society}}, \textbf{\textcolor{984EA3}{country}}, \textbf{\textcolor{984EA3}{relationship}}, \textbf{\textcolor{984EA3}{market}}, \textbf{\textcolor{984EA3}{international}}, view, person, point, \textbf{\textcolor{984EA3}{strategy}}, interest, \textbf{\textcolor{984EA3}{china}}, concept, reduction, term, principle, regulation
& \textbf{\textcolor{FF7F00}{정보}}, 데이터, \textbf{\textcolor{FF7F00}{장치}}, 지역, \textbf{\textcolor{FF7F00}{연구}}, \textbf{\textcolor{FF7F00}{경제}}, \textbf{\textcolor{FF7F00}{기업}}, 분석, 국가, \textbf{\textcolor{FF7F00}{중국}}, \textbf{\textcolor{FF7F00}{입력}}, \textbf{\textcolor{FF7F00}{결과}}, \textbf{\textcolor{FF7F00}{시스템}}, \textbf{\textcolor{FF7F00}{한국}}, \textbf{\textcolor{FF7F00}{정부}}, \textbf{\textcolor{FF7F00}{기술}}, \textbf{\textcolor{FF7F00}{사회}}, \textbf{\textcolor{FF7F00}{구성}}, \textbf{\textcolor{FF7F00}{미국}}, 영향
& \textbf{\textcolor{FF7F00}{information}}, unit, \textbf{\textcolor{FF7F00}{device}}, image, \textbf{\textcolor{FF7F00}{research}}, \textbf{\textcolor{FF7F00}{company}}, \textbf{\textcolor{FF7F00}{economy}}, \textbf{\textcolor{FF7F00}{china}}, method, display, \textbf{\textcolor{FF7F00}{input}}, \textbf{\textcolor{FF7F00}{result}}, \textbf{\textcolor{FF7F00}{system}}, \textbf{\textcolor{FF7F00}{korea}}, \textbf{\textcolor{FF7F00}{government}}, \textbf{\textcolor{FF7F00}{technology}}, \textbf{\textcolor{FF7F00}{unitedstates}}, object, \textbf{\textcolor{FF7F00}{society}}, \textbf{\textcolor{FF7F00}{composition}} \\
\end{longtable}
\end{landscape}
\restoregeometry
\clearpage

\subsection{Inter-Metric Correlation Analysis}

\begin{figure}[h!]
    \centering
    \includegraphics[width=\linewidth]{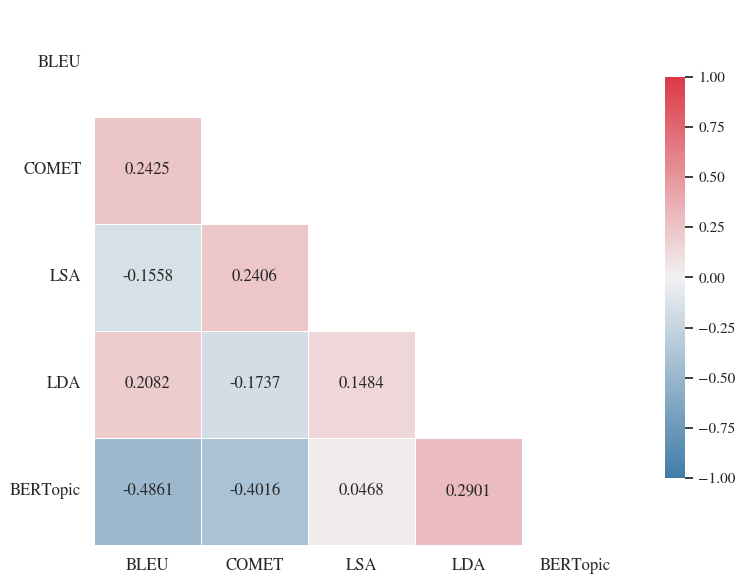}
    \caption{Pearson correlation matrix of BLEU, CometKiwi, LSA, LDA, and BERTopic across datasets.}
    \label{fig:Pearson_Results}
\end{figure}

Figure~\ref{fig:Pearson_Results} shows the Pearson correlation coefficients among BLEU, COMETKiwi, LSA, LDA, and BERTopic across 14 domain-specific datasets ($N=14$). BLEU and BERTopic exhibit a moderate negative correlation ($r=-0.49$), indicating that domains receiving higher BERTopic scores tend to receive lower BLEU scores. COMETKiwi similarly shows a moderate negative correlation with BERTopic ($r=-0.40$), suggesting that the two metrics rank domains differently despite both relying on pretrained representations. In contrast, BLEU and COMETKiwi share a weak positive correlation ($r=0.24$), implying limited alignment between exact $N$-gram overlap and learned adequacy estimation. LSA shows weak correlations overall: a weak negative correlation with BLEU ($r=-0.16$), a weak positive correlation with COMETKiwi ($r=0.24$), and a near-zero correlation with BERTopic ($r=0.05$), highlighting its relatively independent, frequency-driven perspective. Finally, LDA correlates weakly to moderately positively with BERTopic ($r=0.29$) but weakly negatively with COMETKiwi ($r=-0.17$), indicating partial alignment with BERTopic's topical structure but divergence from adequacy-oriented scoring.

These correlation patterns can be understood by considering what each metric primarily measures. Because the analysis is based on 14 domains, the reported correlations should be interpreted as descriptive tendencies within this dataset. BLEU is a surface-form metric that computes corpus-level 4-gram precision against references. It rewards exact token overlap and penalizes lexical variation, which can lead to weak or negative associations with approaches that are less sensitive to exact word choice. LSA applies singular value decomposition to the term-document matrix, capturing frequency-driven co-occurrence structure without directly referencing translation adequacy, which is consistent with its near-zero correlations with both BLEU and COMETKiwi. LDA estimates Dirichlet-multinomial topic mixtures over raw word counts, providing bag-of-words topical structure that can partially align with BERTopic’s document-level clustering, while remaining less aligned with adequacy-oriented semantic scoring. COMETKiwi fine-tunes a multilingual transformer to predict human judgments of adequacy and fluency, emphasizing meaning preservation and contextual appropriateness rather than exact phrasing. BERTopic clusters sentence embeddings to infer coherent topical structure at the document level, abstracting away from token repetition and thereby showing reduced sensitivity to domain-specific lexical realizations. Although both COMETKiwi and BERTopic rely on pretrained representations, they are intended to capture different properties. COMETKiwi is trained to predict adequacy and fluency judgments, while BERTopic focuses on estimating topical coherence from embedding-based clustering, which can lead to non-identical domain rankings. In this view, the negative correlation between COMETKiwi and BERTopic ($r=-0.40$) indicates that sharing pretrained representations does not guarantee similar domain-level behavior, because the two metrics are designed to capture different aspects of translation quality. By contrast, BLEU’s negative correlations primarily reflect its reliance on rigid $N$-gram matching.

Collectively, the analysis of the correlation matrix suggests that the five metrics assess translation quality from multifaceted perspectives rather than measuring a single, unified concept of “good translation.” Specifically, BLEU relies primarily on surface-form overlap, whereas COMETKiwi emphasizes semantic adequacy and fluency. In contrast, the topic modeling-based metrics consider the preservation and consistent representation of thematic content at the document level as a core evaluation criterion. Since each metric targets distinct evaluation objectives, high agreement among them is not inherently expected, and a high score on a specific metric does not automatically guarantee qualitative excellence in other aspects. Therefore, within the context of the 14 domains examined in this study, these findings substantiate the multi-dimensional nature of translation quality. Consequently, determining the qualitative merit of translation outputs in a dichotomous manner based solely on a single metric or score should be avoided, necessitating a cautious interpretation.

\FloatBarrier
\section{Conclusions}\label{sec:Conclusions}
This study presents a topic-based, reference-free framework for document-level machine translation evaluation, designed to produce interpretable scores without relying on human reference translations. For each Korean source document and its English translation, we extract latent topics, identify the dominant representative topic for the document, and construct topic representations using the top-ranked tokens and their associated weights. These representations are then aligned via a Korean–English bilingual dictionary, and thematic consistency is quantified using cosine similarity in the aligned space. Beyond reporting a scalar score, the proposed framework provides transparent evidence by explicitly indicating which salient tokens are matched or absent across languages, thereby supporting document-level interpretability. Experiments on a large-scale Korean–English parallel corpus spanning 14 domains, with comparisons to BLEU and CometKiwi which is a reference-free pretrained metric, demonstrate that the proposed approach offers a complementary perspective and token-level explanations for assessing translation quality beyond conventional metrics.

The experimental results indicate that the evaluation metrics exhibit different sensitivities to domain variation. BLEU produced generally low scores with substantial dispersion across domains, suggesting that it strongly reflects domain-specific difficulty, whereas COMETkiwi yielded a relatively stable distribution with scores concentrated in a narrow, high range. Among the topic-based measures, LSA showed moderate variability, LDA displayed the largest dispersion including extremely low minima, and BERTopic produced a compressed distribution similar to COMETkiwi, thereby tending to attenuate inter-domain differences. Importantly, the proposed framework goes beyond reporting score magnitudes by providing token-level evidence: using the top bilingual tokens extracted for each domain, it enables explicit interpretation of topic preservation and topic drift through concrete lexical cues. These findings suggest that our approach can serve as an auxiliary source of qualitative interpretation and validation for the outcomes of conventional evaluation metrics.

This study’s Korean–English token matching relies on dictionary-based one-to-one alignment over only the top 20 tokens, which can miss polysemy, context-dependent meaning, idioms, multiword expressions and compound nouns, asymmetric correspondences, and domain-specific out-of-vocabulary terminology. In addition, token weights are not directly comparable across topic models, and score behavior can vary with preprocessing choices and topic-number selection. Future work should move beyond lexical matching by adopting phrase- or sentence-level alignment or embedding-based alignment, strengthen validation against document-level human judgments of topic continuity and coherence, and benchmark accuracy and computational cost across document types and lengths to derive practical metric-selection guidelines.

\bibliographystyle{plainnat}
\bibliography{references}

@article{bahdanau2014neural,
  title={Neural machine translation by jointly learning to align and translate},
  author={Bahdanau, Dzmitry},
  journal={arXiv preprint arXiv:1409.0473},
  year={2014}
}

@article{vaswani2017attention,
  title={Attention is all you need},
  author={Vaswani, A},
  journal={Advances in Neural Information Processing Systems},
  year={2017}
}

@article{hassan2018achieving,
  title={Achieving human parity on automatic chinese to english news translation},
  author={Hassan, Hany and Aue, Anthony and Chen, Chang and Chowdhary, Vishal and Clark, Jonathan and Federmann, Christian and Huang, Xuedong and Junczys-Dowmunt, Marcin and Lewis, William and Li, Mu and others},
  journal={arXiv preprint arXiv:1803.05567},
  year={2018}
}

@article{miculicich2018document,
  title={Document-level neural machine translation with hierarchical attention networks},
  author={Miculicich, Lesly and Ram, Dhananjay and Pappas, Nikolaos and Henderson, James},
  journal={arXiv preprint arXiv:1809.01576},
  year={2018}
}

@article{sun2020rethinking,
  title={Rethinking document-level neural machine translation},
  author={Sun, Zewei and Wang, Mingxuan and Zhou, Hao and Zhao, Chengqi and Huang, Shujian and Chen, Jiajun and Li, Lei},
  journal={arXiv preprint arXiv:2010.08961},
  year={2020}
}

@article{zhang2022smdt,
  title={Smdt: Selective memory-augmented neural document translation},
  author={Zhang, Xu and Yang, Jian and Huang, Haoyang and Ma, Shuming and Zhang, Dongdong and Li, Jinlong and Wei, Furu},
  journal={arXiv preprint arXiv:2201.01631},
  year={2022}
}

@article{kang2021enhancing,
  title={Enhancing lexical translation consistency for document-level neural machine translation},
  author={Kang, Xiaomian and Zhao, Yang and Zhang, Jiajun and Zong, Chengqing},
  journal={Transactions on Asian and Low-Resource Language Information Processing},
  volume={21},
  number={3},
  pages={1--21},
  year={2021},
  publisher={ACM New York, NY}
}

@inproceedings{chen2022effective,
  title={Effective Graph Context Representation for Document-level Machine Translation.},
  author={Chen, Kehai and Yang, Muyun and Utiyama, Masao and Sumita, Eiichiro and Wang, Rui and Zhang, Min},
  booktitle={IJCAI},
  pages={4079--4085},
  year={2022}
}

@article{wang2023document,
  title={Document-level machine translation with large language models},
  author={Wang, Longyue and Lyu, Chenyang and Ji, Tianbo and Zhang, Zhirui and Yu, Dian and Shi, Shuming and Tu, Zhaopeng},
  journal={arXiv preprint arXiv:2304.02210},
  year={2023}
}

@inproceedings{papineni2002bleu,
    title = "{B}leu: a Method for Automatic Evaluation of Machine Translation",
    author = "Papineni, Kishore  and
      Roukos, Salim  and
      Ward, Todd  and
      Zhu, Wei-Jing",
    editor = "Isabelle, Pierre  and
      Charniak, Eugene  and
      Lin, Dekang",
    booktitle = "Proceedings of the 40th Annual Meeting of the Association for Computational Linguistics",
    month = jul,
    year = "2002",
    address = "Philadelphia, Pennsylvania, USA",
    publisher = "Association for Computational Linguistics",
    url = "https://aclanthology.org/P02-1040/",
    doi = "10.3115/1073083.1073135",
    pages = "311--318"
}

@inproceedings{banerjee2005meteor,
    title = "{METEOR}: An Automatic Metric for {MT} Evaluation with Improved Correlation with Human Judgments",
    author = "Banerjee, Satanjeev  and
      Lavie, Alon",
    editor = "Goldstein, Jade  and
      Lavie, Alon  and
      Lin, Chin-Yew  and
      Voss, Clare",
    booktitle = "Proceedings of the {ACL} Workshop on Intrinsic and Extrinsic Evaluation Measures for Machine Translation and/or Summarization",
    month = jun,
    year = "2005",
    address = "Ann Arbor, Michigan",
    publisher = "Association for Computational Linguistics",
    url = "https://aclanthology.org/W05-0909/",
    pages = "65--72"
}

@inproceedings{soricut2010trustrank,
  title={Trustrank: Inducing trust in automatic translations via ranking},
  author={Soricut, Radu and Echihabi, Abdessamad},
  booktitle={Proceedings of the 48th Annual Meeting of the Association for Computational Linguistics},
  pages={612--621},
  year={2010}
}

@article{laubli2018has,
  title={Has machine translation achieved human parity? a case for document-level evaluation},
  author={L{\"a}ubli, Samuel and Sennrich, Rico and Volk, Martin},
  journal={arXiv preprint arXiv:1808.07048},
  year={2018}
}

@inproceedings{scarton2014document,
  title={Document-level translation quality estimation: exploring discourse and pseudo-references},
  author={Scarton, Carolina and Specia, Lucia},
  booktitle={Proceedings of the 17th Annual conference of the European Association for Machine Translation},
  pages={101--108},
  year={2014}
}

@article{sellam2020bleurt,
  title={BLEURT: Learning robust metrics for text generation},
  author={Sellam, Thibault and Das, Dipanjan and Parikh, Ankur P},
  journal={arXiv preprint arXiv:2004.04696},
  year={2020}
}

@article{zhang2019bertscore,
  title={Bertscore: Evaluating text generation with bert},
  author={Zhang, Tianyi and Kishore, Varsha and Wu, Felix and Weinberger, Kilian Q and Artzi, Yoav},
  journal={arXiv preprint arXiv:1904.09675},
  year={2019}
}

@article{kocmi2023large,
  title={Large language models are state-of-the-art evaluators of translation quality},
  author={Kocmi, Tom and Federmann, Christian},
  journal={arXiv preprint arXiv:2302.14520},
  year={2023}
}

@article{rei2020comet,
  title={COMET: A neural framework for MT evaluation},
  author={Rei, Ricardo and Stewart, Craig and Farinha, Ana C and Lavie, Alon},
  journal={arXiv preprint arXiv:2009.09025},
  year={2020}
}

@inproceedings{freitag2021results,
  title={Results of the WMT21 metrics shared task: Evaluating metrics with expert-based human evaluations on TED and news domain},
  author={Freitag, Markus and Rei, Ricardo and Mathur, Nitika and Lo, Chi-kiu and Stewart, Craig and Foster, George and Lavie, Alon and Bojar, Ond{\v{r}}ej},
  booktitle={Proceedings of the Sixth Conference on Machine Translation},
  pages={733--774},
  year={2021}
}

@inproceedings{rei2021references,
  title={Are references really needed? unbabel-IST 2021 submission for the metrics shared task},
  author={Rei, Ricardo and Farinha, Ana C and Zerva, Chrysoula and van Stigt, Daan and Stewart, Craig and Ramos, Pedro and Glushkova, Taisiya and Martins, Andr{\'e} FT and Lavie, Alon},
  booktitle={Proceedings of the Sixth Conference on Machine Translation},
  pages={1030--1040},
  year={2021}
}

@inproceedings{rei2022comet,
  title={COMET-22: Unbabel-IST 2022 submission for the metrics shared task},
  author={Rei, Ricardo and De Souza, Jos{\'e} GC and Alves, Duarte and Zerva, Chrysoula and Farinha, Ana C and Glushkova, Taisiya and Lavie, Alon and Coheur, Luisa and Martins, Andr{\'e} FT},
  booktitle={Proceedings of the Seventh Conference on Machine Translation (WMT)},
  pages={578--585},
  year={2022}
}

@article{guerreiro2024xcomet,
  title={xcomet: Transparent machine translation evaluation through fine-grained error detection},
  author={Guerreiro, Nuno M and Rei, Ricardo and Stigt, Daan van and Coheur, Luisa and Colombo, Pierre and Martins, Andr{\'e} FT},
  journal={Transactions of the Association for Computational Linguistics},
  volume={12},
  pages={979--995},
  year={2024},
  publisher={MIT Press 255 Main Street, 9th Floor, Cambridge, Massachusetts 02142, USA~…}
}

@article{blei2003latent,
  title={Latent dirichlet allocation},
  author={Blei, David M and Ng, Andrew Y and Jordan, Michael I},
  journal={Journal of machine Learning research},
  volume={3},
  number={Jan},
  pages={993--1022},
  year={2003}
}

@article{grootendorst2022bertopic,
  title={BERTopic: Neural topic modeling with a class-based TF-IDF procedure},
  author={Grootendorst, Maarten},
  journal={arXiv preprint arXiv:2203.05794},
  year={2022}
}

@article{shen2023research,
  title={Research on high-performance English translation based on topic model},
  author={Shen, Yumin and Guo, Hongyu},
  journal={Digital Communications and Networks},
  volume={9},
  number={2},
  pages={505--511},
  year={2023},
  publisher={Elsevier}
}

@inproceedings{su2015context,
  title={A context-aware topic model for statistical machine translation},
  author={Su, Jinsong and Xiong, Deyi and Liu, Yang and Han, Xianpei and Lin, Hongyu and Yao, Junfeng and Zhang, Min},
  booktitle={Proceedings of the 53rd Annual Meeting of the Association for Computational Linguistics and the 7th International Joint Conference on Natural Language Processing (Volume 1: Long Papers)},
  pages={229--238},
  year={2015}
}

@article{kuang2017modeling,
  title={Modeling coherence for neural machine translation with dynamic and topic caches},
  author={Kuang, Shaohui and Xiong, Deyi and Luo, Weihua and Zhou, Guodong},
  journal={arXiv preprint arXiv:1711.11221},
  year={2017}
}

@misc{aihubportal,
  author       = {{National Information Society Agency (NIA)}},
  title        = {AI Hub (Portal)},
  howpublished = {\url{https://www.aihub.or.kr/}},
  year         = {2025},
}

@misc{googlebleueval,
  author       = {{Google Cloud}},
  title        = {Evaluate models (AutoML Translation Advanced) — Cloud Translation},
  howpublished = {\url{https://cloud.google.com/translate/docs/advanced/automl-evaluate}},
  year         = {2025},
}

@article{park2021empirical,
  title={Empirical analysis of Korean public AI hub parallel corpora and in-depth analysis using LIWC},
  author={Park, Chanjun and Shim, Midan and Eo, Sugyeong and Lee, Seolhwa and Seo, Jaehyung and Moon, Hyeonseok and Lim, Heuiseok},
  journal={arXiv preprint arXiv:2110.15023},
  year={2021}
}

@article{eo2022quak,
  title={QUAK: A synthetic quality estimation dataset for korean-english neural machine translation},
  author={Eo, Sugyeong and Park, Chanjun and Moon, Hyeonseok and Seo, Jaehyung and Kim, Gyeongmin and Lee, Jungseob and Lim, Heuiseok},
  journal={arXiv preprint arXiv:2209.15285},
  year={2022}
}

@book{comrie1989language,
  title={Language universals and linguistic typology: Syntax and morphology},
  author={Comrie, Bernard},
  year={1989},
  publisher={University of Chicago press}
}

@book{song2006korean,
  title={The Korean language: Structure, use and context},
  author={Song, Jae Jung},
  year={2006},
  publisher={Routledge}
}

@misc{unbabel2020wmt20cometqeda,
  author       = {Rei, Ricardo and Martins, Andr{\'e} and Unbabel},
  title        = {Unbabel/wmt20-comet-qe-da},
  year         = {2020},
  howpublished = {\url{https://huggingface.co/Unbabel/wmt20-comet-qe-da}},
  note         = {Hugging Face model checkpoint}
}

@inproceedings{martin2015more,
  title={More efficient topic modelling through a noun only approach},
  author={Martin, Fiona and Johnson, Mark},
  booktitle={Proceedings of the Australasian Language Technology Association Workshop 2015},
  pages={111--115},
  year={2015}
}

@inproceedings{park2014konlpy,
  title={KoNLPy: Korean natural language processing in Python},
  author={Park, Eunjeong L and Cho, Sungzoon},
  booktitle={Proceedings of the 26th annual conference on human \& cognitive language technology},
  volume={6},
  pages={133--136},
  year={2014},
  organization={Korean Institute of Information Scientists and Engineers, The Korean Society~…}
}

@book{vasiliev2020natural,
  title={Natural language processing with Python and spaCy: A practical introduction},
  author={Vasiliev, Yuli},
  year={2020},
  publisher={No Starch Press}
}

@article{savoy2013authorship,
  title={Authorship attribution based on a probabilistic topic model},
  author={Savoy, Jacques},
  journal={Information Processing \& Management},
  volume={49},
  number={1},
  pages={341--354},
  year={2013},
  publisher={Elsevier}
}

@article{landauer1998introduction,
  title={An introduction to latent semantic analysis},
  author={Landauer, Thomas K and Foltz, Peter W and Laham, Darrell},
  journal={Discourse processes},
  volume={25},
  number={2-3},
  pages={259--284},
  year={1998},
  publisher={Taylor \& Francis}
}

@article{park2021klue,
  title={Klue: Korean language understanding evaluation},
  author={Park, Sungjoon and Moon, Jihyung and Kim, Sungdong and Cho, Won Ik and Han, Jiyoon and Park, Jangwon and Song, Chisung and Kim, Junseong and Song, Yongsook and Oh, Taehwan and others},
  journal={arXiv preprint arXiv:2105.09680},
  year={2021}
}

@article{reimers2019sentence,
  title={Sentence-bert: Sentence embeddings using siamese bert-networks},
  author={Reimers, Nils and Gurevych, Iryna},
  journal={arXiv preprint arXiv:1908.10084},
  year={2019}
}

@inproceedings{niekler2012matching,
  title={Matching results of latent dirichlet allocation for text},
  author={Niekler, Andreas and J{\"a}hnichen, Patrick},
  booktitle={Proceedings of ICCM},
  pages={317--322},
  year={2012}
}

@article{reber2019overcoming,
  title={Overcoming language barriers: Assessing the potential of machine translation and topic modeling for the comparative analysis of multilingual text corpora},
  author={Reber, Ueli},
  journal={Communication methods and measures},
  volume={13},
  number={2},
  pages={102--125},
  year={2019},
  publisher={Taylor \& Francis}
}

@misc{lee2025mendeley,
  author       = {Lee, Hyeokmin and Kim, Youngkyu and Yoo, Byounghyun},
  title        = {Domain-specific Topic Modeling Configurations for Explainable Document-level Translation Evaluation},
  year         = {2025},
  howpublished = {Mendeley Data, V2},
  doi          = {10.17632/k65v3vxwhb.2},
}

@inproceedings{satopaa2011finding,
  title={Finding a" kneedle" in a haystack: Detecting knee points in system behavior},
  author={Satopaa, Ville and Albrecht, Jeannie and Irwin, David and Raghavan, Barath},
  booktitle={2011 31st international conference on distributed computing systems workshops},
  pages={166--171},
  year={2011},
  organization={IEEE}
}

@book{kuo2023handbook,
  title={The Handbook of NLP with Gensim: Leverage topic modeling to uncover hidden patterns, themes, and valuable insights within textual data},
  author={Kuo, Chris},
  year={2023},
  publisher={Packt Publishing Ltd}
}

@inproceedings{newman2010automatic,
  title={Automatic evaluation of topic coherence},
  author={Newman, David and Lau, Jey Han and Grieser, Karl and Baldwin, Timothy},
  booktitle={Human language technologies: The 2010 annual conference of the North American chapter of the association for computational linguistics},
  pages={100--108},
  year={2010}
}

\appendix
\section{Domain-specific dataset links}
\label{appendix:a}

The following table lists the official AI-Hub download pages for each Korean--English parallel corpus used in this study:

\begin{flushleft}
\begin{itemize}
  \item Basic science:\\ \url{https://www.aihub.or.kr/aihubdata/data/view.do?dataSetSn=71496}
  \item Broadcast content:\\ \url{https://www.aihub.or.kr/aihubdata/data/view.do?dataSetSn=71382}
  \item Conversation:\\ \url{https://www.aihub.or.kr/aihubdata/data/view.do?dataSetSn=126}
  \item Daily life:\\ \url{https://www.aihub.or.kr/aihubdata/data/view.do?dataSetSn=71265}
  \item Humanities:\\ \url{https://www.aihub.or.kr/aihubdata/data/view.do?dataSetSn=71498}
  \item Korean culture:\\ \url{https://www.aihub.or.kr/aihubdata/data/view.do?dataSetSn=126}
  \item Municipal Websites:\\ \url{https://www.aihub.or.kr/aihubdata/data/view.do?dataSetSn=126}
  \item News:\\ \url{https://www.aihub.or.kr/aihubdata/data/view.do?dataSetSn=126}
  \item Professional Field:\\ \url{https://www.aihub.or.kr/aihubdata/data/view.do?dataSetSn=111}
  \item Regulations:\\ \url{https://www.aihub.or.kr/aihubdata/data/view.do?dataSetSn=126}
  \item Social science:\\ \url{https://www.aihub.or.kr/aihubdata/data/view.do?dataSetSn=125}
  \item Spoken:\\ \url{https://www.aihub.or.kr/aihubdata/data/view.do?dataSetSn=126}
  \item Technical science1:\\ \url{https://www.aihub.or.kr/aihubdata/data/view.do?dataSetSn=124}
  \item Technical science2:\\ \url{https://www.aihub.or.kr/aihubdata/data/view.do?dataSetSn=71266}
\end{itemize}
\end{flushleft}
\end{document}